\documentclass{aa}  

\usepackage{graphicx}
\usepackage{txfonts}
\usepackage[T1]{fontenc}
\usepackage{siunitx}
\usepackage{hyperref}
\usepackage{wrapfig}
\usepackage{xcolor}

\usepackage{subfigure}
\usepackage{subcaption}
\usepackage{soul}
\setstcolor{red}

\begin{document}

\title{Hierarchical Bayesian inference with compositional score modeling for stellar streams}

\author{Giuseppe Viterbo \inst{1, 2} \and
    Jonas Arruda \inst{3,4} \and
    Tobias Buck \inst{1,2}
}

\institute{Universität Heidelberg, Interdisziplinäres Zentrum für Wissenschaftliches Rechnen (IWR), Im Neuenheimer Feld 205, 69120 Heidelberg,
    Germany\\
    \email{giuseppe.viterbo@iwr.uni-heidelberg.de}
    \and
    Universität Heidelberg, Zentrum für Astronomie, Institut für Theoretische Astrophysik, Albert-Ueberle-Straße 2, D-69120 Heidelberg, Germany\\
    \email{tobias.buck@iwr.uni-heidelberg.de}
    \and
    University of Bonn, Bonn Center for Mathematical Life Sciences (BCML), Endenicher Allee 64, D-53115 Bonn, Germany\\
    \and
    University of Bonn, Life \& Medical Sciences Institute (LIMES), Carl-Troll-Straße 31, D-53115 Bonn, Germany\\
    \email{jonas.arruda@uni-bonn.de}
}

\date{Received Month, XXXX; accepted Month Day, XXXX}

\abstract
{Stellar streams trace the Milky Way's gravitational potential over a wide range of Galactocentric radii. Since different streams sample different regions of the Galaxy, combining several of them can constrain the global mass distribution more tightly than modeling any single stream in isolation. Most of the existing multi-stream analyses rely on likelihood-based methods that require a new inference run whenever additional streams or kinematic measurements become available.}
{We aim to infer the Milky Way potential from multiple stellar streams combined with an additional constraint through the Galactic circular velocity curve within a single hierarchical framework.}
{We model the problem hierarchically, separating parameters that are common to all streams from parameters that are specific to each progenitor. 
We train score-based neural posterior estimators on a library of simulated streams and combine information from different streams through compositional score modeling as a post-training step.
}
{Tests on independent simulations show that the inferred posteriors are well calibrated and accurate. Combining several streams reduces the uncertainties on the global potential parameters relative to single-stream analyses. Applied to Gaia data, the model favours a mildly oblate dark matter halo with axis ratio $q_{\mathrm{NFW}}=0.77$, scale radius $a_{\mathrm{NFW}}=9.3$ kpc, a disk mass of $4.3\times10^{10}\,M_\odot$, and a local dark matter density $\rho_{\mathrm{NFW},\odot}=0.01153 M_\odot\,\mathrm{pc}^{-3}$ consistent with recent stream-based studies. The inferred virial mass is lower than estimates based on streams reaching larger Galactocentric distances.
}
{This hierarchical framework provides a practical way to combine information from multiple stellar streams without repeating the full inference procedure for each new dataset. The method is adaptable to new datasets, like future Gaia DR4, or new spectroscopic surveys, with minimal computational cost.}

\maketitle
\nolinenumbers

\section{Introduction}
\label{sec:intro}

The mass distribution of the Milky Way (MW), and in particular the 
density profile of its dark matter halo, records the assembly history of the
Galaxy and provides a local test of the cold dark matter paradigm. Most of the
Galactic mass is dark and recovering the three-dimensional distribution of matter is possible only through its gravitational effect \citep{Green2023}. Several classes of dynamical tracers
have been used for this task. The circular velocity curve constrains the
enclosed mass over a wide radial range
\citep{Huang2016, Eilers2019, Zhou_2023}; equilibrium models of halo stars and
globular clusters constrain the force field under the assumption of a
stationary tracer population \citep{Wegg2019, Callingham2019}; and global
parametric models combine many such constraints into a single mass model
\citep{McMillan2017, Cautun_2020}. Each of these tracers samples a limited
region of the Galaxy or rests on an equilibrium assumption that the outer halo
does not satisfy.

Stellar streams are among the most sensitive probes of the potential in the
stellar halo \citep{BonacaHogg2018}. A stream forms when tidal forces strip
stars from a globular cluster or a dwarf galaxy. The stripped stars escape
near the inner and outer Lagrange points of the progenitor with a small spread
in energy and angular momentum, and they spread into a leading and a trailing
arm on orbits close to that of the progenitor \citep{PriceWhelan2014}. A
kinematically cold stream therefore traces an extended portion of an orbit, and
both its track on the sky and its velocities are set by the acceleration
field it has moved through. Streams can sample the potential away from the disk
plane and over tens of kiloparsecs, where few other tracers are available.

The methods that turn this sensitivity into a measurement of the density distribution fall
into a few families. The earliest fit a single orbit to the stream track
\citep{Koposov2010, Malhan2019}, which is computationally light but biased,
since the stars of a stream do not all share one orbit. Generative models
instead reproduce the full debris distribution, either by spraying test
particles from the progenitor's Lagrange points
\citep{Bonaca2014, Bowden2015, Kuepper2015} or by describing the debris
analytically in action-angle coordinates \citep{Bovy2014, Bovy2016}. More
recent work reconstructs the local acceleration field directly from the stream
track without assuming a functional form for the potential
\citep{Nibauer2022, NibauerBonaca2025}, or by using a likelihood-free approach
\citep{Viterbo_2026}.

Applied to individual streams, these methods have returned a range of values
for the inner halo. The Pal~5 and GD-1 streams favour a nearly spherical or
mildly flattened halo \citep{Koposov2010, Bovy2016, Malhan2019, Reino2021}; the
Sagittarius stream has been read as evidence for a strongly triaxial halo
\citep{LawMajewski2010} and used to weigh the Galaxy \citep{Gibbons2014}, but
accounting for the pull of the infalling Large Magellanic Cloud, which also
bends other halo streams \citep{Erkal2019}, weakened the triaxiality conclusion
\citep{Vasiliev2021} in favour of a radially-varying shape and orientation of the Galactic halo \citep[see also][]{Zhu2026}.  Part of this
spread has a structural origin: a single cold stream constrains mainly the
shape of the potential and the enclosed mass near its own orbit, so the mass
inferred from one stream can be biased when the halo departs from a smooth,
static, analytic form \citep{Bonaca2014, BonacaHogg2018}.

Combining streams that sample different Galactocentric radii reduces these
biases and constrains the global mass distribution more tightly than any single
stream \citep{BonacaHogg2018, Reino2021}. The number of known streams has grown
into the hundreds \citep{Mateu2023}, and global mass models fit to large stream
samples are now available \citep{Ibata_2024}. These multi-stream analyses are
predominantly likelihood-based, and they explore the joint posterior with 
Markov Chain Monte Carlo (MCMC) runs \citep{Palau_2023, Ibata_2024} whose cost is paid again whenever a stream is added or new kinematic measurements arrive.

We aim to provide an alternative using simulation-based inference (SBI) \citep{Cranmer2020, deistler2025simulation-based}. After an upfront
simulation and training cost, an amortized posterior estimator returns the
posterior for any new observation through forward passes of a neural network,
with no further model simulation.
Moreover, diffusion-based SBI additionally permits post hoc modification and composition of posterior estimates at inference time \citep{Arruda_2025}.
SBI has recently gained traction in astrophysics and has been employed to a range of inverse problems, including galactic chemical evolution \citep{Guenes2025,buck_2025,Viterbo_2024}, galactic archaeology \citep{CASBI_2024, Sante2026}, protoplanetary disk \citep{Ruzza2026}, gravitational-wave astronomy \citep{Dax_2025}, cosmology \citep[e.g.][]{Lemos2023}, galaxy evolution \citep{Lovell2025}, galactic dynamics \citep{Viterbo_2026}, binary star inference \citep{Knoell2026} or supernova inference \citep{SN_SBI}.

Compositional score modeling
\citep{Geffner2022, linhart2026diffusion, gloeckler2025compositional, Compositional_arrruda_2025} extends this idea to several observations, leveraging the additive properties of reverse diffusion scores that are summed at inference time to guide the posterior sampling. 
This matches the structure of a population of streams that move in one common
Galactic potential. 

In this work we infer the MW potential from observations of three
stellar streams together with an additional independent constraint given by the circular velocity curve within a single
hierarchical, amortized SBI framework. We separate the global potential
parameters, shared by all streams, from the progenitor parameters specific to
each stream, and we aggregate the streams at inference time through
compositional score modeling. We apply the method to the cold globular cluster stellar streams Pal~5, NGC~3201, and M~68
together with the circular velocity curve of \citet{Zhou_2023}, the same three
streams modelled by \citet{Palau_2023}. The analysis builds on
\citet{Viterbo_2026}, where we introduced a single-stream SBI treatment of GD-1; here we move from one stream to a
population, and from a single posterior to their compositional combination.
Section~\ref{sec:method} describes the potential model, the simulator, the
observational model, and the hierarchical inference; Section~\ref{sec:results}
validates the posteriors on simulations and applies the method to the Gaia
data; Section~\ref{sec:Discussion}  compares with literature results and discusses limitations and future prospects of our method. Finally,~\ref{sec:Conclusion} presents our conclusion and summarizes our findings.

\section{Method}\label{sec:method}

As described in Sec.~\ref{sec:intro}, dynamically cold stellar streams
naturally record in their morphology the gravitational field that
stripped their stars from the globular cluster progenitor. Aggregating the information carried by several such structures constrains the common underlying potential more tightly than any single stream can \citep{Palau_2023, Ibata_2024}. Both of
these analyses, however, rely on dedicated MCMC runs that, whenever new data
arrive, whether in the form of new detected stream or new star phase-space components
(e.g., line-of-sight velocities), must be re-run from scratch to update the posterior.
In addition, \citet{Ibata_2024} carries a high computational budget,
driven by its restricted $N$-body modeling, while \citet{Palau_2023} is limited in the
assumption that each stream can be approximated by the orbit of its
progenitor.

In this work, we propose an amortized approach, i.e. after an upfront training cost,
the posterior for any new observation is obtained by evaluating a neural posterior estimator, without further model
simulations. We model the potential reconstruction
problem hierarchically, splitting the parameters into two levels. The
\emph{global} level holds the Milky Way potential parameters
$\boldsymbol{\eta}_{\mathrm{MW}}$, shared by all observables; the
\emph{local} level holds the progenitor present day phase-space parameters
$\boldsymbol{\theta}_j$ of each stream $j \in \{ 1, 2, 3 \}$, respectively  Pal~5, NGC~3201, and M~68. We build on recent work on compositional score modeling for hierarchical models
\citep{Compositional_arrruda_2025} to aggregate the information on
multiple streams at inference time.

We first define the parametric Milky Way potential model
(Sec.~\ref{sec:potential}) and the progenitor model of the three
globular cluster streams with their local parameters (Sec.~\ref{sec:progenitor}). We
then describe the simulator (Sec.~\ref{sec:forward_sim}) and
the observational model that forward models the training set to Gaia-like data,
producing the stream observables $\mathbf{y}_{\mathrm{stream}}^{j}$
(Sec.~\ref{sec:obs_model}); the circular velocity curve
$\mathbf{y}_{\mathrm{rc}}$, used as a complementary constraint
throughout, is introduced in Sec.~\ref{sec:rotcurve}. Finally, we
formalize the hierarchical Bayesian model (Sec.~\ref{sec:hierarchical})
and present the score-based inference machinery
(Sec.~\ref{sec:diffusion}) together with its compositional aggregation
across streams (Sec.~\ref{sec:composition}), and describe the training pipeline in Sec~\ref{sec:training}.

\subsection{Milky Way potential model}
\label{sec:potential}

We model the Milky Way's gravitational potential with three components:
a dark matter halo, a disk (combining the gas, the thin and the thick components), and a central bulge.

\paragraph{The halo.}
We adopt a generalized Navarro--Frenk--White (NFW) profile \citep{Jing2002}, with an
axisymmetric flattening applied to the density distribution:
\begin{equation}
    \rho_{\mathrm{h}}(s) \;=\; \rho_{\mathrm{NFW}}
    \left(\frac{s}{a_{\mathrm{NFW}}}\right)^{-\gamma_{\mathrm{NFW}}}
    \!\left(1 + \frac{s}{a_{\mathrm{NFW}}}\right)^{\gamma_{\mathrm{NFW}}-3},
    \label{eq:halo}
\end{equation}
where $s^2 = R^2 + (z / q_{\mathrm{NFW}})^2$ is the ellipsoidal radius in
cylindrical coordinates $(R, z)$, $a_{\mathrm{NFW}}$ is the scale radius,
$\gamma_{\mathrm{NFW}}$ is the inner logarithmic slope (a
standard NFW profile corresponds to $\gamma_{\mathrm{NFW}} = 1$), 
$q_{\mathrm{NFW}}$ is the minor-to-major axis ratio of the
iso-density ellipsoids, and the normalization
$\rho_{\mathrm{NFW}}$ sets the characteristic density. All four halos
parameters $(\rho_{\mathrm{NFW}}, \gamma_{\mathrm{NFW}}, a_{\mathrm{NFW}},
    q_{\mathrm{NFW}})$ are treated as free parameters and will be inferred from data.

\paragraph{The disk.}
The disk is modelled as an exponential density profile:
\begin{equation}
    \rho_D(R, z) \;=\; \frac{\Sigma_D}{2\,z_D}\,
    \exp\left(-\frac{R}{r_D} - \frac{|z|}{z_D}\right),
    \label{eq:disk}
\end{equation}
where $\Sigma_D$ is the central surface mass density, $r_D$ is the radial
scale length, and $z_D$ is the vertical scale height. The total disk mass
follows as $M_D = 2\pi \Sigma_D r_D^2$. 
We modelled the axisymmetric components of the Milky Way (thin disk, thick disk, and gas disk) with a single exponential profile. In early experiments, we tested a multi-disk component model, but the inference left many parameters poorly constrained, with posteriors that did not narrow appreciably around them. This is likely because stellar streams do not provide enough leverage to break the degeneracies between multiple axisymmetric components. 


\paragraph{The bulge.}
The bulge is represented by an axisymmetric spheroidal density profile
with an exponential cut-off, following \cite{Cautun_2020}, with all of
its parameters fixed to the best-fitting values of that work. Since the
three streams considered here orbit at Galactocentric distances of
$\sim$$5$--$20\,\mathrm{kpc}$, they carry little information on the
    innermost few kiloparsecs of the Galaxy, and we therefore do not attempt
    to constrain the bulge.


    \paragraph{Global parameter vector.}
    The free parameters of the MW potential are collected into the global
    vector
    \begin{equation}
        \boldsymbol{\eta}_{\mathrm{MW}} = \bigl\{\rho_{\mathrm{NFW}},\;
        \gamma_{\mathrm{NFW}},\; a_{\mathrm{NFW}},\; q_{\mathrm{NFW}},\;
        r_D,\; z_D,\; \Sigma_D\; \bigr\}.
        \label{eq:global_params}
    \end{equation}
    The priors adopted for each component of
$\boldsymbol{\eta}_{\mathrm{MW}}$ are listed in
    Table~\ref{tab:priors_global}. This vector constitutes the global level
    of the hierarchical model of Sec.~\ref{sec:hierarchical} and it is shared
    by all three streams and by the circular velocity curve.

    \begin{table}
        \caption{Priors on the global MW potential parameters
        $\boldsymbol{\eta}_{\mathrm{MW}}$ (Eq.~\ref{eq:global_params}).}
        \label{tab:priors_global}
        \centering
        \begin{tabular}{l l l}
            \hline\hline
            Parameter               & Unit                         & Prior                                 \\
            \hline
            $\rho_{\mathrm{NFW}}$   & $\mathrm{M_\odot\,kpc^{-3}}$ & $\mathcal{U}(10^6,\; 1.5 \cdot 10^8)$ \\
            $\gamma_{\mathrm{NFW}}$ & ---                          & $\mathcal{U}(-2,\; 2)$                \\
            $a_{\mathrm{NFW}}$      & kpc                          & $\mathcal{U}(1,\;30)$                 \\
            $q_{\mathrm{NFW}}$      & ---                          & $\mathcal{U}(0.5,\; 1.5)$             \\
            $r_D$                   & kpc                          & $\mathcal{N}(2.6,\;0.5)$              \\
            $z_D$                   & kpc                          & $\mathcal{N}(0.3,\;0.05)$             \\
            $\Sigma_D$              & $\mathrm{M_\odot\,pc^{-2}}$  & $\mathcal{U}(10^7,\;1.5 \cdot 10^9)$  \\
            \hline
        \end{tabular}
    \end{table}


    \subsection{Progenitor model and local parameters}
    \label{sec:progenitor}

    We analyse the streams of the globular clusters Pal~5, NGC~3201, and
    M~68. Each of these stream
    originates from the ongoing tidal disruption of a globular cluster whose
    present-day six-dimensional phase-space position is observable. Although the
    right ascension $\alpha^{j}$ and declination $\delta^{j}$ are well
    constrained, several
    kinematic properties of the progenitor carry significant
    observational uncertainties and must therefore be treated as free
    parameters of the inference. The stream-specific local parameter vector
    is
    \begin{equation}
        \boldsymbol{\theta}_j \;=\;
        \bigl\{\,V_R^{j},\; R^{j},\;
        \mu_{\alpha*}^{j},\; \mu_\delta^{j}\,\bigr\},
        \label{eq:local_params}
    \end{equation}
    collecting the present-day heliocentric line-of-sight velocity ($V_R^{j}$),
    heliocentric distance ($R^{j}$), and equatorial proper motion ($\mu_{\alpha*}^{j}, \mu_\delta^{j}$) of
    progenitor $j$ . Although the sky position and the proper motions are
    part of the Gaia astrometry, the line-of-sight velocity $V_R^{j}$ can
    only be measured through spectroscopic observations, which are generally
    expensive. These four quantities govern where the progenitor sits in
    phase-space at the moment of observation, and therefore which orbit it
    traces backwards in time (Sec.~\ref{sec:forward_sim}).
    We assign independent Gaussian priors on each component, reported in Table~\ref{tab:priors_local}, adopting the
    means and standard deviations compiled in Table~1 of \citet{Palau_2023}. 
    
    

    \begin{table}
        \caption{Stream-specific Gaussian mean and standard deviations of the priors $p(\boldsymbol{\theta}_j \mid j)$ on the local progenitor
            parameters $\boldsymbol{\theta}_j$ (Eq.~\ref{eq:local_params}). Line-of-sight velocities are from
            \citet{Baumgardt2019}, distances from \citet{Harris1996,Harris2010} and 
            \citet{PriceWhelan2019} for Pal~5, and proper motions from
            \citet{Vasiliev2019b}.}
        \label{tab:priors_local}
        \centering
        \small
        \setlength{\tabcolsep}{4pt}
        \begin{tabular}{l c c c}
            \hline\hline
            Parameter                                 & Pal~5              & NGC~3201           & M~68                \\
            \hline
            $V_R$ [$\mathrm{km\,s^{-1}}$]             & $-58.6 \pm 0.21$   & $494.34 \pm 0.14$  & $-92.99 \pm 0.22$  \\
            $R$ [kpc]                                 & $20.6 \pm 0.2$     & $4.9 \pm 0.11$     & $10.3 \pm 0.52$    \\
            $\mu_{\alpha*}$ [$\mathrm{mas\,yr^{-1}}$] & $-2.736 \pm 0.064$ & $8.324 \pm 0.044$  & $-2.752 \pm 0.054$ \\
            $\mu_\delta$ [$\mathrm{mas\,yr^{-1}}$]    & $-2.646 \pm 0.064$ & $-1.991 \pm 0.044$ & $1.762 \pm 0.053$  \\
            \hline
        \end{tabular}
    \end{table}

    The progenitor potential is modeled as a Plummer sphere characterized by its total
    mass $M_j$ and scale radius $a_j$
    \citep{Plummer1911}. The gravitational potential of the Plummer sphere,
    \begin{equation}
        \Phi_j(r) \;=\;
        -\frac{G M_{j}}{\sqrt{r^2 + a_{j}^2}},
        \label{eq:plummer}
    \end{equation}
    sets both the escape speed profile, hence the energy distribution of
    stripped stars and the spatial extent of the initial conditions. The two
    progenitor structure parameters, $(M_j, a_j)$, are fixed per stream to the values inferred
    from the observed cluster properties \citep{Harris1996, Sollima2017};
    they enter the simulation as fixed auxiliary inputs rather than as part
    of $\boldsymbol{\theta}_j$.

    \begin{figure}
        \centering
        \includegraphics[width=0.99\linewidth]{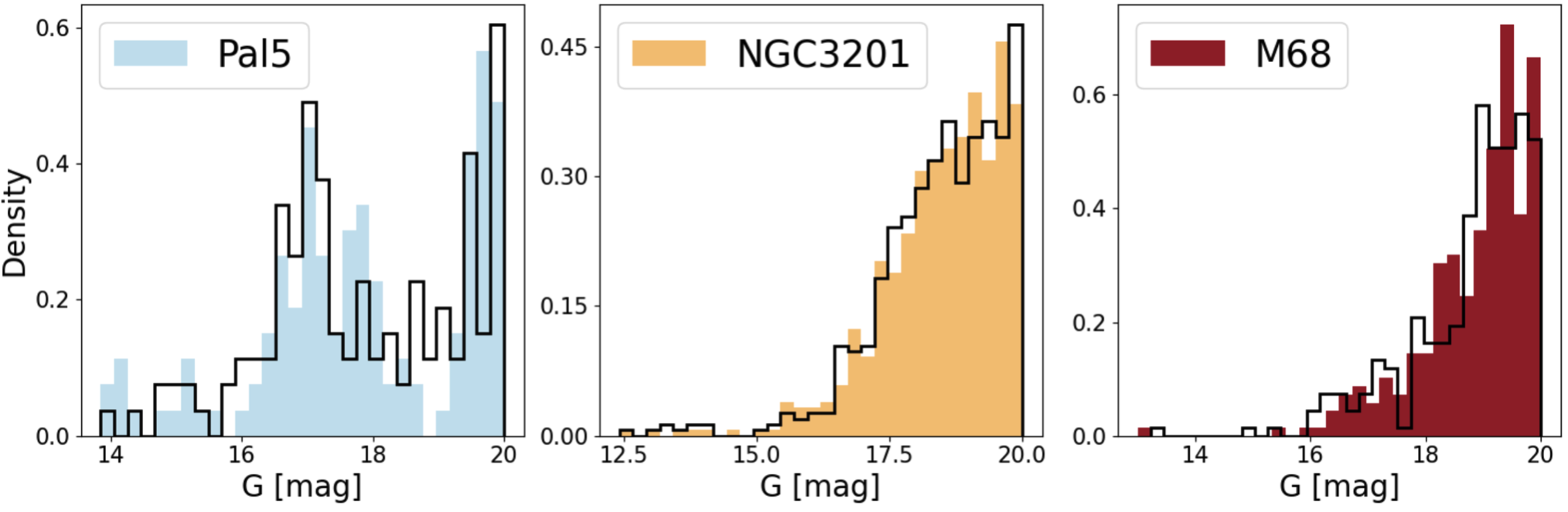}
        \caption{Magnitude distribution in the Gaia G-band of the observed stars from the three streams. The black line shows samples obtained by fitting a KDE to the data.}
        \label{fig:magnitude}
    \end{figure}

    \subsection{Forward simulation:\textsc{Agama}}
    \label{sec:forward_sim}

    Simulation-based inference does not require an explicit likelihood instead it
    only requires a stochastic simulator that maps a parameter draw into a
    synthetic observation. Our specific observational model is described in detail in
    Sec.~\ref{sec:obs_model}, while for the dynamical part we use \textsc{Agama}
    \citep{Agama_vasiliev}.  For a given draw
$(\boldsymbol{\eta}_{\mathrm{MW}}, \boldsymbol{\theta}_j)$
    \textsc{Agama} instantiates the total MW potential of
    Sec.~\ref{sec:potential}, converting the parametrized density profiles
    into a potential through spherical-harmonic expansion, and then generating a synthetic stream snapshot in two steps.

    \paragraph{Progenitor orbit integration.}
    The present-day phase-space position of progenitor $j$ is specified by
$\boldsymbol{\theta}_j$, together with the fixed, stream-dependent right
    ascension and declination $\alpha^{j}, \delta^{j}$ taken from
    \citet{Palau_2023}. We first transform on sky positions and velocities to Galactocentric Cartesian
    coordinates $(\mathbf{x}_f^{j}, \mathbf{v}_f^{j})$, then we 
    integrate the progenitor orbit backwards in the MW potential for a total
    integration time $t_{\mathrm{end}}^{j}$ and assigning a total mass $M_j$ and scale radius $a_j$\footnote{The values  $t_{\mathrm{end}}^{j}, M_j, a_j$ are taken from \citealt{Palau_2023}.} to recover the initial phase-space position
$(\mathbf{x}_0^{j}, \mathbf{v}_0^{j})$ from which the stream formed.

    \paragraph{Particle spray.}
    Starting from $(\mathbf{x}_0^{j}, \mathbf{v}_0^{j})$, we populate
    the stream by evolving $N = 1000$ tracer particles forward in time with
    the particle spray method of \citet{Chen_2025}, as implemented in
    \textsc{Agama}. In particle spray methods the stripped stars are not
    followed through a full $N$-body integration of the progenitor's
    self-gravity while still bound; instead, they are ejected near the two Lagrange points of
    the cluster at a series of discrete time steps along the progenitor
    orbit, and subsequently evolved as test particles in the combined
    potential of the MW and of the orbiting progenitor. This choice is a deliberate trade-off with respect to our previous
    work. In Paper~I \citep{Viterbo_2026}, we used the
    differentiable $N$-body integrator \textsc{Odisseo} \citep{Viterbo_2025}
    as the forward simulator, which self-consistently resolves close encounters
    and the coupling between the cluster's internal dynamics and its tidal
    disruption. In this work, however, we extend the scope of the inference to
    exploit the hierarchical nature of the problem via partial pooling
    (Sec.~\ref{sec:hierarchical}), and we decided to extend our 
    simulation budget to $10^6$ simulations equally divided among the three
    streams, a scale at which the computationally cheaper particle
    spray approach becomes the natural choice, letting us establish the methodology without being limited by data volume.
\begin{figure}
        \centering
        \vspace{-2em}
        \includegraphics[width=0.9\linewidth]{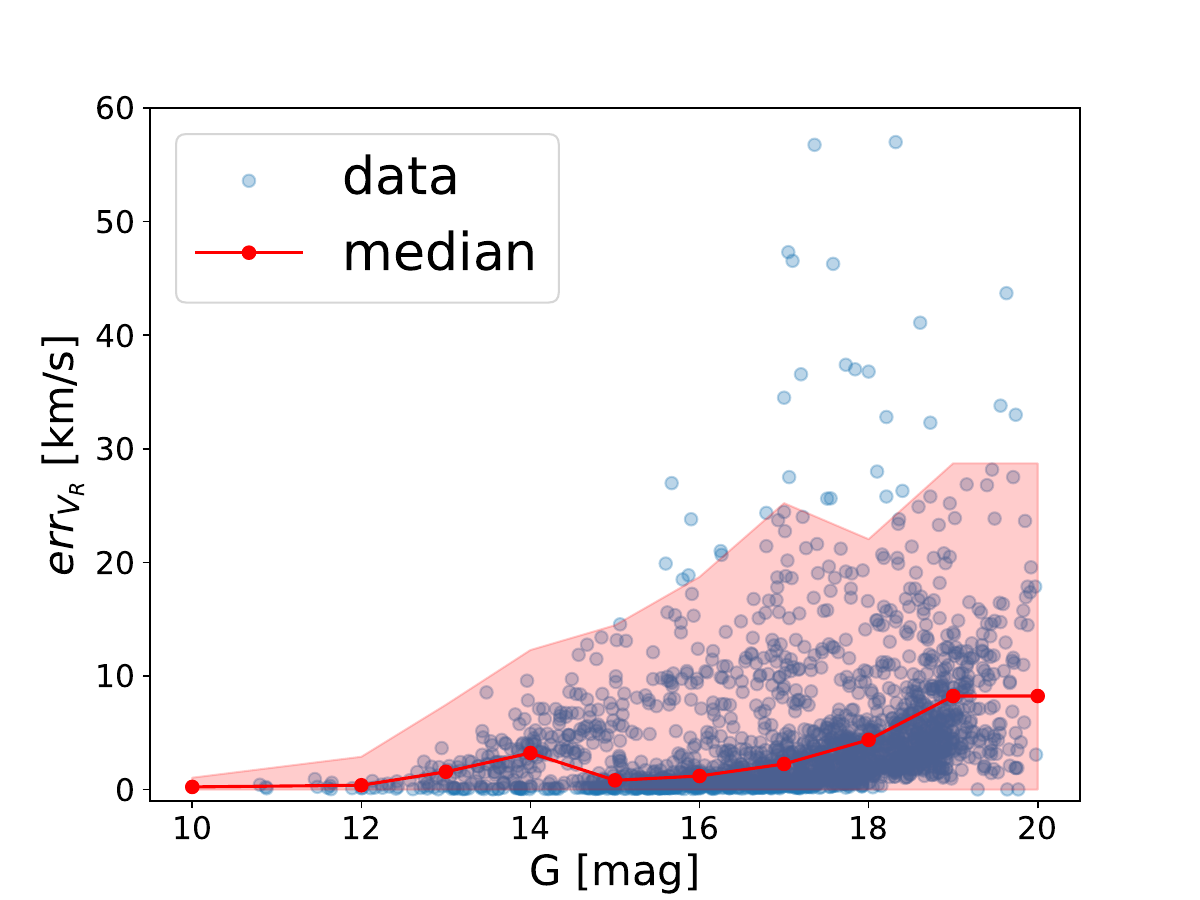}        \caption{Empirical magnitude-dependent line-of-sight velocity error function. After binning the data from \citet{Ibata_2024} into ten bins of $G$-band magnitude, we compute the median and standard deviation of the reported $\sigma_{V_{\mathrm{R}}}$ in each bin, which is used to assign magnitude-dependent line-of-sight velocity uncertainties to the simulated stars. The red error band displays 3 standard deviations.}
        \label{fig:vR_errorfunction_magnitude}
    \end{figure}
    The complete simulation pipeline for a single training instance therefore is: 
    \begin{enumerate}
        \item Sample $(\boldsymbol{\eta}_{\mathrm{MW}}, \boldsymbol{\theta}_j)$ from the prior.
        \item Instantiate the \textsc{Agama} potential.
        \item Integrate the progenitor orbit backwards in time.
        \item Spray the $N$ tracer particles along the orbit.
        \item Evolve the $N$ particles forward in time.
    \end{enumerate}
    The resulting configuration is then passed through the observational model that we are going to describe in the next Section, which turns it into the
    synthetic observable $\mathbf{y}_{\mathrm{stream}}^{j}$.

    \subsection{Observational model}
    \label{sec:obs_model}

    The simulated streams are processed through a forward observational
    model to produce synthetic data directly comparable to the Gaia DR3
    catalogues. We first convert the Galactocentric Cartesian positions and
    velocities of the tracer particles back into the equatorial observables
$(\alpha, \delta, \pi, \mu_{\alpha*}, \mu_\delta, V_{\mathrm{R}})$, respectively
    right ascension, declination, parallax, the two proper motions
    components, and the line-of-sight velocity, using the standard
    \texttt{astropy} coordinate transformations \citep{astropy_2022}.

    \begin{figure*}
        \centering
        \includegraphics[width=0.9\linewidth]{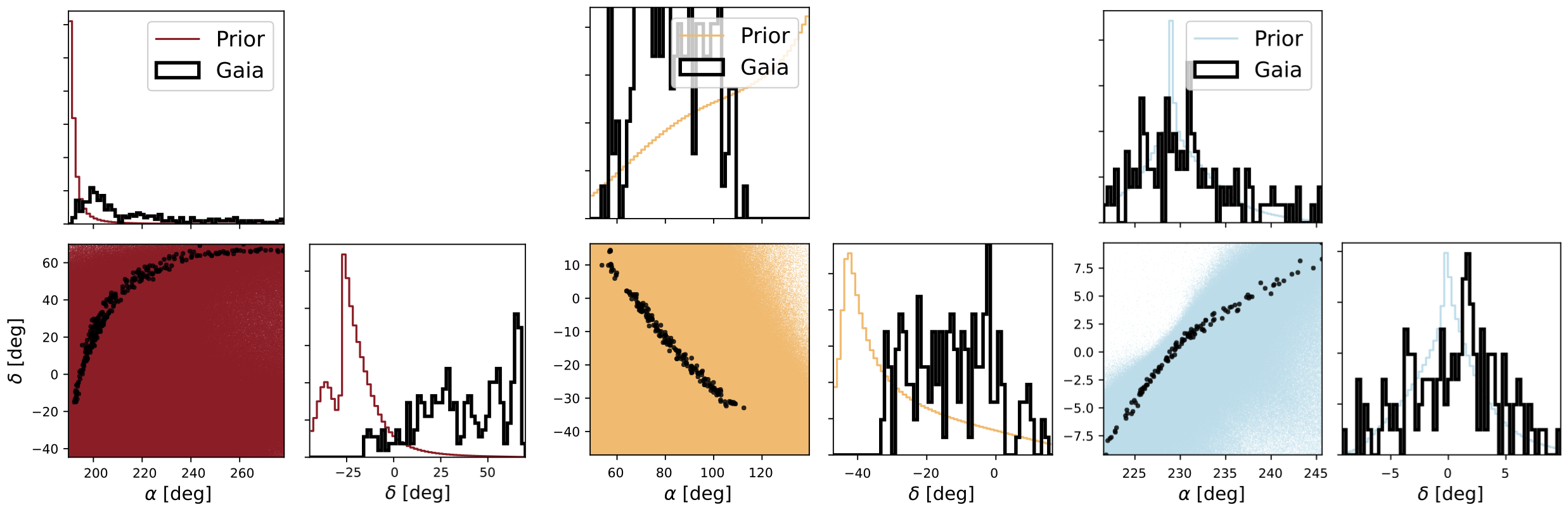}
        \caption{Prior predictive checks for the three streams M~68 (left panel) NGC~3201(centre panel) and Pal~5(right panel): synthetic
            observables generated from the forward modelled prior samples (coloured)
            compared to the observed Gaia data (black). The stream corner plots on the full phase space are shown in Appendix~\ref{sec:appendix_predictive}.}
        \label{fig:prior_predictive_checks_combined}
    \end{figure*}
    
    \subsubsection{Magnitude dependent noise}\label{sec:magnitude_dep_noise}
    Gaia astrometry, and the spectroscopic surveys that complement it, are
    affected by observational uncertainties that depend strongly on the apparent
    magnitude. Simulated stream stars are therefore first assigned $G$-band
    magnitudes, drawn from a kernel density estimate (KDE) of the empirical
    magnitude distribution of each observed stream in \citet{Ibata_2024}
    (Fig.~\ref{fig:magnitude}). For the line-of-sight velocity we construct
    an empirical, magnitude-dependent uncertainty model
$\sigma_{V_{\mathrm{R}}}(G)$ by computing the median and standard
    deviation of the reported $\sigma_{V_{\mathrm{R}}}$ in ten bins of
$G$-band magnitude over the $\sim$24\,000 stars of the
    \citet{Ibata_2024} catalogue; with this empirical approach we do not
    model each spectroscopic survey separately, but rather an ensemble of
    plausible error values, leaving survey-specific error handling to future
    work. The resulting radial velocity error function is shown in
    Fig.~\ref{fig:vR_errorfunction_magnitude}. 
     For the magnitude-dependent
    uncertainties of the astrometric components on the other hand we interpolate over the value  of \citet[][Table~4]{GaiaDR3} 
    . Once a magnitude $G_i$ is assigned to star
$i$, Gaussian noise is added to each of its phase-space components, with
    standard deviation given by the corresponding magnitude-dependent error
    functions evaluated at $G_i$. 

    \subsubsection{Selection effects and missing line-of-sight velocity}\label{sec:selection_effects}
    Real stellar stream catalogues are subject to two distinct,
    stream-dependent selection effects that the forward model must reproduce
    faithfully for the trained networks to be applicable to actual Gaia data \citep{SN_SBI, arruda2026overcoming}. 
    First, each observed stream occupies a specific region of the sky,
    and only stars within that region can plausibly be assigned to it as
    members. Second, spectroscopic follow-up is far from complete: only a
    stream-dependent fraction $f_j$ of the identified member stars has a
    measured line-of-sight velocity $V_{\mathrm{R}}$. We reproduce both
    effects in the following steps:
    \begin{enumerate}
        \item Observational window: after the observational noise has been applied to all
              $N=1000$ tracer particles, we discard any particle that
              falls outside the bounding box
              $[\alpha_{\min}^{j}, \alpha_{\max}^{j}] \times
                  [\delta_{\min}^{j}, \delta_{\max}^{j}]$ defined by the extrema of the
              observed member-star catalogue of stream $j$.
              We then randomly sub-sample the remaining particles down to a pool of
              $N_{\mathrm{pool}} = 300$ stars. This fixed pool size is chosen to exceed
              the largest observed stream membership ($N_{\mathrm{obs}}^{j} \leq 300$
              for all three streams) while keeping the computational cost manageable.

        \item Attention mask: From the pool of 300 stars we draw, without replacement,     exactly $N_{\mathrm{obs}}^{j}$ indices to form the \emph{visible set} for that
              training instance, where $N_{\mathrm{obs}}^{j}$ is the number of member
              stars in the observed catalogue of stream $j$ (a fixed, stream-specific
              integer). The remaining $300 - N_{\mathrm{obs}}^{j}$ stars are masked from the attention heads of the \texttt{SetTransformer} encoder that embeds the stream (Sec.~\ref{sec:gfn} for the architectural details). This is possible because in \texttt{Transformer}-based architectures the attention scores of selected entries can be set to $-\infty$, effectively preventing the model from using the information carried by those particles \citep{lee_set_2019}. 
              Because the same network
              must handle all three streams whose observed memberships differ substantially,
              this procedure ensures that the \texttt{SetTransformer} always receives a set of the same cardinality as the real observation it will be evaluated
              on at inference time, while the stochastic choice of which pool stars enter
              the visible set acts as a data-augmentation mechanism that prevents the
              network from over-fitting to a fixed membership.
        \item Incomplete line-of-sight velocity: Within the visible set of $N_{\mathrm{obs}}^{j}$ stars, a stream-dependent fraction $f_j$ is designated as \emph{spectroscopically observed}. We randomly select $N_{\mathrm{los}}^{j} = f_j N_{\mathrm{obs}}^{j} $ stars to retain their simulated (and later
              noise-corrupted) line-of-sight velocity $V_{\mathrm{R},i}$, which
              carries a magnitude-dependent uncertainty $\sigma_{V_{\mathrm{R}},i}$, with $i \in \mathcal{S}_{\mathrm{los}}^{j}$, the index set of the
              spectroscopically observed stars.
              For each of the remaining $N_{\mathrm{obs}}^{j} - N_{\mathrm{los}}^{j}$
              stars that lack spectroscopic information, the line-of-sight velocity entry
              is replaced by the mean of the $N_{\mathrm{los}}^{j}$ available
              velocities and the associated uncertainty is set to the
              standard deviation of those available velocities:
              \begin{equation}
                  \begin{split}
                      V_{\mathrm{R},k \notin \mathcal{S}_{\mathrm{los}}^{j}}                  & = \bar{V}_{\mathrm{R}} =
                      \frac{1}{N_{\mathrm{los}}^{j}}
                      \sum_{i \in \mathcal{S}_{\mathrm{los}}^{j}} V_{\mathrm{R},i},                                       \\
                      \tilde{\sigma}_{V_{\mathrm{R}},k \notin \mathcal{S}_{\mathrm{los}}^{j}} & =
                      \sqrt{\frac{1}{N_{\mathrm{los}}^{j}} \sum_{i \in \mathcal{S}_{\mathrm{los}}^{j}} (V_{\mathrm{R},i} - \bar{V}_{\mathrm{R}})^2}
                      ,
                  \end{split}
                  \label{eq:vlos_handling}
              \end{equation}

              which is by construction independent of the $G$-band magnitude of star $k$
              and therefore decoupled from the astrometric noise model. This mean-filling
              strategy preserves the bulk kinematic information carried by the stream
              without introducing an artificial $V_{\mathrm{R}}$ placeholder. We report the values of $N^j_{\text{obs}}$ and $f_j$ in Tab. \ref{tab:observed_stars}.

        \begin{table}
        \centering
        \caption{Number of observed stars and percentage with line-of-sight velocity measurements for each stream.}
        \begin{tabular}{lccc}
            \hline
            & Pal~5 & NGC~3201 & M~68 \\
            \hline
            $N^j_{\rm obs}$ & 129 & 195 & 297 \\
            $f_j$ (\%) & 53.5 & 19.0 & 9.8 \\
            \hline
        \end{tabular}
        
        \label{tab:observed_stars}
        \end{table}

        \item Flagging:
              To allow the \texttt{SetTransformer} to distinguish between stars with genuine
              spectroscopic measurements and stars whose velocity entry is the imputed
              mean, every star $i$ in the visible set carries a binary flag:
              \begin{equation}
                  \zeta_i \;=\;
                  \begin{cases}
                      1 & \text{if } i \in \mathcal{S}_{\mathrm{los}}^{j}    \\
                      0 & \text{if } i \notin \mathcal{S}_{\mathrm{los}}^{j}
                  \end{cases}
                  \label{eq:epsilon}
              \end{equation}
              This flag is concatenated to the per-star feature vector alongside the other observables, so the encoder can learn to weight true measurements and incomplete entries appropriately without hard-coding any assumption about their relative information content.

    \end{enumerate}

    All the steps of Sects.~\ref{sec:magnitude_dep_noise} and \ref{sec:selection_effects} are applied independently at every batch during training. As a result, no two training instances present the same membership selection or the same $V_{\mathrm{R}}$
    assignment, even when they share the same underlying simulation. This
    stochastic data augmentation matches the structured incompleteness of stellar stream observations, and it forces
    the SBI pipeline to learn posterior representations that are robust to the
    precise membership realization rather than overfitting to a single catalogue.
    Hence, the full observable vector for stream $j$, presented to the \texttt{SetTransformer} encoder (Sec.~\ref{sec:gfn} for the specific architecture), is:
    \begin{equation}
        \begin{split}
            \mathbf{y}_{\mathrm{stream}}^{j} \;=\;
            \Bigl\{\,
             & \alpha_i,\;\delta_i,\;\pi_i,\;\mu_{\alpha*,i},\;\mu_{\delta,i},\;
            \;V_{\mathrm{R},i},\;                                                                                                                            \\
             & \sigma_{\alpha_i},\; \sigma_{\delta_i}, \; \sigma_{\pi_i}, \; \sigma_{\mu_{\alpha*_i}}, \;\sigma_{\mu_{\delta_i}}, \sigma_{V_{\mathrm{R}},i}, \\
             & \zeta_i, G_i, j
            \,\Bigr\}_{i=1}^{N_{pool}},
            \label{eq:observable_vector}
        \end{split}
    \end{equation}
    where we have extended the phase space to include also its magnitude-dependent error, following the adaptive noise-level strategy described in \citet{gebhard2024_atmosphere}.
    Of these $N_{pool}$ entries, exactly $N_{\mathrm{obs}}^{j}$ are visible
    to the encoder through the attention mask described in
    Sec.~\ref{sec:selection_effects}, matching the cardinality of the
    corresponding real Gaia catalogue to which the trained network is applied
    at inference time. 

    As a sanity check of the full forward model, we verify that the prior
    predictive distribution of the simulated observables brackets the observed
    member-star catalogues of the three streams in Fig.~\ref{fig:prior_predictive_checks_combined}.

    \subsection{Circular velocity constraint}
    \label{sec:rotcurve}

    In addition to the three stellar streams, we include the MW
    circular velocity curve as a complementary kinematic constraint. Each
    stream is mostly sensitive to the gravitational field along its own
    orbit, while the rotation curve instead pins the enclosed mass in the
    Galactic plane over a wide radial range. We use the
$N_{\mathrm{rc}} = 38$ measurements of luminous red giant stars from \citet{Zhou_2023}, which span
$R \in [5, 30]\,\mathrm{kpc}$. For any draw of
$\boldsymbol{\eta}_{\mathrm{MW}}$, the predicted circular velocity is
    computed analytically from the \textsc{Agama} potential as
$V_c(R) = \sqrt{R\,|\partial_R \Phi|_{z=0}}$, and the synthetic
    rotation-curve data vector
$\mathbf{y}_{\mathrm{rc}} = \{V_c(R_i)\}_{i=1}^{N_{\mathrm{rc}}}$
    is obtained by adding Gaussian noise to the observational
    uncertainties $\sigma_{V_c}(R_i)$ reported by \citet{Zhou_2023}.
    Note that, unlike the stream observables, the rotation curve depends on
    the global parameters only, it is shared by all three streams, and it
    enters the inference as a conditioning input of every network that will be introduced in Sec.~\ref{sec:diffusion}.

    \subsection{Hierarchical Bayesian model}
    \label{sec:hierarchical}

\begin{figure}
        \centering
        \includegraphics[width=0.8\linewidth, trim=25 0 1000 0, clip]{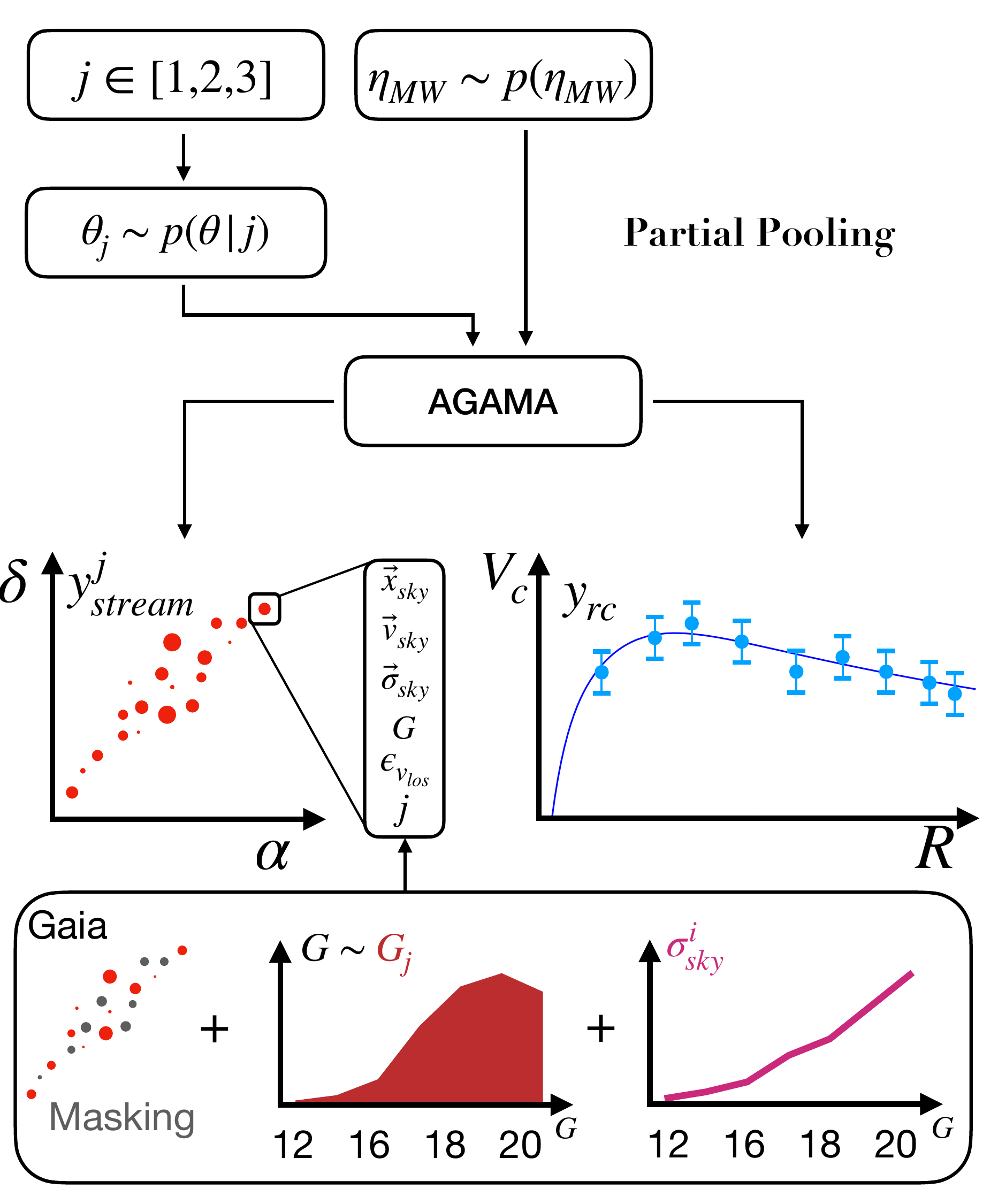}
        \caption{Overview of the hierarchical inference pipeline. We draw global parameters $\boldsymbol{\eta}_{\mathrm{MW}}$ and stream-specific progenitor parameters $\boldsymbol{\theta}_j$ from their respective priors and feed them to \textsc{Agama} to produce synthetic stream stars which are then corrupted with the noise and selection effects described in Sec.~\ref{sec:obs_model}. The circular velocity curve is generated by evaluating the potential on a grid of radii and adding Gaussian noise with the observational uncertainties from \citet{Zhou_2023}.
        }
        \label{fig:training}
    \end{figure}

    The inference has a natural hierarchical structure,  each
    observed stream $j$ has its own progenitor parameters
$\boldsymbol{\theta}_j \sim p(\boldsymbol{\theta}_j \mid j)$, while
    all streams and the rotation curve share the global potential parameters
$\boldsymbol{\eta}_{\mathrm{MW}} \sim p(\boldsymbol{\eta}_{\mathrm{MW}})$.
    The corresponding graphical model is illustrated in
    Fig.~\ref{fig:training}. For each stream $j$ we collect the per-star stream
    observables together with the circular rotation curve into a single
    observation group
$\mathbf{Y}^{j} \equiv (\mathbf{y}_{\mathrm{stream}}^{j},
\mathbf{y}_{\mathrm{rc}})$, and write $\mathbf{Y} \equiv
(\mathbf{Y}^{1}, \mathbf{Y}^{2}, \mathbf{Y}^{3})$ for the full set of
    observations. Our task is to estimate the joint posterior distribution
    \begin{equation}
        \begin{split}
            p(\boldsymbol{\eta}_{\mathrm{MW}}, \boldsymbol{\theta}_{1:3} \mid \mathbf{Y})
            \propto\; &
            p(\boldsymbol{\eta}_{\mathrm{MW}})
            \prod_{j=1}^{3}
            p\bigl(\mathbf{Y}^{j} \mid \boldsymbol{\theta}_j,
            \boldsymbol{\eta}_{\mathrm{MW}}, j\bigr)\,
            p(\boldsymbol{\theta}_j \mid j),
        \end{split}
        \label{eq:joint_posterior}
    \end{equation}
    where $\boldsymbol{\theta}_{1:3} \equiv (\boldsymbol{\theta}_1,
\boldsymbol{\theta}_2, \boldsymbol{\theta}_3)$, $p(\boldsymbol{\eta}_\mathrm{MW})$ encodes the global prior and $p(\boldsymbol{\theta}_j \mid j)$ local stream-dependent prior. The per-group factorization in
    Eq.~(\ref{eq:joint_posterior}) follows from the independence of
    the stream observations given $(\boldsymbol{\theta}_j,
\boldsymbol{\eta}_{\mathrm{MW}})$ and of the rotation curve given
$\boldsymbol{\eta}_{\mathrm{MW}}$. This is the mathematical structure that the
    compositional aggregation of Sec.~\ref{sec:composition} exploits.

    Instead of relying on the various analytic forms that the literature
    offers for the stream likelihood
$p(\mathbf{Y}^{j} \mid \boldsymbol{\theta}_j,
\boldsymbol{\eta}_{\mathrm{MW}}, j)$, our approach relies entirely on the
    implicit likelihood defined by the \textsc{Agama} simulator. In the next section (Sec.~\ref{sec:diffusion}) we also discuss its advantages over
    gold-standard MCMC approaches, in particular the amortised compositional
    property presented in Sec.~\ref{sec:composition}.

    \subsection{Score-based diffusion models for SBI}
    \label{sec:diffusion}

    We adopt score-based diffusion models \citep{Song2020, sharrock2024sequential, Arruda_2025} as our
    posterior approximators. Diffusion models recast sampling from a complex
    target distribution as learning to reverse a forward noising process. In
    the following, $\boldsymbol{z}$ denotes a generic parameter vector,
    standing for either the global parameters $\boldsymbol{\eta}_{\mathrm{MW}}$
    or the local parameters $\boldsymbol{\theta}_j$, while $\mathbf{y}$ denotes the conditioning observables.
    Starting from a clean parameter sample $\boldsymbol{z}_0 \sim p(\boldsymbol{z})$,
    the forward process gradually corrupts the signal:
    \begin{equation}
        \boldsymbol{z}_t = \alpha_t\,\boldsymbol{z}_0 + \sigma_t\,\boldsymbol{\epsilon},
        \qquad \boldsymbol{\epsilon} \sim \mathcal{N}(\mathbf{0}, \mathbf{I}),
        \label{eq:forward_process}
    \end{equation}
    where $t \in [0, 1]$ is the diffusion time, $\alpha_t$ interpolates from $\alpha_0 = 1$ (clean signal) to $\alpha_1 \approx 0$
    (pure noise), and together with $\sigma_t$ defines the noise schedule $\lambda_t = \log(\alpha_t^2 / \sigma_t^2)$. We adopt
    the EDM schedule as in \cite{Karras2022}. The
    reverse process denoises the corrupted sample by following the score
$\nabla_{\boldsymbol{z}_t}\!\log p_t(\boldsymbol{z}_t \mid \mathbf{y}, t)$,
    the gradient of the log marginal density at noise level $t$ conditioned on
    the conditioning $\mathbf{y}$, which also defines the bridging densities $p_t(\boldsymbol{z}_t \mid \mathbf{y})$ of the forward process.
    The score is approximated by a neural network $s_\phi(\boldsymbol{z}_t, t \mid \mathbf{y})$ trained on samples from the simulator using
    the denoising score matching objective described in \ref{sec:dsm}.

    \subsubsection{Denoising score matching loss}
    \label{sec:dsm}

    We train neural networks $s_\phi(\boldsymbol{z}_t, t \mid \mathbf{y})$
    to approximate the conditional score
$\nabla_{\boldsymbol{z}_t}\!\log p_t(\boldsymbol{z}_t \mid \mathbf{y}, t)$
    using the \emph{denoising score matching} (DSM) objective
    \citep{Vincent2011, Song2020}. The key insight is that, while the score of
    the noisy marginal $p_t(\boldsymbol{z}_t \mid \mathbf{y})$ is intractable,
    the score of the forward perturbation kernel $\nabla_{\boldsymbol{z}_t}\!\log p_t(\boldsymbol{z}_t \mid
\boldsymbol{z}_0)$ is available in closed form 
    given 
$p_t(\boldsymbol{z}_t \mid \boldsymbol{z}_0) = \mathcal{N}(\alpha_t
\boldsymbol{z}_0,\, \sigma_t^2 \mathbf{I})$.
    The DSM loss is therefore
    \begin{align}
        \mathcal{L}_{\mathrm{DSM}}(\phi)
         & = \mathbb{E}_{t,\;(\boldsymbol{z}_0, \mathbf{y}),\;\boldsymbol{\epsilon}\sim\mathcal{N}(\mathbf{0},\mathbf{I})}
        \!\left[\,\omega_t\,
            \left\| s_\phi\!\left(\alpha_t\boldsymbol{z}_0 + \sigma_t\boldsymbol{\epsilon},\;
            t \mid \mathbf{y}\right)
            + \frac{\boldsymbol{\epsilon}}{\sigma_t} \right\|_2^2\,\right],
        \label{eq:dsm_loss}
    \end{align}
    where the expectation over $(\boldsymbol{z}_0, \mathbf{y})$ runs over the
    joint distribution of parameters and synthetic observations produced by
    the simulator. 
    Many equivalent reparametrizations of the DSM are possible, effectively changing the type of prediction the neural network is making (e.g. noise prediction \citep{ho_denoising_2020}, velocity prediction \citep{Salimans_2022}) and the stability of the training (see \citep{Arruda_2025}). In our case, we selected the EDM prediction and the EDM weighting $\omega_t$ \citep{Karras2022}\footnote{This is the default choice in \texttt{BayesFlow}.}.

    Once trained, posterior samples are obtained by integrating the
    reverse-time stochastic differential equation (SDE) associated with the forward process from $t=1$ (pure
    Gaussian noise) to $t=0$ (posterior sample), with the learned score
$s_\phi$ in place of the unknown true score. We use the adaptive
    step-size integration scheme of \citet{jolicoeur-martineau2021gotta, Compositional_arrruda_2025}, which
    automatically selects the discretisation of the reverse process.

    \begin{figure}
        \centering
        \includegraphics[width=0.8\linewidth, trim=0 0 950 0, clip]{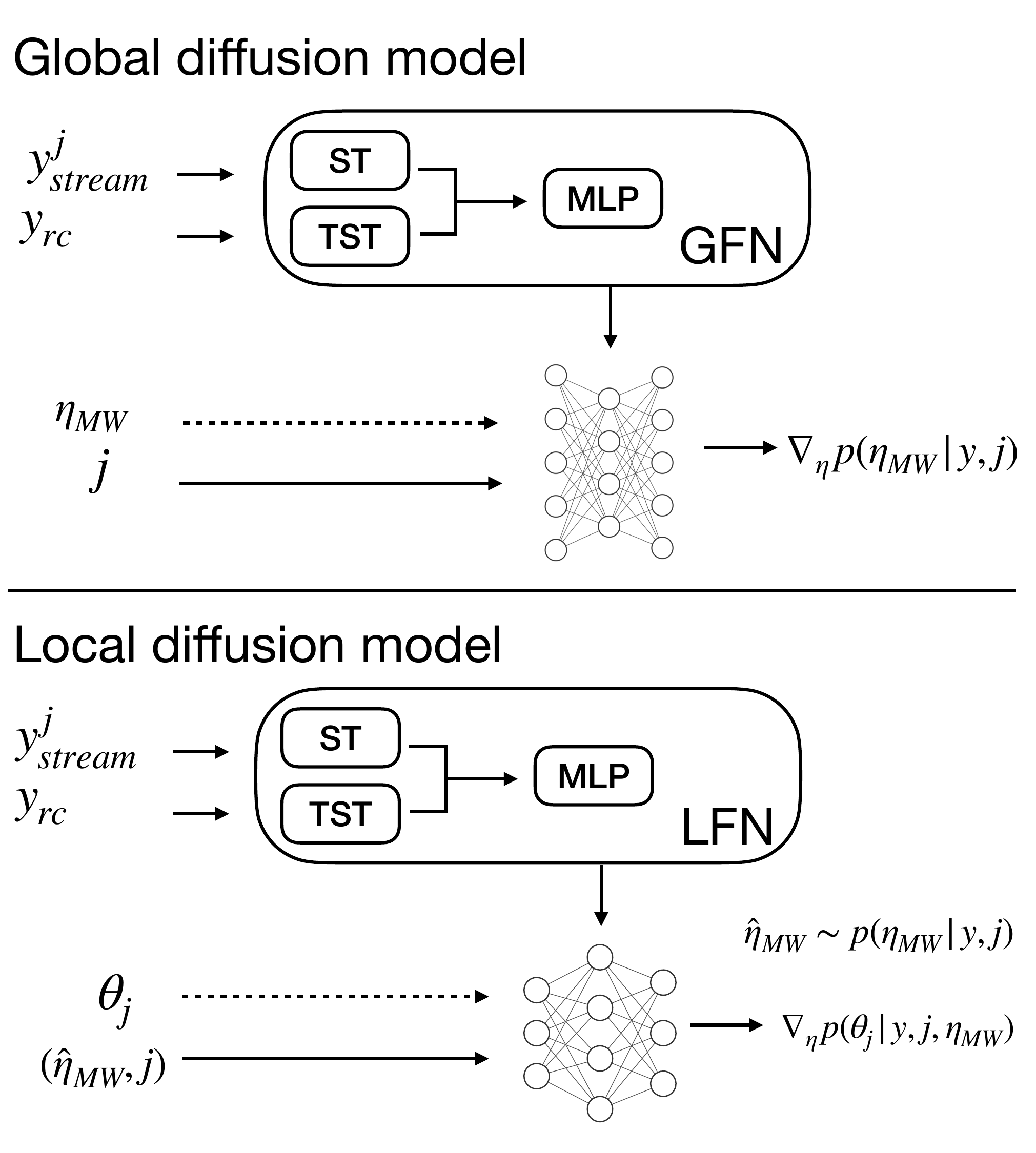}
        \caption{
        Training pipeline. Upper panel: simulated stream stars in $y_{\rm{stream}}^j$ are encoded by a \texttt{SetTransformer} (ST) and the circular-velocity curve $y_{\rm{rc}}$ by a time-series transformer encoder (TST).
        The resulting embeddings condition is then fed to an MLP forming the Global Fusion Network which conditions the diffusion model that learns the score of the per-stream posterior over the potential parameters $\boldsymbol{\eta}_{\mathrm{MW}}$.
        Lower panel: the Local Fusion Network mirrors the GFN in its input and conditions the local diffusion model that learns the score of the
        progenitor-parameter posterior conditioned additionally on
        $\boldsymbol{\eta}_{\mathrm{MW}}$. The dashed arrows indicate what is used only at training time, while the solid arrows indicate what is used at both training and inference time.
        }
        \label{fig:global_local_network}
    \end{figure}
    
    \subsubsection{Global network }
    \label{sec:gfn}

    The global network approximates the score of the posterior of
    the MW potential parameters conditioned on a \emph{single} stream
    observation and on the rotation curve:
    \begin{equation}
        s_\phi(\boldsymbol{\eta}_t, t \mid \mathbf{y}_{\mathrm{stream}}^{j},
        \mathbf{y}_{\mathrm{rc}}, j)
        \;\approx\;
        \nabla_{\boldsymbol{\eta}_t}\!\log p_t(\boldsymbol{\eta}_t
        \mid \mathbf{y}_{\mathrm{stream}}^{j}, \mathbf{y}_{\mathrm{rc}}, j).
        \label{eq:gfn_score}
    \end{equation}
    Training on one stream at a time (also known as partial pooling), rather than on the joint multi-stream
    observation (complete pooling), is a deliberate design choice. It keeps the simulation budget
    linear in the number of streams, and the single-stream posterior scores
    are exactly the objects that the compositional aggregation of
    Sec.~\ref{sec:composition} recombines at inference time.

    The stream observation $\mathbf{y}_{\mathrm{stream}}^{j}$ is a set of
    variable cardinality (one entry per stream member star), requiring a
    permutation-invariant encoder. We use a \texttt{SetTransformer}
    \citep{lee_set_2019}, whose attention mask implements the selection
    effects described in Sec.~\ref{sec:selection_effects}, to produce a
    fixed-dimensional embedding of the stream. The rotation curve data
$\mathbf{y}_{\mathrm{rc}}$, an ordered sequence of circular velocity
    measurements, is encoded by a separate time series transformer.
    The two embeddings are concatenated together to form a single observation embedding, which is then passed to a multi-layer perceptron (MLP), forming the Global Fusion Network (GFN), that will output the final conditioning vector.
    The resulting embeddings, together with the noisy parameter
    vector $\boldsymbol{\eta}_t$, and  the stream index $j$, are passed through another
    separate MLP which constitutes the backbone of the inference networks that
    outputs the score vector (Fig.~\ref{fig:global_local_network}, upper panel).

    A single global network is thus amortised over all three streams:
    the stream identity enters only through the conditioning inputs, not
    through separate per-stream networks.

    \subsubsection{Compositional score modeling}
    \label{sec:composition}

    In score-based methods, products of probability
    densities become sums in score space. The product structure of the
    hierarchical model in Eq.~(\ref{eq:joint_posterior}) implies that the
    per-stream posteriors can be recombined into the compositional distribution
    \begin{equation}
        p(\boldsymbol{\eta}_{\mathrm{MW}}|\mathbf{Y})
        \;\propto\;
        p(\boldsymbol{\eta}_{\mathrm{MW}})^{1-J}
        \prod_{j=1}^{J}
        p(\boldsymbol{\eta}_{\mathrm{MW}} \mid \mathbf{Y}^j, j),
        \label{eq:posterior_product}
    \end{equation}
    with $J=3$, where the factor
$p(\boldsymbol{\eta}_{\mathrm{MW}})^{1-J}$ removes the prior contribution
    that would otherwise be counted once per stream
    \citep{Geffner2022, Compositional_arrruda_2025}. 
    Information from multiple
    streams is thus aggregated entirely at inference time, without retraining
    and without ever simulating the joint multi-stream observation.

\begin{figure}
        \centering
        \includegraphics[width=0.99\linewidth]{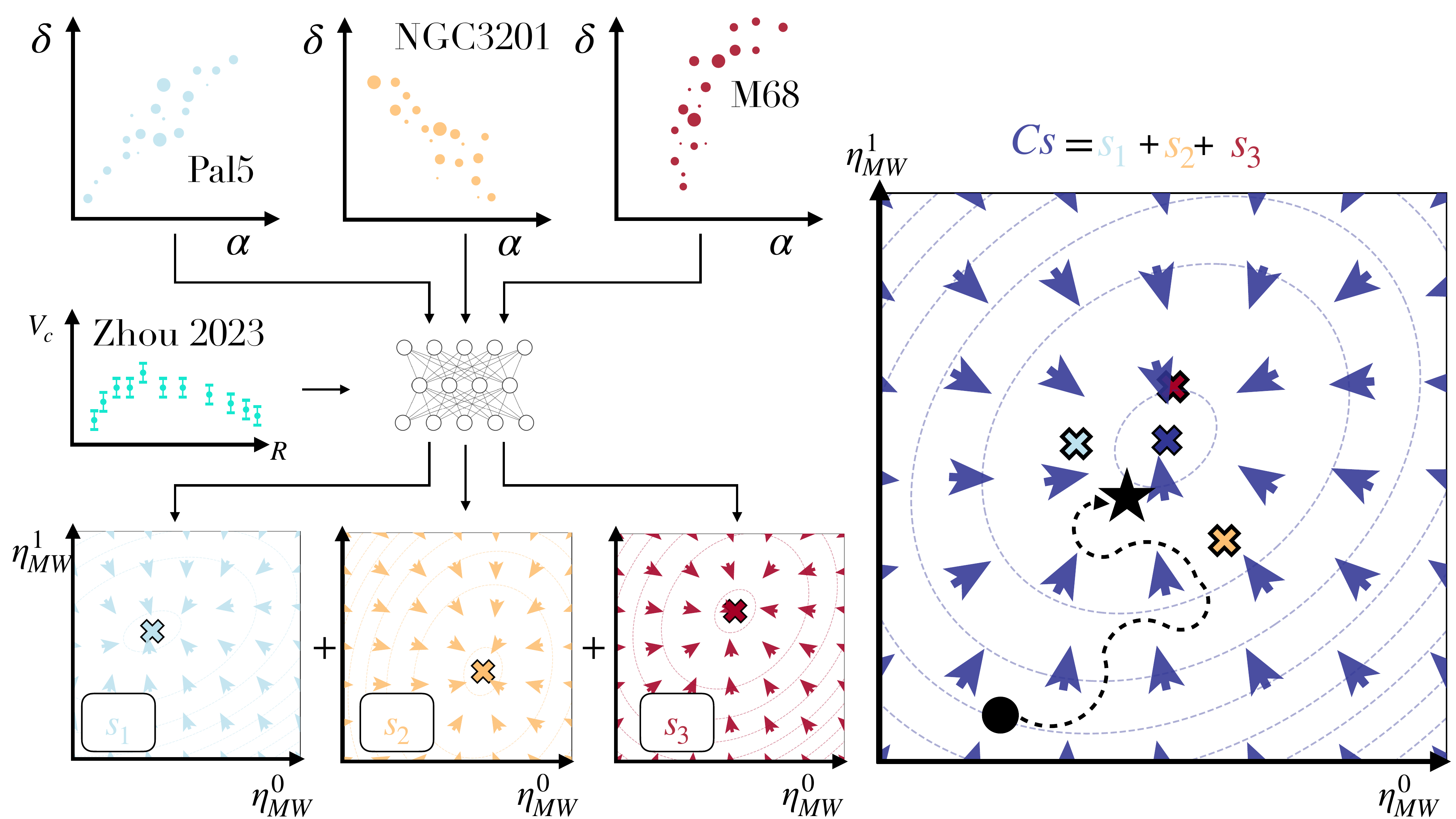}
        \caption{Illustration of compositional score modeling, where we have selected two generic directions $\eta^0_{MW}$ and $\eta^1_{MW}$ of the global parameter space $\eta_{MW}$ . Each stream constrains a different region of the global parameter space (left small panels) given the stream's individual observation and the rotation curve. The
        per-stream posterior scores $s_j$ produced by the global network are combined
        according to Eq.~(\ref{eq:composite_score}) to obtain $C_s = \sum_{j=1}^J s_j$ (right panel). In each panel, the coloured cross indicates the maximum of the posterior and arrows indicate the score vector field direction pointing towards this maximum. In the right panel, the dashed black line exemplifies one single integration path of the reverse diffusion process.}
        \label{fig:compositional_score_modeling}
    \end{figure}

    \paragraph{Compositional score.}
    The factorization of Eq.~(\ref{eq:posterior_product}) holds for the clean
    ($t=0$) distributions. Following \citet{Compositional_arrruda_2025}, we
    combine the score fields at every diffusion time $t$ as
    \begin{equation}
        \begin{split}
            C_s(\boldsymbol{\eta}_t,Y,t)
            \;=\; d(t) \Bigl[\,
             & (1-J)(1-t)\, \nabla_{\boldsymbol{\eta}_t}\!\log p(\boldsymbol{\eta}_t) \\
             & + J \sum_{j=1}^{J} s_\phi(\boldsymbol{\eta}_t, t \mid
                \mathbf{Y}^{j}, j)
                \,\Bigr],
        \end{split}
        \label{eq:composite_score}
    \end{equation}
    where the prior score
$\nabla_{\boldsymbol{\eta}_t}\!\log p(\boldsymbol{\eta}_t)$ is available
    analytically because the prior is known in closed form, the factor
$(1-t)$ anneals the prior correction away toward the pure-noise regime.
    Equation~(\ref{eq:composite_score}) defines a sequence of bridging
    distributions that is exact at the endpoints, the compositional
    distribution $\tilde{p}$ at $t=0$ and the latent Gaussian at $t=1$,
    but only approximate at intermediate times, because the noisy marginal of
    a product of posteriors is not the product of their noisy marginals
    \citep{Geffner2022}; in addition, the approximation errors of the
    individual learned score fields accumulate when the fields are summed.
    The time-dependent damping factor $d(t)= d_0 \cdot \exp(- \ln(d_0/d_1) \cdot t)$ introduced by
    \citet{Compositional_arrruda_2025} mitigates both effects, it decays
    exponentially from $d(0) = 1$, which guarantees that the target compositional
    distribution is recovered as $t \to 0$, to a terminal value
$d(1) = d_1 < 1$ that tempers the compositional score in the high-noise
    regime, where the bridging approximation is poorest. We set
$d_1 = 1/6$, chosen empirically to balance the accuracy of the recovered
    parameters against the calibration of the compositional posterior. Given the
    small number of composed scores ($J = 3$), no further stabilisation, such
    as the mini-batched estimator of \citet{Compositional_arrruda_2025}, is
    required.  Posterior
    samples are drawn by integrating the reverse SDE from
$t=1$ to $t=0$ with $C_{\mathrm{s}}$ in place of the single-stream
    score. The composition is illustrated in
    Fig.~\ref{fig:compositional_score_modeling}. The three stream-specific
    score fields (left small panels) point toward different regions of the parameter space, and
    their combination concentrates probability mass on the intersection of
    the individual posteriors, which correspond to the globally preferred
    potential configuration.

    \subsubsection{Local network}
    \label{sec:lfn}

    The local posterior can be factorized as
    \begin{equation}
        p(\boldsymbol{\boldsymbol{\theta}}_{1:3} \mid \mathbf{Y}, \boldsymbol{\eta}_{\mathrm{MW}})
        \;=\;
        \prod_{j=1}^{3}
        p(\boldsymbol{\theta}_j \mid \mathbf{Y^j}, \boldsymbol{\eta}_{\mathrm{MW}},  j),
        \label{eq:posterior_factorization}
    \end{equation}
    since the local parameters of different streams are conditionally
    independent of one another given the potential. The local network
    approximates the score\footnote{We choose to use a diffusion model as a posterior approximator also for the local parameters, but we could have used any other generative model (e.g., flow matching, normalizing flows).} of the per-stream conditional:
    \begin{equation}
        s_\psi(\boldsymbol{\theta}_t, t \mid \mathbf{y}_{\mathrm{stream}}^{j},
        \mathbf{y}_{\mathrm{rc}}, \boldsymbol{\eta}_{\mathrm{MW}}, j)
        \;\approx\;
        \nabla_{\boldsymbol{\theta}_t}\!\log p_t(\boldsymbol{\theta}_t
        \mid \mathbf{y}_{\mathrm{stream}}^{j}, \mathbf{y}_{\mathrm{rc}},
        \boldsymbol{\eta}_{\mathrm{MW}}, j).
        \label{eq:lfn_score}
    \end{equation}

    Under the hierarchical model of Sec.~\ref{sec:hierarchical} the progenitor
    parameters are \emph{a priori} independent of the potential, so that the prior
$p(\boldsymbol{\theta}_j \mid j)$ carries no dependence on
$\boldsymbol{\eta}_{\mathrm{MW}}$, yet the two blocks of parameters are
    strongly coupled \emph{a posteriori} by the fact that the same observed stream morphology
    implies different progenitor kinematics in different host potentials.
    Conditioning the local network, using a Local Fusion Network (LFN),
    on $\boldsymbol{\eta}_{\mathrm{MW}}$ is therefore
    essential for Eq.~(\ref{eq:posterior_factorization}) to hold, as it lets
    the network resolve, for every candidate potential, which progenitor
    phase-space configurations could have produced the observed stream.
    Conditioning on the rotation curve is, by contrast, formally redundant, 
    given the potential parameters, $\boldsymbol{\theta}_j$ is conditionally
    independent of $\mathbf{y}_{\mathrm{rc}}$ as the relevant information is already encoded in $\boldsymbol{\eta}_{\mathrm{MW}}$, so that
$p(\boldsymbol{\theta}_j \mid \mathbf{y}_{\mathrm{stream}}^{j}, \mathbf{y}_{\mathrm{rc}}, \boldsymbol{\eta}_{\mathrm{MW}}, j)
= p(\boldsymbol{\theta}_j \mid 
\mathbf{y}_{\mathrm{stream}}^{j}, \boldsymbol{\eta}_{\mathrm{MW}}, j)$, and the additional input leaves the
    target of Eq.~(\ref{eq:lfn_score}) mathematically unchanged. We nevertheless retain $\mathbf{y}_{\mathrm{rc}}$ as an
    input for two practical reasons: first, the circular velocity curve gives a
    direct indication of the enclosed mass profile of the potential and
    providing it alongside the parameter vector
$\boldsymbol{\eta}_{\mathrm{MW}}$ helps the network
    identify which progenitor parameters are dynamically compatible with the
    conditioning potential. Second, at inference time the LFN is evaluated at
$\boldsymbol{\eta}_{\mathrm{MW}}$ values sampled from the
    \emph{approximate} compositional posterior obtained by the global network,
    rather than at the true parameters; the redundant conditioning makes
    the local posterior more robust to residual errors in those samples.

    Practically, the LFN mirrors the GFN, with
$\boldsymbol{\eta}_{\mathrm{MW}}$ as additional conditioning input
    (Fig.~\ref{fig:global_local_network}, lower panel). 
    
    One preprocessing choice proved particularly important in practice. The GFN standardizes parameters and observations using statistics computed over the full training set. In contrast, the LFN standardizes the stream observations and the four local parameters separately for each stream, using that stream’s own mean and standard deviation. Posterior samples are then transformed back to physical units using the same per-stream statistics. Empirically, this per-stream standardization substantially improved both the accuracy and calibration of the local posteriors.


    \subsection{Training and hyperparameter tuning}
    \label{sec:training}
    Our dataset is composed of $10^6$ stream simulations, split into $10^4$ used as validation set for the hyperparameter tuning and the remaining for training. The global and local network have been trained for 1000 and 250 epochs, respectively, with the \texttt{AdamW} \citep{adamw_2017} optimizer and an initial learning rate of $5 \cdot10^{-4}$. The hyperparameter tuning has been carried out using \texttt{optuna} \citep{akiba2019optuna} and targeting the Pareto front of mean root mean square error and marginal calibration error over the approximated posterior samples of the validation set.
     All the architecture implementation, training and evaluation were obtained using the \texttt{BayesFlow} package \citep{kuhmichel2026bayesflow}.

    \section{Results}\label{sec:results}

    The networks of Sec.~\ref{sec:method} return, for any observation, an
    amortised approximation to the corresponding Bayesian posterior. Before
    such approximations can be trusted on the real Gaia data, they must be validated as part of any amortised Bayesian workflow, since neither the finite capacity of the networks nor the training procedure guarantees that the learned score fields reproduce the true posterior \citep{li2024amortized}.
    As usual in SBI, we validate the posteriors on held-out simulations, where the true parameters
    are known, with three complementary diagnostics.
    \emph{Calibration} (Sec.~\ref{subsec:calibration}) asks
    whether the posteriors are statistically self-consistent, and in particular whether their
    stated credible intervals have the correct frequentist coverage.
    \emph{Accuracy} (Sec.~\ref{subsec:accuracy})
    asks whether the posteriors actually concentrate around the true
    parameters, i.e.\ whether they are informative.
    \emph{Posterior predictive checks} (Sec.~\ref{subsec:gaia_data} and \ref{subsec:ppc}) ask
    whether the inferred parameters, propagated back through the forward model,
    reproduce the observed data. Lastly we explore model mispecification with a quantitative test in Sec. \ref{subsec:misspecification}.

    Calibration and accuracy are both needed, since a posterior can be perfectly calibrated yet so
    broad as to be useless, and, on the other hand, a sharply peaked posterior can be
    badly biased and severely overconfident. Coverage on simulations is
    necessary but not sufficient for the application to real data, since it is
    measured under the very generative model used for training. A possible stress test for validating our results on real data where the ground truth is not known are posterior predictive checks \citep{gelman2013bayesian}, which can reveal if the assumed simulator
    fails to describe the Galaxy.

    The global model accuracy and calibration are evaluated at \emph{two} levels.
    The first is the network used as trained, so to test the per-stream global posterior
$p(\boldsymbol{\eta}_{\mathrm{MW}}\mid\mathbf{Y}^{j},j)$ produced directly
    by the global network, which we refer to as the \emph{base} model. The second is the
    \emph{compositional} posterior of Eq.~(\ref{eq:composite_score}), which
    aggregates the three streams at inference time. The composition introduces approximations that the base network
    never exercises during training, making this two steps needed. A base network that
    is perfectly calibrated and accurate per stream therefore does not
    guarantee a calibrated, accurate composition, so both must be verified
    independently. Validating the base model in isolation additionally
    confirms that any degradation seen after composition originates in the
    aggregation step rather than in the underlying single-stream estimator.
    We use a held out set of 1000 streams, split into triplets that share the same global parameters
    for all the validation diagnostics of Sects.~\ref{subsec:calibration} and \ref{subsec:accuracy}.

    \begin{figure}
        \centering
        \includegraphics[width=0.99\linewidth]{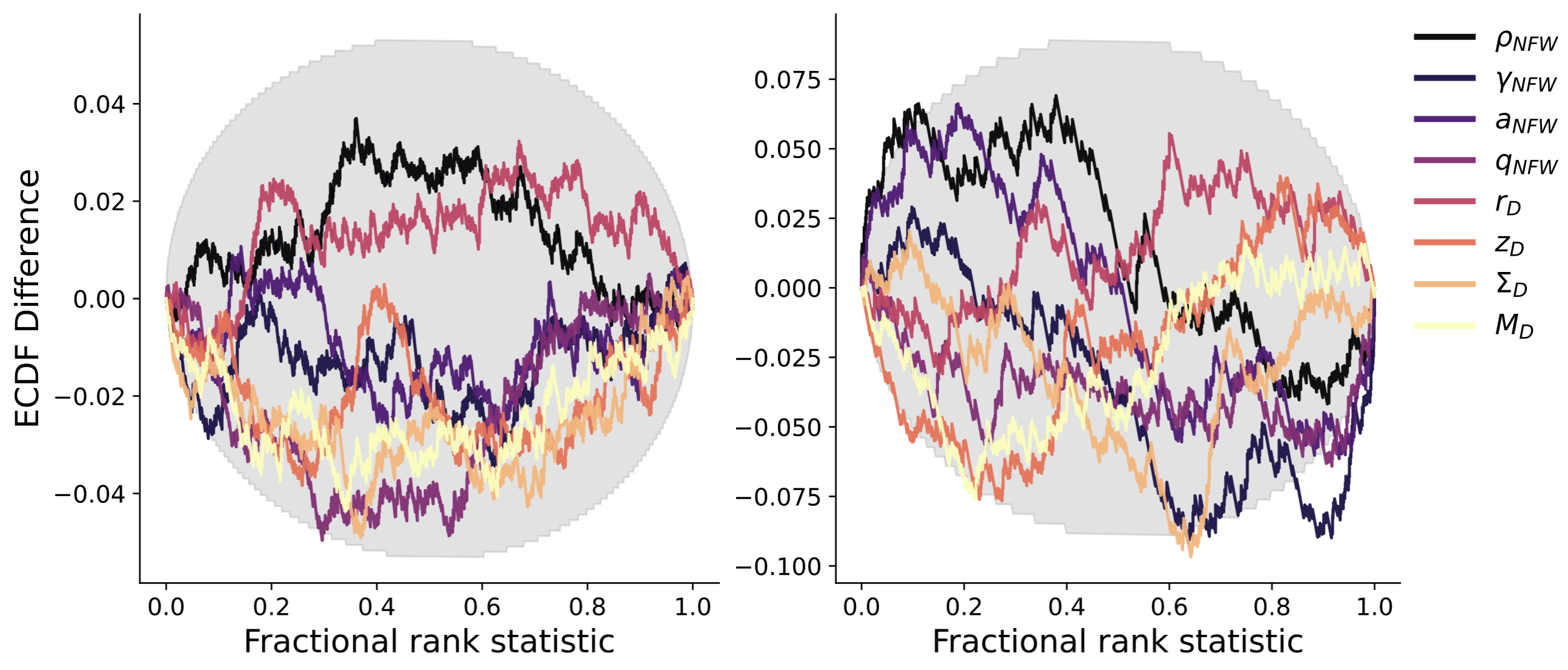}
        \caption{Simulation-based calibration of the global potential
        posterior. Each curve is the empirical CDF of the fractional rank
        statistic of one global parameter $\boldsymbol{\eta}_{\mathrm{MW}}$,
        plotted as a deviation from uniformity; the shaded region is the confidence band.
        Left: the base per-stream posterior of the global network. Right: the
        compositional posterior of Eq.~(\ref{eq:composite_score}). }
        \label{fig:global_PP_plot}
    \end{figure}
    \subsection{Calibration}\label{subsec:calibration}

    We assess calibration with simulation-based calibration
    \citep{Talts2018}, the standard self-consistency test for
    amortised posterior estimators. For each trial, we draw parameters and simulated data from the prior predictive distribution, then infer the posterior for that data. If the posterior is correctly calibrated, the true parameter value should be equally likely to appear anywhere within the posterior samples. Consequently, the rank of the true parameter among the posterior samples follows a uniform distribution. We repeat
    this over a held out set of simulations, and for every one-dimensional
    marginal to compute the fractional rank statistic $r\in[0,1]$, the
    rank of the ground truth value among the posterior draws normalized by
    their number.
    A visual check of uniform ranks is to compare the
    empirical cumulative distribution function (ECDF) of the fractional
    ranks against the CDF of the uniform distribution, as described in
    \citep{Sailynoja2022}.
    A calibrated posterior keeps its ECDF inside the band along its whole length,
    and any excursion outside is significant at the chosen level. This is the diagnostic plotted in
    Figs.~\ref{fig:global_PP_plot} and \ref{fig:local_PP_plot}, where the
    fractional rank ECDF is shown as a deviation from the diagonal so that
    perfect calibration corresponds to a flat, horizontal curve at zero within the band.

    Figure~\ref{fig:global_PP_plot} shows the calibration of the global
    potential posterior for the base model (left) and for the compositional
    one (right), for each global parameter $\boldsymbol{\eta}_{\mathrm{MW}}$.
    The base per-stream posteriors are well calibrated, their ECDFs remain
    within the confidence band for all seven parameters, which confirms that
    the global network has learned an internally consistent single-stream posterior. The
    compositional posterior remains calibrated as well, except for $\gamma_{\mathrm{NFW}}$ where we introduce
    a small deviation around $r=0.9$, underlying overconfidence; the ECDFs of the other parameters remain
    within the band.
    

        \begin{figure*}[t!]
        \centering
        \begin{minipage}[t]{0.49\linewidth}
            \centering
            \includegraphics[width=\linewidth]{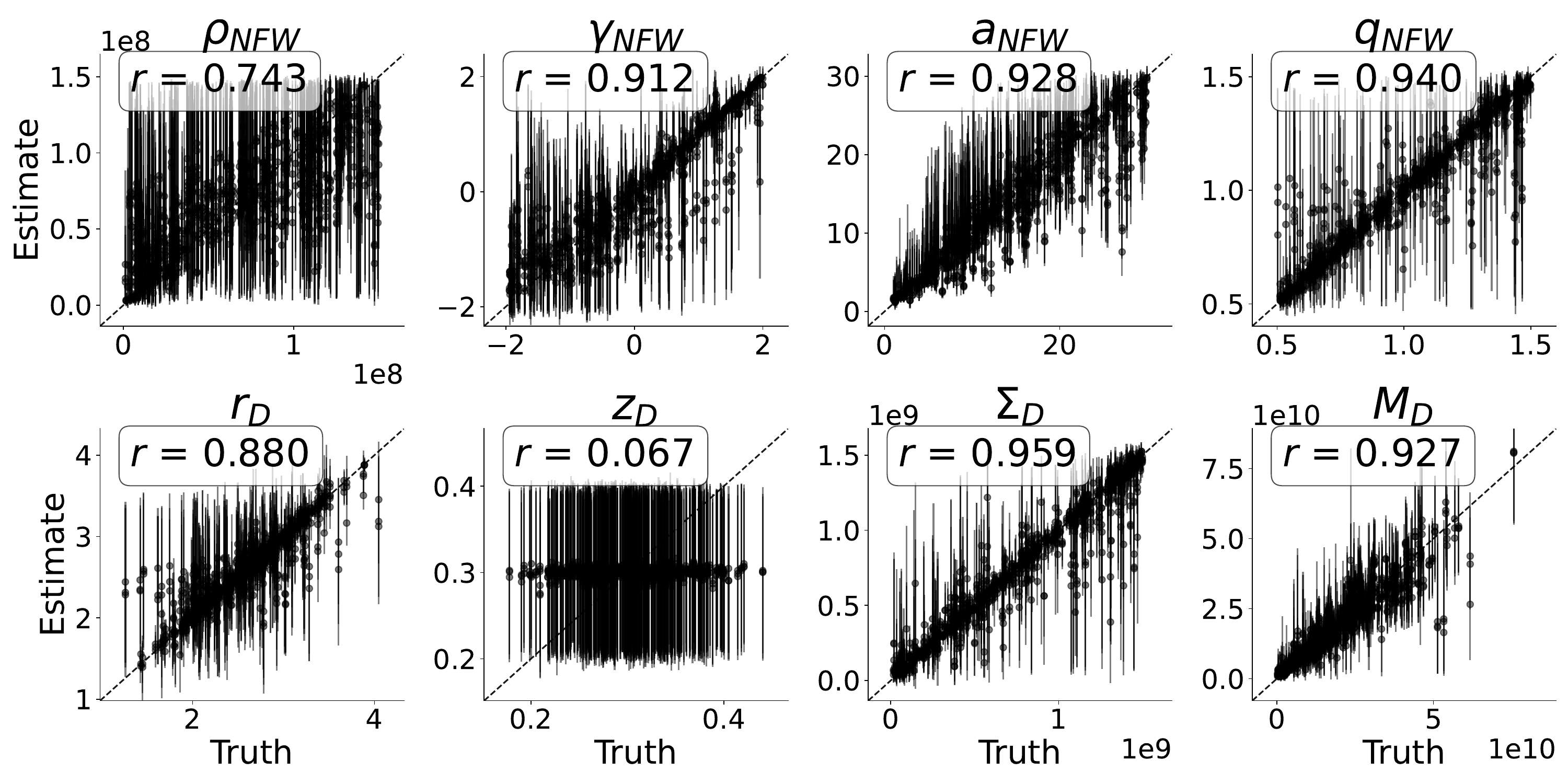}
        \end{minipage}
        \hfill
        \begin{minipage}[t]{0.49\linewidth}
            \centering
            \includegraphics[width=\linewidth]{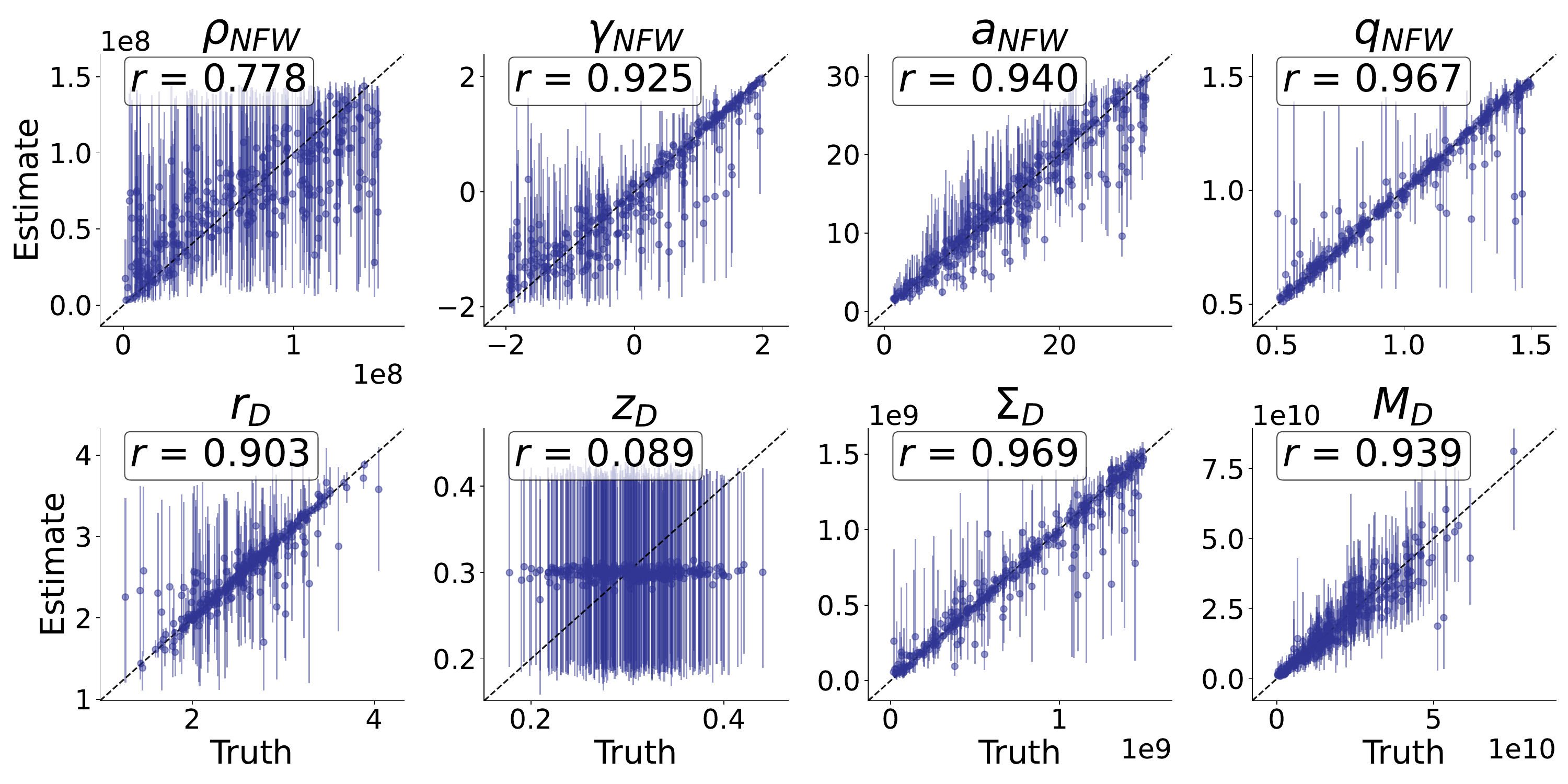}
        \end{minipage}
        \caption{Recovery of the global potential parameters on held-out test
            simulations. We plot the inferred posterior median (point) and
            central 95th percentile interval (error bar) against the
            ground-truth value, with the identity line represented by the dotted line. Left: base
            per-stream model. Right: compositional posterior. The Pearson correlation coefficient $r$ is reported in each panel.}
        \label{fig:TvP}
    \end{figure*}
    Figure~\ref{fig:local_PP_plot} shows that the fractional-rank ECDFs of the
    four local parameters of each stream remain within the confidence band.


    \subsection{Accuracy}\label{subsec:accuracy}

    Calibration certifies that the posteriors are honest about their
    uncertainty, but a calibrated posterior may still be too wide to be
    scientifically useful; the recovery plots, in the case of unimodal distribution, show both how tightly and how
    accurate the posteriors localise the truth. We report recovery plots for on the same
    held out test simulations. For each test set sample, we
    summarize the marginal posterior of every parameter by its median and its central 95\% percentile, and plot this estimate against the
    ground truth value used to generate the data. Perfect inference places all
    points on the identity line, with the credible intervals crossing it; a
    vertical offset reveals bias, while the vertical scatter and the interval
    width measure the residual uncertainty.

    Figure~\ref{fig:TvP} compares the recovery of the global parameters for
    the base model (left) and the compositional posterior (right). This comparison makes the gain from composition explicit, where a single stream constrains only a subspace of
$\boldsymbol{\eta}_{\mathrm{MW}}$ and leaves the remaining directions broad,
    the composition tightens the intervals and reduces the scatter by combining
    the complementary constraints of the three streams, as expected for constraints that sample different phase-space positions in the Galaxy. This is further summarized by the Pearson coefficient $r$ indicated in each panel.  In particular, the flattening of the halo density $q_\mathrm{NFW}$ and the scale radius of the disk $r_D$ benefit the most from this aggregation strategy. 
    We also report the recovery for the mass of the disk $M_D$, a derived quantity from the other disk parameters.

\begin{figure}
        \centering
        \begin{minipage}[t]{0.99\linewidth}
            \centering
            \includegraphics[width=\linewidth]{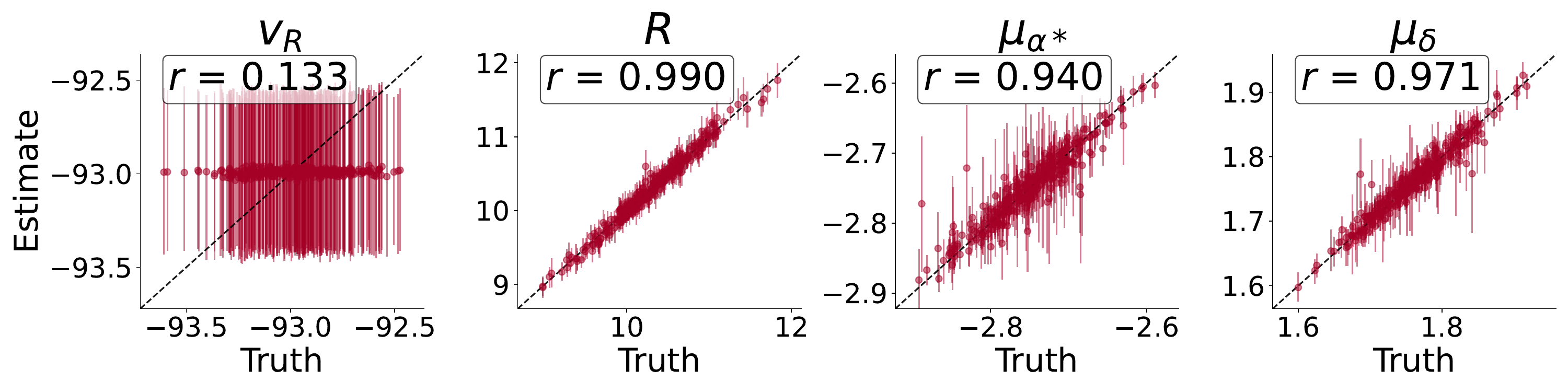}
        \end{minipage}
        \\
        \begin{minipage}[t]{0.99\linewidth}
            \centering
            \includegraphics[width=\linewidth]{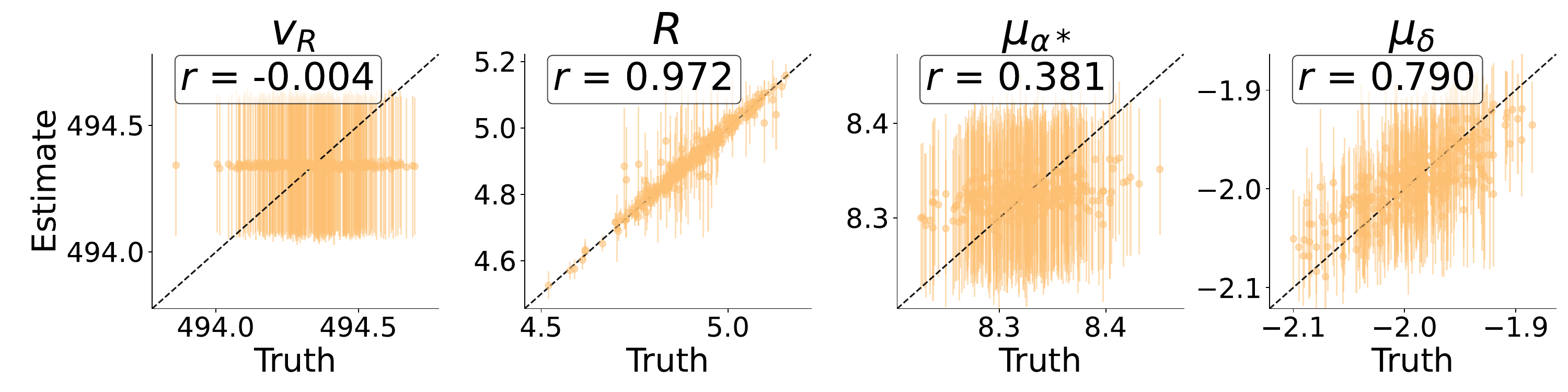}
        \end{minipage}
        \\
        \begin{minipage}[t]{0.99\linewidth}
            \centering
            \includegraphics[width=\linewidth]{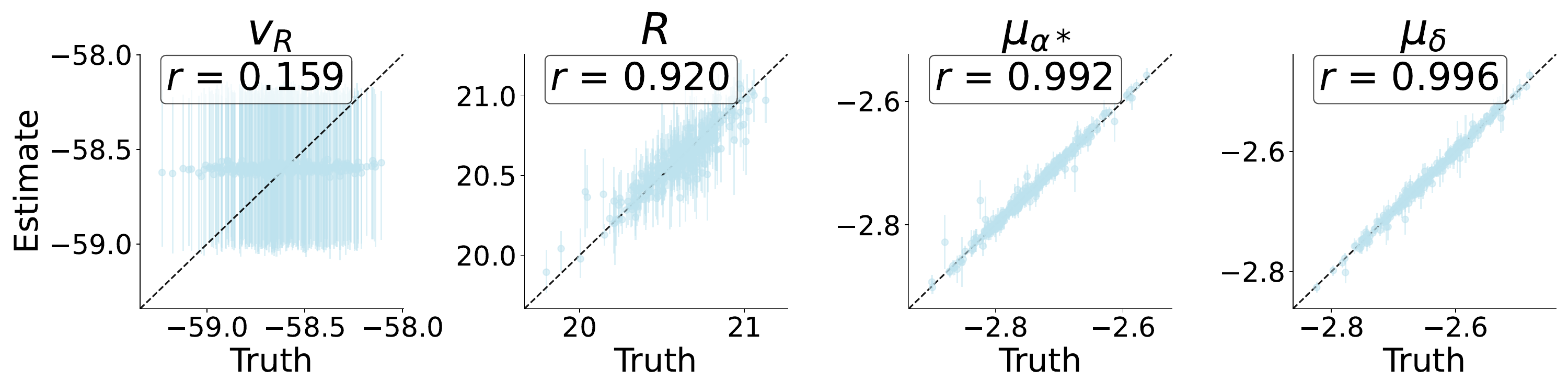}
        \end{minipage}
        \caption{Recovery of the four local progenitor parameters
            $\boldsymbol{\theta}_j$ on held out test simulations. We show one row per
            stream (top to bottom: M~68, NGC~3201, Pal~5). We plot again the posterior median and central 
            $95$th percentile interval against the ground truth, with
            the dotted diagonal reporting the identity line.}
        \label{fig:recovery_local}
    \end{figure}

    The accuracy of the local network must be established as well, since the
    progenitor parameters $\boldsymbol{\theta}_j$ are part of the final
$19$-dimensional inference and feed the posterior predictive streams. The
    per-stream recovery of the four local parameters is shown in
    Fig.~\ref{fig:recovery_local}. The local posteriors are accurate for all three streams. However, a common trend emerges: regardless of which stream is being considered, the observations are uninformative about the progenitor's line-of-sight velocity $V_R$. This is expected given the poor percentage of available coverage of line-of-sight velocity present in the observation, as shown in Tab.~\ref{tab:observed_stars}. We attribute the worst recovery of NGC~3201 to its higher noise level, produced by the higher magnitude of its stars, as shown in Fig. \ref{fig:magnitude}.

    \subsection{Gaia data}\label{subsec:gaia_data}

\begin{figure*}
        \centering
        \includegraphics[width=0.80\linewidth, trim=400 0 450 10, clip]{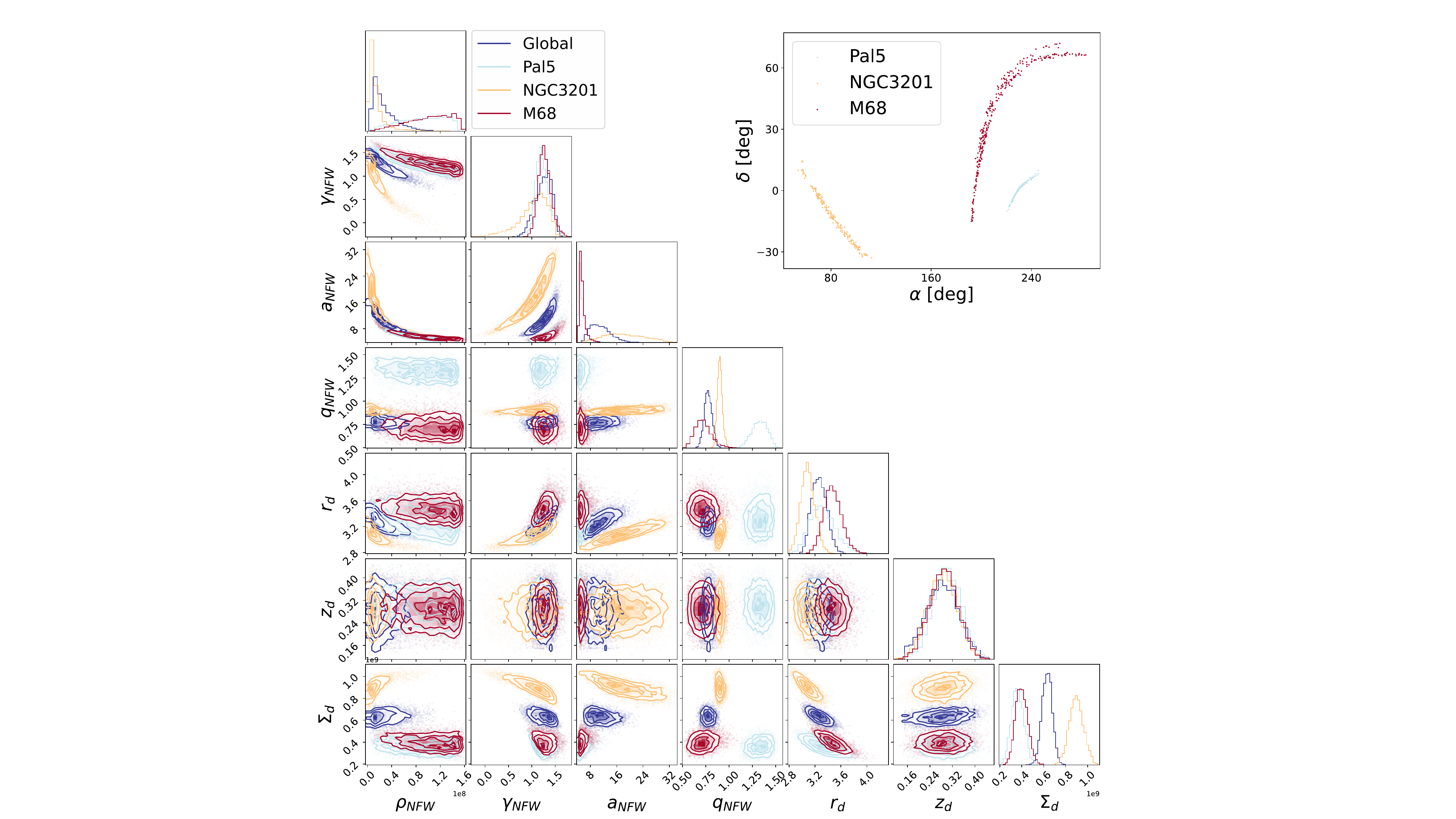}
        \caption{Posterior samples over the seven global 
        parameters $\boldsymbol{\eta}_{\mathrm{MW}}$ inferred from the circular velocity curve and  the Gaia data for the individual streams Pal~5 (cyan), NGC~3201 (yellow), M~68 (red) and all streams together (blue). In the upper-right corner, we report the sky position of the three observed streams' stars.}
        \label{fig:Cornerplot_gaia}
    \end{figure*}

    Having established  that the inference is both calibrated and
    accurate, both at the local and global level, in this section we apply it to
    real data, the Gaia astrometry and the spectroscopic line-of-sight
    velocities of the member stars of Pal~5, NGC~3201, and M~68 from \citet{Ibata_2024}, together with
    the circular velocity measurements of \citet{Zhou_2023}. 
    We run inference on the combined three streams and for each stream individually.
    Figure~\ref{fig:Cornerplot_gaia} shows the
    corner plot of the compositional posterior over the seven global parameters. The marginal posterior constraints on the global potential parameters obtained from each individual stream, and from their compositional combination, are summarized in Table~\ref{tab:results}.
     We can split our findings in single stream and compositional posteriors as follows.
    \paragraph{Single-stream posteriors.}
    The single-stream posteriors in Fig.~\ref{fig:Cornerplot_gaia} show that each
    cluster constrains a different subset of the potential and the three do not
    fully agree. The sharpest disagreement is in the halo flattening
    $q_{\mathrm{NFW}}$ as Pal~5 favours a prolate halo ($q_{\mathrm{NFW}}\approx1.33$),
    M~68 an oblate one ($q_{\mathrm{NFW}}\approx0.71$), and NGC~3201 lies between them
    ($q_{\mathrm{NFW}}\approx0.90$). This pattern is consistent with the streams
    sampling different regions of the Galaxy. In the same way, NGC~3201, the innermost of the three streams, disagrees the most with the other two on   disk surface density $\Sigma_D$, for which
    it returns the highest value and the largest disk mass (Table~\ref{tab:results}),
    and it leaves the halo density $\rho_{\mathrm{NFW}}$ low and the scale radius $a_{\mathrm{NFW}}$ poorly determined. The
    $(\rho_{\mathrm{NFW}}, a_{\mathrm{NFW}})$ panel displays the expected degeneracy
    of the generalized NFW profile, in which a low central density trades against a
    large scale radius in order to generate similar enclosed mass at a given radii;  Pal~5 and M~68 reach larger Galactocentric radii and pin the halo
    density and scale radius more tightly
    ($\rho_{\mathrm{NFW}}\approx10^{8}\,M_\odot\,\mathrm{kpc^{-3}}$,
    $a_{\mathrm{NFW}}\approx5\,\mathrm{kpc}$), yet they disagree on the flattening.
    The vertical scale height $z_D$, as expected from the recovery plot in Fig.\ref{fig:TvP}, is returned at its prior by all three streams and
    is effectively unconstrained by the data, while the radial scale length $r_D$ is consistent across streams. 
    Pal~5 and NGC~3201 agree on a mass of the MW inside 20 kpc to be $\sim 2  \cdot 10^{11} M_\odot$, a value which is higher than the low mass expectation of M~68. 

    \begin{figure*}
        \centering
        \includegraphics[width=0.9\linewidth]{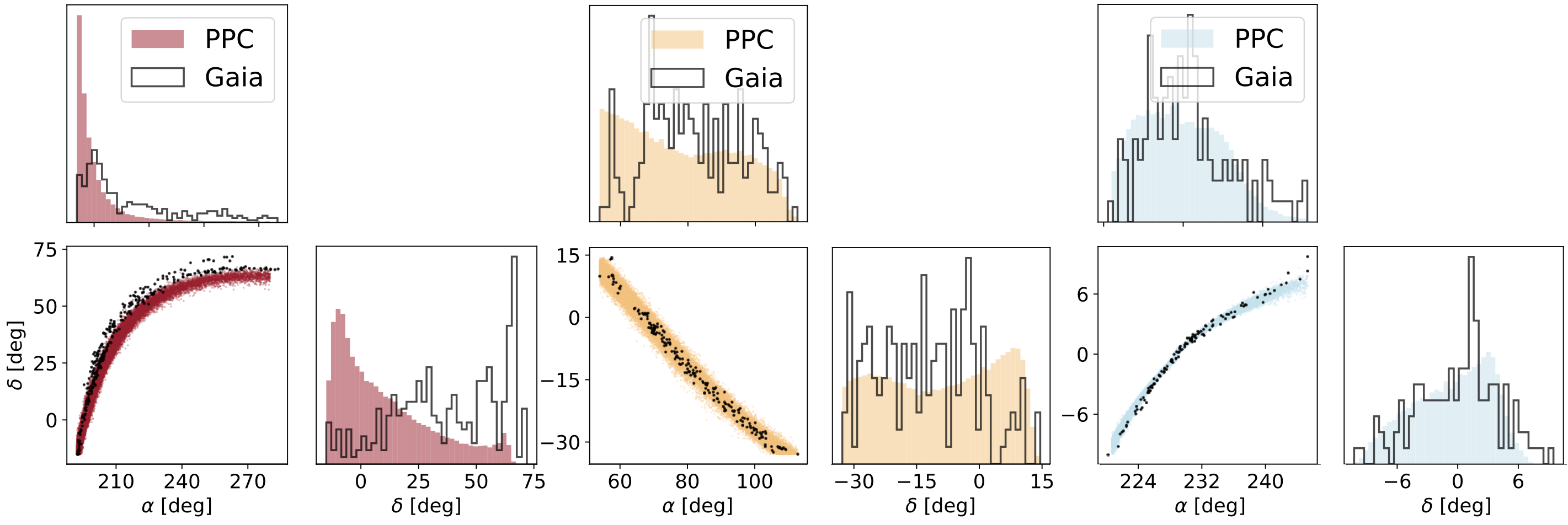}
        \caption{Posterior predictive check of the three streams. Observed
            member stars (scatter) are overlaid with mock streams generated by
            propagating posterior draws of the global and local parameters
            through the forward model and observational pipeline.}
        \label{fig:ppc_combined}
    \end{figure*}

    \paragraph{Compositional posterior.}
    The compositional posterior (blue in Fig.\ref{fig:Cornerplot_gaia}) combines the three single-stream scores and
    settles at the overlap of the individual constraints rather than at any one of
    them. It is tighter than the single-stream posteriors along the directions that
    at least one stream constrains (e.g. $a_{\mathrm{NFW}}$, $q_{\mathrm{NFW}}$), and it does not narrow the directions that all
    three leave broad (e.g. $z_D$). The halo flattening converges to $q_{\mathrm{NFW}}\approx0.77$,
    a mildly oblate value that sits between the NGC~3201 and M~68 solutions and away
    from the prolate Pal~5 preference, so the combination resolves the per-stream
    disagreement in favour of an oblate inner halo. The inner slope
    $\gamma_{\mathrm{NFW}}\approx1.36$ and the scale radius
    $a_{\mathrm{NFW}}\approx9.3\,\mathrm{kpc}$ are intermediate between the compact,
    dense solutions of Pal~5 and M~68 and the extended, low-density solution of
    NGC~3201, and they follow the $(\rho_{\mathrm{NFW}}, a_{\mathrm{NFW}})$ degeneracy
    toward a common value. The disk parameters converge on
    $r_D\approx3.3\,\mathrm{kpc}$ and
    $\Sigma_D\approx6.4\times10^{8}\,M_\odot\,\mathrm{kpc^{-2}}$, giving a total disk
    mass of $M_D = 4.3\times10^{10}\,M_\odot$ that falls between the high estimate of the
    innermost stream, NGC~3201, and the two outer streams. As in the single-stream
    case, $z_D$ stays at its prior, so the composition adds no information where the data are uninformative.
    The mass of the MW inside 20 kpc is tightly constrained around $\sim 1.9 \cdot 10^{11} M_\odot$.
    With the available posterior samples we also computed the posterior over the local parameters $\boldsymbol{\theta}_j$ of each stream, which are also reported in Tab.~\ref{tab:results}. 

    \begin{table*}
        \caption{Posterior constraints on the global Milky Way potential
            parameters from the Gaia data: median and central 68th percentile interval
            of the marginal posterior for each single-stream inference and for the
            compositional ones are reported. The middle rows show the derived total
            disk mass, the total mass within 20 kpc, the virial quantities
            $M_{200}$, $R_{200}$ and $c_{200}=R_{200}/a_{\mathrm{NFW}}$ (defined by
            the radius where the average interior density of potential equals $200\rho_c$ with
            $H_0 = 70\,\mathrm{km\,s^{-1}\,Mpc^{-1}}$) and the local dark-matter
            density $\rho_{\mathrm{NFW},\odot}$ at $R_\odot = 8.122$\,kpc;
            the last rows report the per-stream local parameters.}
        \label{tab:results}
        \centering
        \renewcommand{\arraystretch}{1.1}
        {\scriptsize
        \begin{tabular}{l l l l l}
            \hline\hline
            Parameter                                            & Pal~5 & NGC~3201 & M~68 & Compositional \\[4pt]
            \hline
            $\rho_{\mathrm{NFW}}$ [$\mathrm{M_\odot\,kpc^{-3}}$]& $1.30^{+0.01}_{-0.83}\times10^{8}$ & $6.52^{+15.85}_{-2.48}\times10^{6}$ & $1.06^{+0.34}_{-0.44}\times10^{8}$ & $1.32^{+3.40}_{-0.26}\times10^{7}$ \\[4pt]
            $\gamma_{\mathrm{NFW}}$                             & $1.177^{+0.201}_{-0.084}$ & $1.17^{+0.15}_{-0.44}$ & $1.25^{+0.16}_{-0.11}$ & $1.357^{+0.066}_{-0.302}$ \\[4pt]
            $a_{\mathrm{NFW}}$ [kpc]                            & $4.73^{+1.83}_{-0.20}$ & $15.4^{+9.6}_{-3.3}$ & $5.06^{+1.26}_{-0.24}$ & $9.3^{+4.8}_{-1.3}$ \\[4pt]
            $q_{\mathrm{NFW}}$                                  & $1.330^{+0.083}_{-0.089}$ & $0.900^{+0.027}_{-0.027}$ & $0.708^{+0.097}_{-0.072}$ & $0.767^{+0.050}_{-0.033}$ \\[4pt]
            $r_D$ [kpc]                                         & $3.26^{+0.26}_{-0.13}$ & $3.081^{+0.116}_{-0.098}$ & $3.44^{+0.17}_{-0.11}$ & $3.26^{+0.12}_{-0.11}$ \\[4pt]
            $z_D$ [kpc]                                         & $0.298^{+0.053}_{-0.046}$ & $0.292^{+0.044}_{-0.061}$ & $0.297^{+0.042}_{-0.058}$ & $0.280^{+0.062}_{-0.055}$ \\[4pt]
            $\Sigma_D$ [$\mathrm{M_\odot\,pc^{-2}}$]           & $3.59^{+0.61}_{-0.52}\times10^{8}$ & $8.94^{+0.71}_{-0.61}\times10^{8}$ & $3.85^{+0.68}_{-0.42}\times10^{8}$ & $6.43^{+0.33}_{-0.59}\times10^{8}$ \\[4pt]
            \hline
            $M_{\mathrm{D}}$ [$10^{10}\,\mathrm{M_\odot}$]   & $2.47^{+0.26}_{-0.22}$ & $5.38^{+0.22}_{-0.24}$ & $3.01^{+0.26}_{-0.29}$ & $4.27^{+0.18}_{-0.28}$ \\[4pt]
            $\rho_{\mathrm{NFW},\odot}$ [$\mathrm{M_\odot\,pc^{-3}}$]& $9.51^{+0.93}_{-0.78}\times10^{-3}$ & $9.64^{+2.19}_{-1.00}\times10^{-3}$ & $0.0118^{+0.0017}_{-0.0014}$ & $0.0115^{+0.0007}_{-0.0008}$ \\[4pt]
            $M(<20\,\mathrm{kpc})$ [$10^{11}\,\mathrm{M_\odot}$]& $2.052^{+0.094}_{-0.131}$ & $1.97^{+0.35}_{-0.15}$ & $1.70^{+0.13}_{-0.10}$ & $1.892^{+0.066}_{-0.063}$ \\[4pt]
            $M_{200}$ [$10^{12}\,\mathrm{M_\odot}$]             & $0.578^{+0.050}_{-0.055}$ & $0.92^{+0.36}_{-0.15}$ & $0.408^{+0.051}_{-0.036}$ & $0.587^{+0.063}_{-0.029}$ \\[4pt]
            $R_{200}$ [kpc]                                     & $172.0^{+4.6}_{-5.8}$ & $205^{+19}_{-16}$ & $153.3^{+5.8}_{-5.0}$ & $172.9^{+5.8}_{-3.1}$ \\[4pt]
            $c_{200}$                                           & $36.97^{+0.65}_{-10.65}$ & $11.2^{+5.0}_{-2.3}$ & $30.8^{+1.2}_{-6.6}$ & $14.2^{+7.2}_{-1.5}$ \\[4pt]
            \hline
            V$_R$ [$\mathrm{km\,s^{-1}}$]                       & $-58.66^{+0.23}_{-0.20}$ & $494.368^{+0.097}_{-0.134}$ & $-92.94^{+0.23}_{-0.19}$ & - \\[4pt]
            $R$ [kpc]                                           & $20.392^{+0.233}_{-0.062}$ & $4.353^{+0.074}_{-0.040}$ & $10.16^{+0.11}_{-0.17}$ & - \\[4pt]
            $\mu_{\alpha*}$ [$\mathrm{mas\,yr^{-1}}$]           & $-2.7697^{+0.0067}_{-0.0101}$ & $8.358^{+0.033}_{-0.029}$ & $-2.744^{+0.029}_{-0.022}$ & - \\[4pt]
            $\mu_{\delta}$ [$\mathrm{mas\,yr^{-1}}$]            & $-2.6609^{+0.0088}_{-0.0092}$ & $-1.930^{+0.023}_{-0.032}$ & $1.831^{+0.015}_{-0.026}$ & - \\[4pt]
            \hline
        \end{tabular}
        }
    \end{table*}

    \subsection{Posterior predictive check}\label{subsec:ppc}

    The calibration and accuracy tests of
    Sec.~\ref{subsec:calibration}--\ref{subsec:accuracy} are carried out on
    data drawn from the same generative model that trained the networks, and
    so cannot detect a mismatch between that model and the real Galaxy. On the
    real data an important test is posterior predictive check. 
    We draw global and local parameters from the joint posterior, push them
    back through the\textsc{Agama} forward model and the observational
    pipeline of Sec.~\ref{sec:obs_model}, and ask whether the resulting mock
    streams and rotation curve reproduce the observations. Agreement
    establishes that the inferred potential is dynamically compatible with the
    data; a systematic discrepancy could expose model misspecification.
  Figure~\ref{fig:ppc_combined}\footnote{Just like for prior predictive check, we report the entire observational space in Figs.\ref{fig:M68_PPC_vs_prior}, \ref{fig:NGC3201_PPC_vs_prior}, \ref{fig:Pal5_PPC_vs_prior} in Appendix \ref{sec:appendix_predictive}.} overlays the posterior predictive streams on
    the observed member stars. The predicted tracks follow the observed sky
    positions and kinematics closely, narrowing our prediction over the prior predictive checks, except for M~68 for which we consistently underestimate the star's declination $\delta$. We expect this behaviour to be the result of M~68 being always at the edge of our prior predictive check (Fig. \ref{fig:M68_PPC_vs_prior}) in the entire observation space, and we intend to investigate more on what are the possible causes of this training set discrepancy (e.g., prior boundaries on local/global parameters, galaxy model, type of simulator). This comparison is, however, only
    qualitative, since visual agreement in data space cannot by itself certify
    that the generative model is well specified. We therefore complement it with
    a quantitative test of model misspecification based on the maximum mean
    discrepancy (MMD) in latent space between training and real data, reported in Sec.~\ref{subsec:misspecification}.
    
\begin{figure}
        \centering
        \includegraphics[width=0.90\linewidth]{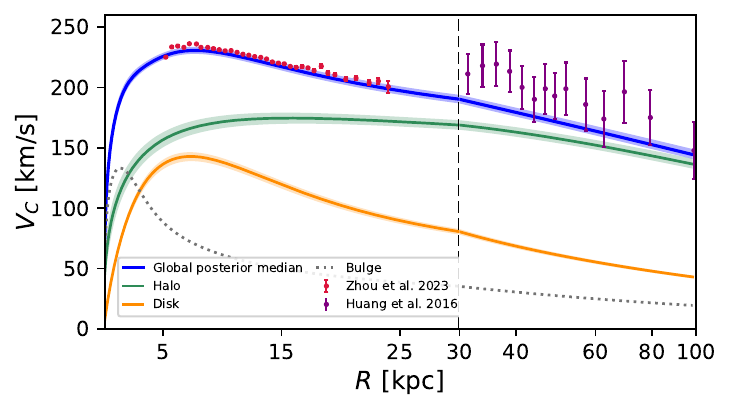}
        \caption{Posterior predictive circular velocity curve compared
            with the \citet{Zhou_2023} data used in the inference and with the
            independent \citet{Huang2016} measurements that extend to larger
            Galactocentric radii. The solid line is the prediction obtained by the median of the posterior with colored bands indicating the 68th posterior percentile. }
        \label{fig:rotation_curve}
    \end{figure}

\begin{figure}
        \centering
        \includegraphics[width=0.90\linewidth]{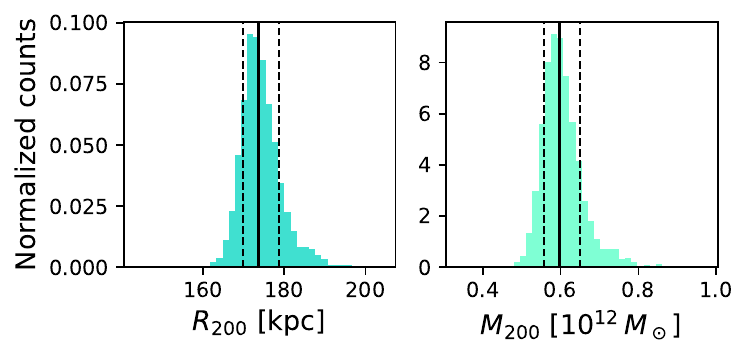}
        \caption{Posterior on the virial radius $R_{200}$ and virial mass
            $M_{200}$ derived from the compositional global posterior for the MW. The vertical dashed line indicates the 16th-84th percentiles, while the solid line indicates the median.}
        \label{fig:R200_M200}
    \end{figure}

    Moreover, we have run predictive check using the circular-velocity curve. The posterior potential predicts a rotation curve which we compare to the
    \citet{Zhou_2023} measurements used in the inference and, as an independent
    test, to the determination of \citet{Huang2016}, which was not used
    in training and is used here solely as an extrapolation test to larger radii (Fig.~\ref{fig:rotation_curve}). We are in mild agreement with observation at R$<30$ kpc, with a slight underestimation\footnote{We report that 62\% of the data land in the central 68-th percentile of the forward model posterior samples.} that becomes clearer at the extrapolation to R$>30$ kpc. 
    We can understand this discrepancy in the light of the virial quantities 
    $M_{200}$ and $R_{200}$ obtained for the MW,  Tab.~\ref{tab:results} and Fig.~\ref{fig:R200_M200}, whose distribution is slightly on the lower end of the literature results, as will be described in Sec. \ref{sec:comparison_literature}.
    
    \subsection{Model misspecification}\label{subsec:misspecification}
\begin{figure}
        \centering
        \includegraphics[width=0.9\linewidth]{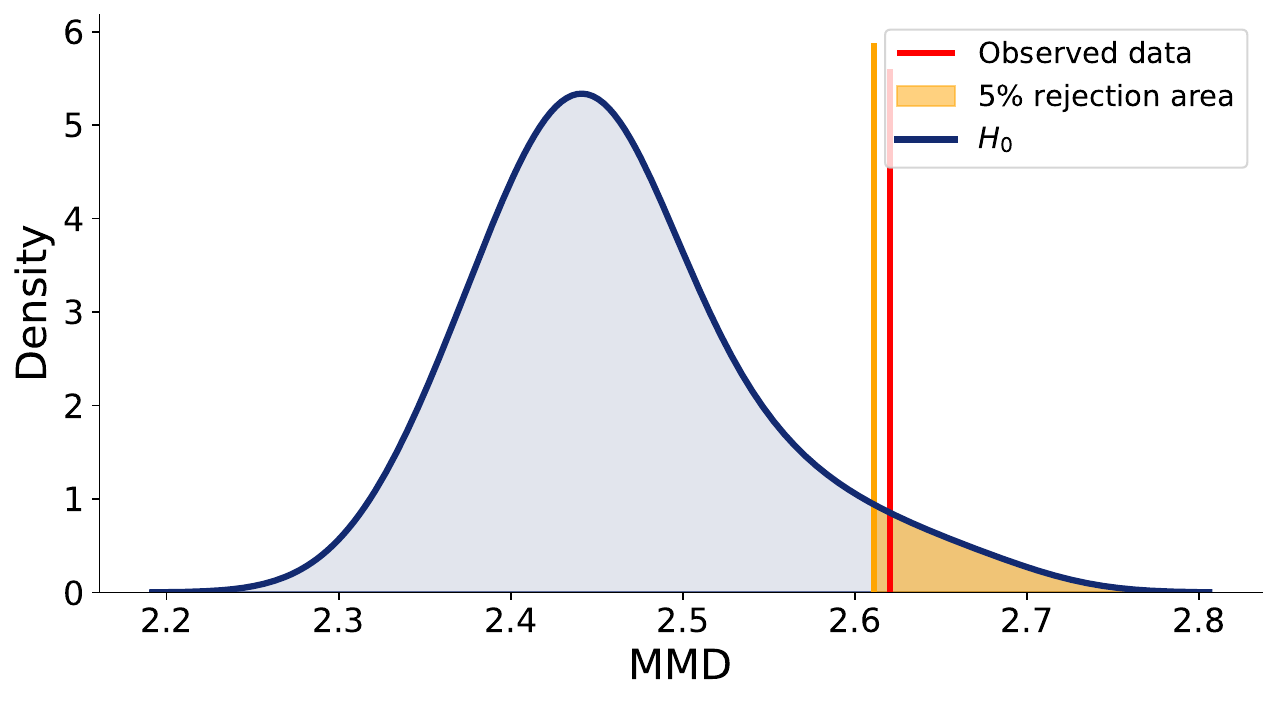}
        \caption{Model-misspecification check in the GFN summary space. The MMD
            between the observed data and the simulated test set (vertical line) is
            compared with the null distribution of the MMD obtained from held-out
            simulations under a well-specified model.}
        \label{fig:MMD_model_mispecification}
    \end{figure}

    We adopted the method presented in \citet{Schmitt2021} to test for model
    misspecification. The test is based on the MMD 
    between simulated test set members and the real data after both are projected
    onto the latent summary space learned by the GFN. If the model was well
    specified, the summary statistics of the observed streams would follow the
    same distribution as those of the simulations. We build the reference
    distribution of the MMD by repeatedly computing the discrepancy between
    independent batches of held out simulations, and we locate the MMD between
    the real MW data and the simulated test set within it
    (Fig.~\ref{fig:MMD_model_mispecification}). The observed value lies in the
    tail of this reference distribution, so we reject the null hypothesis that the data and the simulations share the same distribution.
    This mismatch means that the observed streams occupy a region of the summary
    space that the simulator does not populate well enough, i.e., that the real observations lie
    outside the support of our generative model (as we spotted for M~68 in Sec.~\ref{subsec:ppc}). We will discuss further the possible cause of this mismatched in Sec.\ref{sec:Discussion}.

    \section{Discussion}\label{sec:Discussion}

    The validation on simulations (Sec.~\ref{subsec:calibration} and
    \ref{subsec:accuracy}) reveals two properties of the compositional
    posterior. The base per-stream posteriors are well calibrated, but the
    composition develops a small calibration deviation for $\gamma_{\mathrm{NFW}}$
    around $r=0.9$ (Fig.~\ref{fig:global_PP_plot}). We associate this
    discrepancy between the base and the compositional performance with the
    compounding error of each individual score introduced by the aggregation step. At the same time, the composition behaves conservatively where the data are
    uninformative, i.e. for parameters that a single stream cannot constrain (e.g.\
$z_D$), the compositional posterior does not hallucinate knowledge that we
    would not expect to emerge (Fig.~\ref{fig:TvP}), tightening only those
    directions that the individual streams already constrain.
    The posterior predictive checks of Sec.~\ref{subsec:ppc} are only qualitative.
    The mock streams reproduce the observed sky positions and kinematics, but
    visual agreement cannot rule out a systematic mismatch between the simulator
    and the Galaxy. We therefore complement them with a quantitative test presented in Sec.~\ref{subsec:misspecification}, which confirmed our conclusion from the visual inspection. Several modeling choices could be
    responsible for this mismatch, starting from the choices of our prior boundaries. But, in our opinion, the most likely culprit is the simplified parametrization of the Milky Way potential, whose handful of global parameters cannot capture
    the full structural complexity of the real Galaxy (e.g., a radially
    varying halo, spiral arms, a multi-component disk). A
    second possibility lies in the forward model itself, as the particle-spray
    prescription is only an approximation to the self-consistent tidal
    disruption that a full $N$-body simulation would resolve, and it may distort the predicted stream morphology and kinematics. Moreover, residual
    inaccuracies in the observational pipeline (e.g., the noise model, the
    selection function, or the treatment of the spectroscopic completeness)
    could shift the data away from the simulated distribution. Lastly, during the hyperparameter optimization, we observed that the posterior prediction on Gaia data with similar performance would return incompatible posteriors, largely because of a collapse for some parameters toward the edge of the prior, which is a common failure mode of SBI methods. We therefore selected the hyperparameters that returned a more conservative posterior, which could be less accurate but more robust to model misspecification.
    In practice, the discrepancy most plausibly arises from the
    combination of these modeling choices. Disentangling them is left to future work, where we plan to
    probe the sensitivity of the MMD to each ingredient in turn by selectively
    perturbing the potential parametrization, the disruption model, and the
    observational pipeline. 
    In the next section, we proceeded to compare our results with the literature values.

    \paragraph{Comparison to classical analyses}\label{sec:comparison_literature}
    A useful external benchmark for our global posterior is the STREAMFINDER
    global mass model of \citet{Ibata_2024}, which fit a flexible Galactic potential
    (fixed bulge and gaseous disks, free halo, thin and thick disks) to a sample of
$29$ streams with a restricted $N$-body forward model and an MCMC exploration.
    We compare our compositional results of
    Table~\ref{tab:results} with their fiducial fit in which the halo outer
    power-law slope is held fixed at $\beta_h = 3$ (their Table~2). We decided to compare with this work because the observational data have been directly taken from their analysis. We recover a mildly oblate halo with axis ratio $q_{\mathrm{NFW}} =
0.767^{+0.050}_{-0.033}$, in excellent agreement with their tightly
    constrained mass flattening $q_{m,h} = 0.749^{+0.026}_{-0.030}$. Our total disk
    mass, $M_{D} = 4.27^{+0.18}_{-0.28}\times10^{10}\,M_\odot$, matches
    their combined thin plus thick disk mass $M_{d+t} =
4.20^{+0.44}_{-0.53}\times10^{10}\,M_\odot$. The inner halo slope for the combination, $\gamma_{\mathrm{NFW}} =
1.357^{+0.066}_{-0.302}$, has a median higher than their near-NFW value $\gamma_h =
0.97^{+0.17}_{-0.21}$, while remaining consistent above our predicted 16th percentile. Our halo scale radius $a_{\mathrm{NFW}} = 9.3^{+4.8}_{-1.3}\,\mathrm{kpc}$ is
    shorter than their $r_{0,h} = 14.7^{+4.7}_{-1.0}\,\mathrm{kpc}$, the
    two being still compatible at the level of the central 68th percentile. This median shift is the root cause of our underestimation of the virial quantities, hence our underestimate of the circular velocities. We note good agreement between the local dark matter density $\rho_{\mathrm{NFW}, \odot} = 0.01153^{+0.0072}_{-0.0077},M_\odot\,\mathrm{pc}^{-3}$ and their $\rho_{h,\odot} = 0.0114^{+0.0007}_{-0.0007},M_\odot\,\mathrm{pc}^{-3}$, which is a direct consequence of the agreement in the inner halo parameters. We also underestimate the total integrated mass up to $R=20$ kpc $M(<20\, \text{kpc})= 1.892^{+0.066}_{-0.063}$ to be slightly lower than the value $M_{20} = (2.17 \pm 0.21) \times 10^{11} \, M_\odot$ obtained by recent work on Gaia data in local stellar halo \citep{Dillamore2025} used to investigate also the bar pattern speed $\Omega_b$.
    The principal discrepancy is in the virial mass, as we obtain $M_{200} =
0.59^{+0.06}_{-0.03}\times10^{12}\,M_\odot$, roughly half of \citep{Ibata_2024}
$M_{200} = 1.09^{+0.19}_{-0.14}\times10^{12}\,M_\odot$. One possible cause of this discrepancy it  that the three
    streams probe the potential only out to $R \lesssim 20\,\mathrm{kpc}$, so the
    virial mass is an extrapolation well beyond the constrained regime and is
    sensitive to the assumed profile shape, whereas \citet{Ibata_2024} anchor the
    outer halo with streams reaching larger Galactocentric radii. To alleviate this problem \cite{Palau_2023} incorporate an additional likelihood component that includes the estimate on the total mass of the MW obtained by \citet{Callingham2019}. As a separate validation, our predicted ($\gamma_{\mathrm{NFW}}, M_{200}$) is in agreement with dark matter halo mass-inner slope relation calibrated on hydrodynamic simulation of MW-like galaxies from \citep{Tollet2016}.
    We are planning to extend this approach in future work using additional probes of the density distribution (e.g., HI thermal velocity curve as a function of angle \citep{McClure-Griffiths_2016}, measurement of the vertical force above the galactic plane \citep{Kuijken1991}), pre-training (transfer learning \citep{Saoulis_2025}) and post-training approaches by leveraging the possibility of adding additional vector field on top of the compositional score to guide the inference, like presented in \citet{Arruda_2025}.
    The local parameters are in accordance with \citet{Palau_2023}, except for the radius $R$ of NGC~3201 which seems to be the only one outside the $3\sigma$ range of the literature value in \citep{Palau_2023}.


    \section{Conclusion}\label{sec:Conclusion}

    We have presented a hierarchical, amortised simulation-based inference
    framework for reconstructing the MW gravitational potential from a
    population of stellar streams together with additional constraints such as the circular velocity curve observation. The
    method couples hierarchical models with compositional score
    modeling \citep{Compositional_arrruda_2025}, i.e. a global network learns the
    per-stream posterior over the seven potential parameters
$\boldsymbol{\eta}_{\mathrm{MW}}$, and streams are
    aggregated at inference time through the compositional score of
    Eq.~(\ref{eq:composite_score}).
    Then, a local network learns the four progenitor
    parameters $\boldsymbol{\theta}_j$ of each stream.  On held-out simulations, the posteriors are
    both calibrated and accurate, and the composition tightens the
    global constraints by combining the complementary information of the three
    streams, while behaving conservatively along the directions that no single
    stream constraints. Applied to the Gaia data of Pal~5, NGC~3201 and M~68, the
    combined posterior recovers a mildly oblate halo, a disk mass and a local
    dark matter density in good agreement with the STREAMFINDER mass model of
    \citet{Ibata_2024}; the virial mass is lower due to an underestimate of the virial radius $R_{200}$, and we interpret this being due to  the three
    streams constrain the potential only out to $R\lesssim20\,\mathrm{kpc}$ and
$M_{200}$ is an extrapolation beyond that range. The MMD
    test (Sec.~\ref{subsec:misspecification}) shows that the present generative
    model is mildly misspecified with respect to the real Galaxy, a limitation that motivates further improvements of our model as discussed below.

    The defining advantage of this framework is amortisation. The classical
    analyses of the same streams \citep{Palau_2023, Ibata_2024} rely on bespoke
    MCMC explorations whose cost is paid afresh whenever the dataset changes. In
    our approach that sampling cost is paid once during training, after which the
    posterior for any observation is a forward evaluation of the networks. This
    changes qualitatively how the analysis responds to the steady arrival of new
    data:
    \begin{itemize}
        \item Updated observations of a stream already in the training set:
              additional member stars (obtained by running STREAMFINDER on new data, or by changing the clustering algorithm \citep{Oliver2024}), newly measured line-of-sight velocities, or the
              improved astrometry expected from Gaia~DR4 generally call for
              retraining, because they alter the statistics that the observational model
              (Sec.~\ref{sec:obs_model}) imposes during training, like the magnitude
              distribution of the members, the fraction of stars carrying spectroscopic
              line-of-sight velocities, and the fraction of stars that are flagged as members to
              the stream. This retraining is nonetheless cheap, as these effects enter only
              through the observational layer applied on the fly to the pre-simulated
              library. The\textsc{Agama} simulations are reused unchanged and
              only the augmentation and the network optimization are repeated. The full
              training of both networks took roughly $20$~GPU\, hours on a single NVIDIA
              A100 40~GB and could be trivially parallelized on larger hardware. Moreover, the inference network weight could be frozen and the summary network could be fine tuned to adapt to the new data. 
        \item Newly discovered streams: newly discovered structures are the one case in which the amortization
              is reduced. A stream that is not part of the training set has no
              presimulated counterpart, so the simulator must be run again to populate the
              library with mock realizations of the new progenitor, and the networks must
              be retrained on the enlarged set.
    \end{itemize}
    This scalability is what makes the method well matched to the next generation
    of data. The WEAVE, 4MOST and DESI spectroscopic campaigns will multiply the
    number of velocity-confirmed stream members,
    and Gaia~DR4 will both sharpen the astrometry and reveal new structures. Each such increment tightens the compositional posterior at the marginal cost of an sde sampling step, rather than of a new Monte Carlo sampling. But, as for the non-amortized approaches, if the galaxy model is changed (Sec.\ref{sec:potential}), then the dataset must be created again. 

    Several improvements are possible. The detected model misspecification could point 
    to the rigidity of the potential parametrization, which could be relaxed by using a radially varying or
    non-axisymmetric halo, a multi-component disk, a misalignment of halo and disk plane \citep{Zhu2026} or a non-parametric
    basis-function model. The particle-spray forward model could be replaced by the
    fully differentiable, GPU-native, $N$-body code \textsc{Odisseo} \citep{Viterbo_2025} to
capture tidal disruption self-consistently. Finally, as described in Sec \ref{sec:comparison_literature}, the under constrained virial regime could be pinned by additional observational constraints along the circular velocities (e.g., HI thermal velocity curve or vertical force above the disk plane), or by guiding the compositional score with an external total mass constraint
\citep{Callingham2019} through post-training guidance \citep{Arruda_2025},
extending to the amortised setting the strategy adopted by
\citet{Palau_2023}. Together, these developments would turn the present
proof of concept into a tool capable of exploiting the full stream population
of the Milky Way to probe its mass distribution holistically.

\section{Code availability}\label{sec:appendix_code}

We publicly release our code to reproduce all the training, evaluation, hyperparameter tuning and the figures via Github: \url{https://github.com/vepe99/diffusion-experiments/tree/stable_bf}.

\begin{acknowledgements}
    This project was made possible by funding from the Carl-Zeiss-Stiftung.
    This work was supported by the Deutsche Forschungsgemeinschaft (DFG, German Research Foundation) – 573580088. 
    This work was supported by the Deutsche Forschungsgemeinschaft (DFG, German Research Foundation) under Germany’s Excellence Strategy EXC 2181/1 - 390900948 (the Heidelberg STRUCTURES Excellence Cluster). We acknowledge the usage of the AI-clusters Tom and Jerry funded by the Field of Focus 2 of Heidelberg University.
\end{acknowledgements}

\bibliographystyle{aa}
\bibliography{bibtex/bib}

@ARTICLE{Knoell2026,
       author = {{Kn{\"o}ll}, Niklas and {Buck}, Tobias and {Branca}, Lorenzo and {Viterbo}, Giuseppe},
        title = "{Amortized Simulation-Based Inference of Colliding-Wind Binaries from Short, Noisy Image Time Series}",
      journal = {arXiv e-prints},
         year = 2026,
        month = jun,
          eid = {arXiv:2606.10762},
        pages = {arXiv:2606.10762},
          doi = {10.48550/arXiv.2606.10762},
archivePrefix = {arXiv},
       eprint = {2606.10762},
 primaryClass = {astro-ph.SR},
       adsurl = {https://ui.adsabs.harvard.edu/abs/2026arXiv260610762K}
}

@ARTICLE{Lemos2023,
       author = {{Lemos}, Pablo and {Cranmer}, Miles and {Abidi}, Muntazir and {Hahn}, ChangHoon and {Eickenberg}, Michael and {Massara}, Elena and {Yallup}, David and {Ho}, Shirley},
        title = "{Robust simulation-based inference in cosmology with Bayesian neural networks}",
      journal = {Machine Learning: Science and Technology},
         year = {2023a},
        month = mar,
       volume = {4},
       number = {1},
          eid = {01LT01},
        pages = {01LT01},
          doi = {10.1088/2632-2153/acbb53},
archivePrefix = {arXiv},
       eprint = {2207.08435},
 primaryClass = {astro-ph.CO},
       adsurl = {https://ui.adsabs.harvard.edu/abs/2023MLS&T...4aLT01L}
}

@ARTICLE{Zhu2026,
       author = {{Zhu}, Ling and {Cai}, Runsheng and {Kang}, Xi and {Xue}, Xiang-Xiang and {Yang}, Chengqun and {Zhang}, Lan and {Mao}, Shude and {Liu}, Chao},
        title = "{A vertically orientated dark matter halo marks a flip of the Galactic disc}",
      journal = {\aap},
         year = 2026,
        month = feb,
       volume = {706},
          eid = {A193},
        pages = {A193},
          doi = {10.1051/0004-6361/202557613},
archivePrefix = {arXiv},
       eprint = {2510.08684},
 primaryClass = {astro-ph.GA},
       adsurl = {https://ui.adsabs.harvard.edu/abs/2026A&A...706A.193Z}
}

@ARTICLE{SN_SBI,
       author = {{Boyd}, Benjamin M. and {Mandel}, Kaisey S. and {Grayling}, Matthew and {Mitra}, Ayan and {Kessler}, Richard and {Autenrieth}, Maximilian and {Do}, Aaron and {Ginolin}, Madeleine and {Kelsey}, Lisa and {Narayan}, Gautham and {O'Callaghan}, Matthew and {Sarin}, Nikhil and {Thorp}, Stephen},
        title = "{FlowSN: Neural Simulation-Based Inference under Realistic Selection Effects applied to Supernova Cosmology}",
      journal = {arXiv e-prints},
         year = 2026,
        month = mar,
          eid = {arXiv:2603.11165},
        pages = {arXiv:2603.11165},
          doi = {10.48550/arXiv.2603.11165},
archivePrefix = {arXiv},
       eprint = {2603.11165},
 primaryClass = {astro-ph.CO},
       adsurl = {https://ui.adsabs.harvard.edu/abs/2026arXiv260311165B}
}

@article{arruda2026overcoming,
  title={Overcoming Selection Bias in Statistical Studies With Amortized Bayesian Inference},
  author={Arruda, Jonas and Chervet, Sophie and Staudt, Paula and Wieser, Andreas and Hoelscher, Michael and Sermet-Gaudelus, Isabelle and Binder, Nadine and Opatowski, Lulla and Hasenauer, Jan},
  journal={arXiv preprint arXiv:2604.18319},
  year={2026}
}

@inproceedings{sharrock2024sequential,
  author    = {Sharrock, Louis and Simons, Jack and Liu, Song and Beaumont, Mark},
  title     = {Sequential neural score estimation: likelihood-free inference with conditional score based diffusion models},
  booktitle = {Proceedings of the 41st International Conference on Machine Learning},
  year      = {2024},
  publisher = {JMLR.org},
  series    = {ICML'24},
  location  = {Vienna, Austria},
  url       = {https://dl.acm.org/doi/10.5555/3692070.3693884},
  articleno = {1814},
  numpages  = {38},
}

@article{jolicoeur-martineau2021gotta,
  author  = {Jolicoeur-Martineau, Alexia and Li, Ke and Pich{\'e}-Taillefer, R{\'e}mi and Kachman, Tal and Mitliagkas, Ioannis},
  title   = {Gotta go fast when generating data with score-based models},
  journal = {arXiv preprint arXiv:2105.14080},
  year    = {2021},
  doi     = {10.48550/arxiv.2105.14080},
}

@book{gelman2013bayesian,
  author    = {Gelman, Andrew and Carlin, John B and Stern, Hal S and Dunson, David B and Vehtari, Aki and Rubin, Donald B},
  title     = {{B}ayesian Data Analysis (3rd Edition)},
  year      = {2013},
  publisher = {Chapman and Hall/CRC},
  doi       = {10.1201/b16018},
}

@article{Oliver2024,
  author        = {{Oliver}, William H. and {Elahi}, Pascal J. and {Lewis}, Geraint F. and {Buck}, Tobias},
  title         = {{The hierarchical structure of galactic haloes: differentiating clusters from stochastic clumping with ASTROLINK}},
  journal       = {\mnras},
  year          = 2024,
  month         = may,
  volume        = {530},
  number        = {3},
  pages         = {2637-2647},
  doi           = {10.1093/mnras/stae1029},
  archiveprefix = {arXiv},
  eprint        = {2312.14632},
  primaryclass  = {astro-ph.GA},
  adsurl        = {https://ui.adsabs.harvard.edu/abs/2024MNRAS.530.2637O}
}

@article{Guenes2025,
  author        = {{Gunes}, Berkay and {Buder}, Sven and {Buck}, Tobias},
  title         = {{A COMPASS to Model Comparison and Simulation-Based Inference in Galactic Chemical Evolution}},
  journal       = {arXiv e-prints},
  year          = 2025,
  month         = jul,
  eid           = {arXiv:2507.05060},
  pages         = {arXiv:2507.05060},
  doi           = {10.48550/arXiv.2507.05060},
  archiveprefix = {arXiv},
  eprint        = {2507.05060},
  primaryclass  = {astro-ph.GA},
  adsurl        = {https://ui.adsabs.harvard.edu/abs/2025arXiv250705060G}
}

@article{buck_2025,
  author        = {{Buck}, Tobias and {G{\"u}nes}, Berkay and {Viterbo}, Giuseppe and {Oliver}, William H. and {Buder}, Sven},
  title         = {{Inferring Galactic parameters from chemical abundances with simulation-based inference}},
  journal       = {\aap},
  year          = 2025,
  month         = oct,
  volume        = {702},
  eid           = {A184},
  pages         = {A184},
  doi           = {10.1051/0004-6361/202554306},
  archiveprefix = {arXiv},
  eprint        = {2503.02456},
  primaryclass  = {astro-ph.GA},
  adsurl        = {https://ui.adsabs.harvard.edu/abs/2025A&A...702A.184B}
}

@article{Viterbo_2024,
  author        = {{Viterbo}, Giuseppe and {Buck}, Tobias},
  title         = {{CASBI -- Chemical Abundance Simulation-Based Inference for Galactic Archeology}},
  journal       = {arXiv e-prints},
  year          = 2024,
  month         = nov,
  eid           = {arXiv:2411.17269},
  pages         = {arXiv:2411.17269},
  doi           = {10.48550/arXiv.2411.17269},
  archiveprefix = {arXiv},
  eprint        = {2411.17269},
  primaryclass  = {astro-ph.GA},
  adsurl        = {https://ui.adsabs.harvard.edu/abs/2024arXiv241117269V}
}

@article{Dax_2025,
  author        = {{Dax}, Maximilian and {Green}, Stephen R. and {Gair}, Jonathan and {Gupte}, Nihar and {P{\"u}rrer}, Michael and {Raymond}, Vivien and {Wildberger}, Jonas and {Macke}, Jakob H. and {Buonanno}, Alessandra and {Sch{\"o}lkopf}, Bernhard},
  title         = {{Real-time inference for binary neutron star mergers using machine learning}},
  journal       = {\nat},
  year          = 2025,
  month         = mar,
  volume        = {639},
  number        = {8053},
  pages         = {49-53},
  doi           = {10.1038/s41586-025-08593-z},
  archiveprefix = {arXiv},
  eprint        = {2407.09602},
  primaryclass  = {gr-qc},
  adsurl        = {https://ui.adsabs.harvard.edu/abs/2025Natur.639...49D}
}

@article{Saoulis_2025,
  author        = {{Saoulis}, Alex A. and {Piras}, Davide and {Jeffrey}, Niall and {Mancini}, Alessio Spurio and {Ferreira}, Ana M.~G. and {Joachimi}, Benjamin},
  title         = {{Transfer learning for multifidelity simulation-based inference in cosmology}},
  journal       = {\mnras},
  year          = 2025,
  month         = sep,
  doi           = {10.1093/mnras/staf1436},
  archiveprefix = {arXiv},
  eprint        = {2505.21215},
  primaryclass  = {astro-ph.CO},
  adsurl        = {https://ui.adsabs.harvard.edu/abs/2025MNRAS.tmp.1394S}
}

@article{Geffner2022,
  author        = {{Geffner}, Tomas and {Papamakarios}, George and {Mnih}, Andriy},
  title         = {{Compositional Score Modeling for Simulation-based Inference}},
  journal       = {arXiv e-prints},
  year          = 2022,
  month         = sep,
  eid           = {arXiv:2209.14249},
  pages         = {arXiv:2209.14249},
  doi           = {10.48550/arXiv.2209.14249},
  archiveprefix = {arXiv},
  eprint        = {2209.14249},
  primaryclass  = {cs.LG},
  adsurl        = {https://ui.adsabs.harvard.edu/abs/2022arXiv220914249G}
}

@article{linhart2026diffusion,
  author  = {Julia Linhart and Gabriel Cardoso and Alexandre Gramfort and Sylvain Le Corff and Pedro L. C. Rodrigues},
  title   = {Diffusion posterior sampling for simulation-based inference in tall data settings},
  journal = {Transactions on Machine Learning Research},
  year    = {2026},
  url     = {https://openreview.net/forum?id=cdhfoS6Gyo},
}

@inproceedings{gloeckler2025compositional,
  author    = {Manuel Gloeckler and Shoji Toyota and Kenji Fukumizu and Jakob H. Macke},
  title     = {Compositional simulation-based inference for time series},
  booktitle = {The Thirteenth International Conference on Learning Representations},
  year      = {2025},
  url       = {https://openreview.net/forum?id=uClUUJk05H},
}

@article{deistler2025simulation-based,
  author   = {Deistler, Michael and Boelts, Jan and Steinbach, Peter and Moss, Guy and Moreau, Thomas and Gloeckler, Manuel and Rodrigues, Pedro L. C. and Linhart, Julia and Lappalainen, Janne K. and Miller, Benjamin Kurt and Gon{\c c}alves, Pedro J. and Lueckmann, Jan-Matthis and Schr{\"o}der, Cornelius and Macke, Jakob H.},
  title    = {Simulation-{{Based Inference}}: {{A Practical Guide}}},
  journal  = {arXiv preprint arXiv:2508.12939},
  year     = {2025},
  doi      = {10.48550/arxiv.2508.12939},
}

@article{Cranmer2020,
  author        = {{Cranmer}, Kyle and {Brehmer}, Johann and {Louppe}, Gilles},
  title         = {{The frontier of simulation-based inference}},
  journal       = {Proceedings of the National Academy of Science},
  year          = 2020,
  month         = dec,
  volume        = {117},
  number        = {48},
  pages         = {30055-30062},
  doi           = {10.1073/pnas.1912789117},
  archiveprefix = {arXiv},
  eprint        = {1911.01429},
  primaryclass  = {stat.ML},
  adsurl        = {https://ui.adsabs.harvard.edu/abs/2020PNAS..11730055C}
}

@article{Viterbo_2025,
  author        = {{Viterbo}, Giuseppe and {Buck}, Tobias},
  title         = {{Differentiable N-body code for Galactic Dynamics -- Odisseo}},
  journal       = {arXiv e-prints},
  year          = 2025,
  month         = nov,
  eid           = {arXiv:2511.22468},
  pages         = {arXiv:2511.22468},
  archiveprefix = {arXiv},
  eprint        = {2511.22468},
  primaryclass  = {astro-ph.GA},
  adsurl        = {https://ui.adsabs.harvard.edu/abs/2025arXiv251122468V}
}

@article{Chen_2025,
  author        = {{Chen}, Yingtian and {Valluri}, Monica and {Gnedin}, Oleg Y. and {Ash}, Neil},
  title         = {{Improved Particle Spray Algorithm for Modeling Globular Cluster Streams}},
  journal       = {\apjs},
  year          = 2025,
  month         = feb,
  volume        = {276},
  number        = {2},
  eid           = {32},
  pages         = {32},
  doi           = {10.3847/1538-4365/ad9904},
  archiveprefix = {arXiv},
  eprint        = {2408.01496},
  primaryclass  = {astro-ph.GA},
  adsurl        = {https://ui.adsabs.harvard.edu/abs/2025ApJS..276...32C}
}

@misc{lee_set_2019,
  title      = {Set {Transformer}: {A} {Framework} for {Attention}-based {Permutation}-{Invariant} {Neural} {Networks}},
  shorttitle = {Set {Transformer}},
  url        = {http://arxiv.org/abs/1810.00825},
  doi        = {10.48550/arXiv.1810.00825},
  language   = {en},
  urldate    = {2025-09-04},
  publisher  = {arXiv},
  author     = {Lee, Juho and Lee, Yoonho and Kim, Jungtaek and Kosiorek, Adam R. and Choi, Seungjin and Teh, Yee Whye},
  month      = may,
  year       = {2019},
  note       = {arXiv:1810.00825 [cs]}
}

@misc{ho_denoising_2020,
  title     = {Denoising {Diffusion} {Probabilistic} {Models}},
  url       = {http://arxiv.org/abs/2006.11239},
  doi       = {10.48550/arXiv.2006.11239},
  language  = {en},
  urldate   = {2025-10-12},
  publisher = {arXiv},
  author    = {Ho, Jonathan and Jain, Ajay and Abbeel, Pieter},
  month     = dec,
  year      = {2020},
  note      = {arXiv:2006.11239 [cs]}
}

@misc{Talts2018,
  title     = {Validating {Bayesian} {Inference} {Algorithms} with {Simulation}-{Based} {Calibration}},
  url       = {http://arxiv.org/abs/1804.06788},
  doi       = {10.48550/arXiv.1804.06788},
  author    = {Talts, Sean and Betancourt, Michael and Simpson, Daniel and Vehtari, Aki and Gelman, Andrew},
  year      = {2018},
  publisher = {arXiv},
  note      = {arXiv:1804.06788 [stat]}
}

@article{Sailynoja2022,
  title     = {Graphical test for discrete uniformity and its applications in goodness-of-fit evaluation and multiple sample comparison},
  author    = {S{\"a}ilynoja, Teemu and B{\"u}rkner, Paul-Christian and Vehtari, Aki},
  journal   = {Statistics and Computing},
  volume    = {32},
  number    = {2},
  pages     = {32},
  year      = {2022},
  doi       = {10.1007/s11222-022-10090-6},
  publisher = {Springer}
}

@article{Schmitt2021,
  author        = {{Schmitt}, Marvin and {B{\"u}rkner}, Paul-Christian and {K{\"o}the}, Ullrich and {Radev}, Stefan T.},
  title         = {{Detecting Model Misspecification in Amortized Bayesian Inference with Neural Networks}},
  journal       = {arXiv e-prints},
  year          = 2021,
  month         = dec,
  eid           = {arXiv:2112.08866},
  pages         = {arXiv:2112.08866},
  doi           = {10.48550/arXiv.2112.08866},
  archiveprefix = {arXiv},
  eprint        = {2112.08866},
  primaryclass  = {stat.ME},
  adsurl        = {https://ui.adsabs.harvard.edu/abs/2021arXiv211208866S}
}

@article{li2024amortized,
  author  = {Li, Chengkun and Vehtari, Aki and B{\"u}rkner, Paul-Christian and Radev, Stefan T and Acerbi, Luigi and Schmitt, Marvin},
  title   = {Amortized {B}ayesian workflow},
  journal = {arXiv preprint arXiv:2409.04332},
  year    = {2024},
  doi     = {10.48550/arxiv.2409.04332},
}

@article{Viterbo_2026,
  author        = {{Viterbo}, Giuseppe and {Buck}, Tobias},
  title         = {{The dynamical memory of tidal stellar streams: Joint inference of the Galactic potential and the progenitor of GD-1 with flow matching}},
  journal       = {\aap},
  year          = 2026,
  month         = mar,
  volume        = {707},
  eid           = {A363},
  pages         = {A363},
  doi           = {10.1051/0004-6361/202558358},
  archiveprefix = {arXiv},
  eprint        = {2512.04600},
  primaryclass  = {astro-ph.GA},
  adsurl        = {https://ui.adsabs.harvard.edu/abs/2026A&A...707A.363V}
}

@article{Agama_vasiliev,
  author        = {{Vasiliev}, Eugene},
  title         = {{AGAMA: action-based galaxy modelling architecture}},
  journal       = {\mnras},
  year          = 2019,
  month         = jan,
  volume        = {482},
  number        = {2},
  pages         = {1525-1544},
  doi           = {10.1093/mnras/sty2672},
  archiveprefix = {arXiv},
  eprint        = {1802.08239},
  primaryclass  = {astro-ph.GA},
  adsurl        = {https://ui.adsabs.harvard.edu/abs/2019MNRAS.482.1525V}
}

@article{Palau_2023,
  author        = {{Palau}, Carles G. and {Miralda-Escud{\'e}}, Jordi},
  title         = {{The oblateness of the Milky Way dark matter halo from the stellar streams of NGC 3201, M68, and Palomar 5}},
  journal       = {\mnras},
  year          = 2023,
  month         = sep,
  volume        = {524},
  number        = {2},
  pages         = {2124-2147},
  doi           = {10.1093/mnras/stad1930},
  archiveprefix = {arXiv},
  eprint        = {2212.03587},
  primaryclass  = {astro-ph.GA},
  adsurl        = {https://ui.adsabs.harvard.edu/abs/2023MNRAS.524.2124P}
}

@article{Ibata_2024,
  author        = {{Ibata}, Rodrigo and {Malhan}, Khyati and {Tenachi}, Wassim and {Ardern-Arentsen}, Anke and {Bellazzini}, Michele and {Bianchini}, Paolo and {Bonifacio}, Piercarlo and {Caffau}, Elisabetta and {Diakogiannis}, Foivos and {Errani}, Raphael and {Famaey}, Benoit and {Ferrone}, Salvatore and {Martin}, Nicolas F. and {di Matteo}, Paola and {Monari}, Giacomo and {Renaud}, Florent and {Starkenburg}, Else and {Thomas}, Guillaume and {Viswanathan}, Akshara and {Yuan}, Zhen},
  title         = {{Charting the Galactic Acceleration Field. II. A Global Mass Model of the Milky Way from the STREAMFINDER Atlas of Stellar Streams Detected in Gaia DR3}},
  journal       = {\apj},
  year          = 2024,
  month         = jun,
  volume        = {967},
  number        = {2},
  eid           = {89},
  pages         = {89},
  doi           = {10.3847/1538-4357/ad382d},
  archiveprefix = {arXiv},
  eprint        = {2311.17202},
  primaryclass  = {astro-ph.GA},
  adsurl        = {https://ui.adsabs.harvard.edu/abs/2024ApJ...967...89I}
}

@article{GaiaDR3,
  author        = {{Lindegren}, L. and {Klioner}, S.~A. and {Hern{\'a}ndez}, J. and {Bombrun}, A. and {Ramos-Lerate}, M. and {Steidelm{\"u}ller}, H. and {Bastian}, U. and {Biermann}, M. and {de Torres}, A. and {Gerlach}, E. and {Geyer}, R. and {Hilger}, T. and {Hobbs}, D. and {Lammers}, U. and {McMillan}, P.~J. and {Stephenson}, C.~A. and {Casta{\~n}eda}, J. and {Davidson}, M. and {Fabricius}, C. and {Gracia-Abril}, G. and {Portell}, J. and {Rowell}, N. and {Teyssier}, D. and {Torra}, F. and {Bartolom{\'e}}, S. and {Clotet}, M. and {Garralda}, N. and {Gonz{\'a}lez-Vidal}, J.~J. and {Torra}, J. and {Abbas}, U. and {Altmann}, M. and {Anglada Varela}, E. and {Balaguer-N{\'u}{\~n}ez}, L. and {Balog}, Z. and {Barache}, C. and {Becciani}, U. and {Bernet}, M. and {Bertone}, S. and {Bianchi}, L. and {Bouquillon}, S. and {Brown}, A.~G.~A. and {Bucciarelli}, B. and {Busonero}, D. and {Butkevich}, A.~G. and {Buzzi}, R. and {Cancelliere}, R. and {Carlucci}, T. and {Charlot}, P. and {Cioni}, M.-R.~L. and {Crosta}, M. and {Crowley}, C. and {del Peloso}, E.~F. and {del Pozo}, E. and {Drimmel}, R. and {Esquej}, P. and {Fienga}, A. and {Fraile}, E. and {Gai}, M. and {Garcia-Reinaldos}, M. and {Guerra}, R. and {Hambly}, N.~C. and {Hauser}, M. and {Jan{\ss}en}, K. and {Jordan}, S. and {Kostrzewa-Rutkowska}, Z. and {Lattanzi}, M.~G. and {Liao}, S. and {Licata}, E. and {Lister}, T.~A. and {L{\"o}ffler}, W. and {Marchant}, J.~M. and {Masip}, A. and {Mignard}, F. and {Mints}, A. and {Molina}, D. and {Mora}, A. and {Morbidelli}, R. and {Murphy}, C.~P. and {Pagani}, C. and {Panuzzo}, P. and {Pe{\~n}alosa Esteller}, X. and {Poggio}, E. and {Re Fiorentin}, P. and {Riva}, A. and {Sagrist{\`a} Sell{\'e}s}, A. and {Sanchez Gimenez}, V. and {Sarasso}, M. and {Sciacca}, E. and {Siddiqui}, H.~I. and {Smart}, R.~L. and {Souami}, D. and {Spagna}, A. and {Steele}, I.~A. and {Taris}, F. and {Utrilla}, E. and {van Reeven}, W. and {Vecchiato}, A.},
  title         = {{Gaia Early Data Release 3. The astrometric solution}},
  journal       = {\aap},
  year          = 2021,
  month         = may,
  volume        = {649},
  eid           = {A2},
  pages         = {A2},
  doi           = {10.1051/0004-6361/202039709},
  archiveprefix = {arXiv},
  eprint        = {2012.03380},
  primaryclass  = {astro-ph.IM},
  adsurl        = {https://ui.adsabs.harvard.edu/abs/2021A&A...649A...2L}
}

@article{astropy_2022,
  author        = {{Astropy Collaboration} and {Price-Whelan}, Adrian M. and {Lim}, Pey Lian and {Earl}, Nicholas and {Starkman}, Nathaniel and {Bradley}, Larry and {Shupe}, David L. and {Patil}, Aarya A. and {Corrales}, Lia and {Brasseur}, C.~E. and {N{\"o}the}, Maximilian and {Donath}, Axel and {Tollerud}, Erik and {Morris}, Brett M. and {Ginsburg}, Adam and {Vaher}, Eero and {Weaver}, Benjamin A. and {Tocknell}, James and {Jamieson}, William and {van Kerkwijk}, Marten H. and {Robitaille}, Thomas P. and {Merry}, Bruce and {Bachetti}, Matteo and {G{\"u}nther}, H. Moritz and {Aldcroft}, Thomas L. and {Alvarado-Montes}, Jaime A. and {Archibald}, Anne M. and {B{\'o}di}, Attila and {Bapat}, Shreyas and {Barentsen}, Geert and {Baz{\'a}n}, Juanjo and {Biswas}, Manish and {Boquien}, M{\'e}d{\'e}ric and {Burke}, D.~J. and {Cara}, Daria and {Cara}, Mihai and {Conroy}, Kyle E. and {Conseil}, Simon and {Craig}, Matthew W. and {Cross}, Robert M. and {Cruz}, Kelle L. and {D'Eugenio}, Francesco and {Dencheva}, Nadia and {Devillepoix}, Hadrien A.~R. and {Dietrich}, J{\"o}rg P. and {Eigenbrot}, Arthur Davis and {Erben}, Thomas and {Ferreira}, Leonardo and {Foreman-Mackey}, Daniel and {Fox}, Ryan and {Freij}, Nabil and {Garg}, Suyog and {Geda}, Robel and {Glattly}, Lauren and {Gondhalekar}, Yash and {Gordon}, Karl D. and {Grant}, David and {Greenfield}, Perry and {Groener}, Austen M. and {Guest}, Steve and {Gurovich}, Sebastian and {Handberg}, Rasmus and {Hart}, Akeem and {Hatfield-Dodds}, Zac and {Homeier}, Derek and {Hosseinzadeh}, Griffin and {Jenness}, Tim and {Jones}, Craig K. and {Joseph}, Prajwel and {Kalmbach}, J. Bryce and {Karamehmetoglu}, Emir and {Ka{\l}uszy{\'n}ski}, Miko{\l}aj and {Kelley}, Michael S.~P. and {Kern}, Nicholas and {Kerzendorf}, Wolfgang E. and {Koch}, Eric W. and {Kulumani}, Shankar and {Lee}, Antony and {Ly}, Chun and {Ma}, Zhiyuan and {MacBride}, Conor and {Maljaars}, Jakob M. and {Muna}, Demitri and {Murphy}, N.~A. and {Norman}, Henrik and {O'Steen}, Richard and {Oman}, Kyle A. and {Pacifici}, Camilla and {Pascual}, Sergio and {Pascual-Granado}, J. and {Patil}, Rohit R. and {Perren}, Gabriel I. and {Pickering}, Timothy E. and {Rastogi}, Tanuj and {Roulston}, Benjamin R. and {Ryan}, Daniel F. and {Rykoff}, Eli S. and {Sabater}, Jose and {Sakurikar}, Parikshit and {Salgado}, Jes{\'u}s and {Sanghi}, Aniket and {Saunders}, Nicholas and {Savchenko}, Volodymyr and {Schwardt}, Ludwig and {Seifert-Eckert}, Michael and {Shih}, Albert Y. and {Jain}, Anany Shrey and {Shukla}, Gyanendra and {Sick}, Jonathan and {Simpson}, Chris and {Singanamalla}, Sudheesh and {Singer}, Leo P. and {Singhal}, Jaladh and {Sinha}, Manodeep and {Sip{\H{o}}cz}, Brigitta M. and {Spitler}, Lee R. and {Stansby}, David and {Streicher}, Ole and {{\v{S}}umak}, Jani and {Swinbank}, John D. and {Taranu}, Dan S. and {Tewary}, Nikita and {Tremblay}, Grant R. and {de Val-Borro}, Miguel and {Van Kooten}, Samuel J. and {Vasovi{\'c}}, Zlatan and {Verma}, Shresth and {de Miranda Cardoso}, Jos{\'e} Vin{\'\i}cius and {Williams}, Peter K.~G. and {Wilson}, Tom J. and {Winkel}, Benjamin and {Wood-Vasey}, W.~M. and {Xue}, Rui and {Yoachim}, Peter and {Zhang}, Chen and {Zonca}, Andrea and {Astropy Project Contributors}},
  title         = {{The Astropy Project: Sustaining and Growing a Community-oriented Open-source Project and the Latest Major Release (v5.0) of the Core Package}},
  journal       = {\apj},
  year          = 2022,
  month         = aug,
  volume        = {935},
  number        = {2},
  eid           = {167},
  pages         = {167},
  doi           = {10.3847/1538-4357/ac7c74},
  archiveprefix = {arXiv},
  eprint        = {2206.14220},
  primaryclass  = {astro-ph.IM},
  adsurl        = {https://ui.adsabs.harvard.edu/abs/2022ApJ...935..167A}
}

@article{Zhou_2023,
  author        = {{Zhou}, Yuan and {Li}, Xinyi and {Huang}, Yang and {Zhang}, Huawei},
  title         = {{The Circular Velocity Curve of the Milky Way from 5-25 kpc Using Luminous Red Giant Branch Stars}},
  journal       = {\apj},
  year          = 2023,
  month         = apr,
  volume        = {946},
  number        = {2},
  eid           = {73},
  pages         = {73},
  doi           = {10.3847/1538-4357/acadd9},
  archiveprefix = {arXiv},
  eprint        = {2212.10393},
  primaryclass  = {astro-ph.GA},
  adsurl        = {https://ui.adsabs.harvard.edu/abs/2023ApJ...946...73Z}
}

@article{Arruda_2025,
  author        = {{Arruda}, Jonas and {Bracher}, Niels and {K{\"o}the}, Ullrich and {Hasenauer}, Jan and {Radev}, Stefan T.},
  title         = {{Diffusion Models in Simulation-Based Inference: A Tutorial Review}},
  journal       = {arXiv e-prints},
  year          = 2025,
  month         = dec,
  eid           = {arXiv:2512.20685},
  pages         = {arXiv:2512.20685},
  doi           = {10.48550/arXiv.2512.20685},
  archiveprefix = {arXiv},
  eprint        = {2512.20685},
  primaryclass  = {stat.ML},
  adsurl        = {https://ui.adsabs.harvard.edu/abs/2025arXiv251220685A}
}

@inproceedings{Compositional_arrruda_2025,
  author    = {Jonas Arruda and Vikas Pandey and Catherine Sherry and Margarida Barroso and Xavier Intes and Jan Hasenauer and Stefan T. Radev},
  title     = {Compositional amortized inference for large-scale hierarchical {Bayesian} models},
  booktitle = {The Fourteenth International Conference on Learning Representations},
  year      = {2026},
  url       = {https://openreview.net/forum?id=N3XCVHZGW5},
}

@inproceedings{Song2020,
  author        = {{Song}, Yang and {Sohl-Dickstein}, Jascha and {Kingma}, Diederik P. and {Kumar}, Abhishek and {Ermon}, Stefano and {Poole}, Ben},
  title         = {{Score-Based Generative Modeling through Stochastic Differential Equations}},
  booktitle     = {International Conference on Learning Representations (ICLR)},
  year          = 2021,
  archiveprefix = {arXiv},
  eprint        = {2011.13456},
  primaryclass  = {cs.LG},
  doi           = {10.48550/arXiv.2011.13456}
}

@article{Vincent2011,
  author  = {{Vincent}, Pascal},
  title   = {{A Connection Between Score Matching and Denoising Autoencoders}},
  journal = {Neural Computation},
  volume  = {23},
  number  = {7},
  pages   = {1661-1674},
  year    = 2011,
  doi     = {10.1162/NECO_a_00142}
}

@article{Plummer1911,
  author  = {{Plummer}, H.~C.},
  title   = {{On the problem of distribution in globular star clusters}},
  journal = {\mnras},
  volume  = {71},
  pages   = {460-470},
  year    = 1911,
  month   = mar,
  doi     = {10.1093/mnras/71.5.460}
}

@article{Baumgardt2019,
  author  = {{Baumgardt}, H. and {Hilker}, M. and {Sollima}, A. and {Bellini}, A.},
  title   = {{Mean proper motions, space orbits, and velocity dispersion profiles of Galactic globular clusters derived from Gaia DR2 data}},
  journal = {\mnras},
  volume  = {482},
  pages   = {5138-5155},
  year    = 2019,
  month   = feb,
  doi     = {10.1093/mnras/sty2997}
}

@article{Vasiliev2019b,
  author  = {{Vasiliev}, Eugene},
  title   = {{Proper motions and dynamics of 150 Milky Way globular clusters from Gaia DR2}},
  journal = {\mnras},
  volume  = {484},
  pages   = {2832-2850},
  year    = 2019,
  month   = apr,
  doi     = {10.1093/mnras/stz171}
}

@article{Harris1996,
  author  = {{Harris}, William E.},
  title   = {{A Catalog of Parameters for Globular Clusters in the Milky Way}},
  journal = {\aj},
  volume  = {112},
  pages   = {1487},
  year    = 1996,
  month   = oct,
  doi     = {10.1086/118116}
}

@article{Harris2010,
  author        = {{Harris}, William E.},
  title         = {{A New Catalog of Globular Clusters in the Milky Way}},
  journal       = {arXiv e-prints},
  year          = 2010,
  month         = dec,
  eid           = {arXiv:1012.3224},
  pages         = {arXiv:1012.3224},
  doi           = {10.48550/arXiv.1012.3224},
  archiveprefix = {arXiv},
  eprint        = {1012.3224},
  primaryclass  = {astro-ph.GA},
  adsurl        = {https://ui.adsabs.harvard.edu/abs/2010arXiv1012.3224H}
}

@article{PriceWhelan2019,
  author        = {{Price-Whelan}, Adrian M. and {Mateu}, Cecilia and {Iorio}, Giuliano and {Pearson}, Sarah and {Bonaca}, Ana and {Belokurov}, Vasily},
  title         = {{Kinematics of the Palomar 5 Stellar Stream from RR Lyrae Stars}},
  journal       = {\aj},
  volume        = {158},
  number        = {6},
  pages         = {223},
  year          = 2019,
  month         = dec,
  doi           = {10.3847/1538-3881/ab4cef},
  archiveprefix = {arXiv},
  eprint        = {1910.00595}
}

@article{Sollima2017,
  author  = {{Sollima}, A. and {Baumgardt}, H.},
  title   = {{The global mass functions of 35 Galactic globular clusters - I. Observational data and correlations with cluster parameters}},
  journal = {\mnras},
  volume  = {471},
  pages   = {3668-3679},
  year    = 2017,
  month   = nov,
  doi     = {10.1093/mnras/stx1856}
}

@article{Karras2022,
  author        = {{Karras}, Tero and {Aittala}, Miika and {Aila}, Timo and {Laine}, Samuli},
  title         = {{Elucidating the Design Space of Diffusion-Based Generative Models}},
  journal       = {arXiv e-prints},
  year          = 2022,
  month         = jun,
  eid           = {arXiv:2206.00364},
  pages         = {arXiv:2206.00364},
  doi           = {10.48550/arXiv.2206.00364},
  archiveprefix = {arXiv},
  eprint        = {2206.00364},
  primaryclass  = {cs.CV},
  adsurl        = {https://ui.adsabs.harvard.edu/abs/2022arXiv220600364K}
}

@article{Huang2016,
  author        = {{Huang}, Y. and {Liu}, X.-W. and {Yuan}, H.-B. and {Xiang}, M.-S. and {Zhang}, H.-W. and {Chen}, B.-Q. and {Ren}, J.-J. and {Wang}, C. and {Zhang}, Y. and {Hou}, Y.-H. and {Wang}, Y.-F. and {Cao}, Z.-H.},
  title         = {{The Milky Way's rotation curve out to 100 kpc and its constraint on the Galactic mass distribution}},
  journal       = {\mnras},
  year          = 2016,
  month         = dec,
  volume        = {463},
  number        = {3},
  pages         = {2623-2639},
  doi           = {10.1093/mnras/stw2096},
  archiveprefix = {arXiv},
  eprint        = {1604.01216},
  primaryclass  = {astro-ph.GA},
  adsurl        = {https://ui.adsabs.harvard.edu/abs/2016MNRAS.463.2623H}
}

@article{Callingham2019,
  author        = {{Callingham}, Thomas M. and {Cautun}, Marius and {Deason}, Alis J. and {Frenk}, Carlos S. and {Wang}, Wenting and {G{\'o}mez}, Facundo A. and {Grand}, Robert J.~J. and {Marinacci}, Federico and {Pakmor}, Ruediger},
  title         = {{The mass of the Milky Way from satellite dynamics}},
  journal       = {\mnras},
  year          = 2019,
  month         = apr,
  volume        = {484},
  number        = {4},
  pages         = {5453-5467},
  doi           = {10.1093/mnras/stz365},
  archiveprefix = {arXiv},
  eprint        = {1808.10456},
  primaryclass  = {astro-ph.GA},
  adsurl        = {https://ui.adsabs.harvard.edu/abs/2019MNRAS.484.5453C}
}

@article{Cautun_2020,
  author        = {{Cautun}, Marius and {Ben{\'\i}tez-Llambay}, Alejandro and {Deason}, Alis J. and {Frenk}, Carlos S. and {Fattahi}, Azadeh and {G{\'o}mez}, Facundo A. and {Grand}, Robert J.~J. and {Oman}, Kyle A. and {Navarro}, Julio F. and {Simpson}, Christine M.},
  title         = {{The milky way total mass profile as inferred from Gaia DR2}},
  journal       = {\mnras},
  year          = 2020,
  month         = may,
  volume        = {494},
  number        = {3},
  pages         = {4291-4313},
  doi           = {10.1093/mnras/staa1017},
  archiveprefix = {arXiv},
  eprint        = {1911.04557},
  primaryclass  = {astro-ph.GA},
  adsurl        = {https://ui.adsabs.harvard.edu/abs/2020MNRAS.494.4291C}
}

@ARTICLE{Green2023,
       author = {{Green}, Gregory M. and {Ting}, Yuan-Sen and {Kamdar}, Harshil},
        title = "{Deep Potential: Recovering the Gravitational Potential from a Snapshot of Phase Space}",
      journal = {\apj},
         year = 2023,
        month = jan,
       volume = {942},
       number = {1},
          eid = {26},
        pages = {26},
          doi = {10.3847/1538-4357/aca3a7},
archivePrefix = {arXiv},
       eprint = {2205.02244},
 primaryClass = {astro-ph.GA},
       adsurl = {https://ui.adsabs.harvard.edu/abs/2023ApJ...942...26G}
}

@ARTICLE{Eilers2019,
       author = {{Eilers}, Anna-Christina and {Hogg}, David W. and {Rix}, Hans-Walter and {Ness}, Melissa K.},
        title = "{The Circular Velocity Curve of the Milky Way from 5 to 25 kpc}",
      journal = {\apj},
         year = 2019,
        month = jan,
       volume = {871},
       number = {1},
          eid = {120},
        pages = {120},
          doi = {10.3847/1538-4357/aaf648},
archivePrefix = {arXiv},
       eprint = {1810.09466},
 primaryClass = {astro-ph.GA},
       adsurl = {https://ui.adsabs.harvard.edu/abs/2019ApJ...871..120E}
}

@ARTICLE{Wegg2019,
       author = {{Wegg}, Christopher and {Gerhard}, Ortwin and {Bieth}, Marie},
        title = "{The gravitational force field of the Galaxy measured from the kinematics of RR Lyrae in Gaia}",
      journal = {\mnras},
         year = 2019,
        month = may,
       volume = {485},
       number = {3},
        pages = {3296-3316},
          doi = {10.1093/mnras/stz572},
archivePrefix = {arXiv},
       eprint = {1806.09635},
 primaryClass = {astro-ph.GA},
       adsurl = {https://ui.adsabs.harvard.edu/abs/2019MNRAS.485.3296W}
}

@ARTICLE{McMillan2017,
       author = {{McMillan}, Paul J.},
        title = "{The mass distribution and gravitational potential of the Milky Way}",
      journal = {\mnras},
         year = 2017,
        month = feb,
       volume = {465},
       number = {1},
        pages = {76-94},
          doi = {10.1093/mnras/stw2759},
archivePrefix = {arXiv},
       eprint = {1608.00971},
 primaryClass = {astro-ph.GA},
       adsurl = {https://ui.adsabs.harvard.edu/abs/2017MNRAS.465...76M}
}

@ARTICLE{Koposov2010,
       author = {{Koposov}, Sergey E. and {Rix}, Hans-Walter and {Hogg}, David W.},
        title = "{Constraining the Milky Way Potential with a Six-Dimensional Phase-Space Map of the GD-1 Stellar Stream}",
      journal = {\apj},
         year = 2010,
        month = mar,
       volume = {712},
       number = {1},
        pages = {260-273},
          doi = {10.1088/0004-637X/712/1/260},
archivePrefix = {arXiv},
       eprint = {0907.1085},
 primaryClass = {astro-ph.GA},
       adsurl = {https://ui.adsabs.harvard.edu/abs/2010ApJ...712..260K}
}

@ARTICLE{Malhan2019,
       author = {{Malhan}, Khyati and {Ibata}, Rodrigo A.},
        title = "{Constraining the Milky Way halo potential with the GD-1 stellar stream}",
      journal = {\mnras},
         year = 2019,
        month = jul,
       volume = {486},
       number = {3},
        pages = {2995-3005},
          doi = {10.1093/mnras/stz1035},
archivePrefix = {arXiv},
       eprint = {1807.05994},
 primaryClass = {astro-ph.GA},
       adsurl = {https://ui.adsabs.harvard.edu/abs/2019MNRAS.486.2995M}
}

@ARTICLE{Bonaca2014,
       author = {{Bonaca}, Ana and {Geha}, Marla and {K{\"u}pper}, Andreas H.~W. and {Diemand}, J{\"u}rg and {Johnston}, Kathryn V. and {Hogg}, David W.},
        title = "{Milky Way Mass and Potential Recovery Using Tidal Streams in a Realistic Halo}",
      journal = {\apj},
         year = 2014,
        month = nov,
       volume = {795},
       number = {1},
          eid = {94},
        pages = {94},
          doi = {10.1088/0004-637X/795/1/94},
archivePrefix = {arXiv},
       eprint = {1406.6063},
 primaryClass = {astro-ph.GA},
       adsurl = {https://ui.adsabs.harvard.edu/abs/2014ApJ...795...94B}
}

@ARTICLE{Bowden2015,
       author = {{Bowden}, A. and {Belokurov}, V. and {Evans}, N.~W.},
        title = "{Dipping our toes in the water: first models of GD-1 as a stream}",
      journal = {\mnras},
         year = 2015,
        month = may,
       volume = {449},
       number = {2},
        pages = {1391-1400},
          doi = {10.1093/mnras/stv285},
archivePrefix = {arXiv},
       eprint = {1502.00484},
 primaryClass = {astro-ph.GA},
       adsurl = {https://ui.adsabs.harvard.edu/abs/2015MNRAS.449.1391B}
}

@ARTICLE{Kuepper2015,
       author = {{K{\"u}pper}, Andreas H.~W. and {Balbinot}, Eduardo and {Bonaca}, Ana and {Johnston}, Kathryn V. and {Hogg}, David W. and {Kroupa}, Pavel and {Santiago}, Basilio X.},
        title = "{Globular Cluster Streams as Galactic High-Precision Scales{\textemdash}the Poster Child Palomar 5}",
      journal = {\apj},
         year = 2015,
        month = apr,
       volume = {803},
       number = {2},
          eid = {80},
        pages = {80},
          doi = {10.1088/0004-637X/803/2/80},
archivePrefix = {arXiv},
       eprint = {1502.02658},
 primaryClass = {astro-ph.GA},
       adsurl = {https://ui.adsabs.harvard.edu/abs/2015ApJ...803...80K}
}

@ARTICLE{Bovy2014,

  author  = {Bovy, Jo},

  title   = {{Dynamical Modeling of Tidal Streams}},

  journal = {ApJ},

  year    = {2014},

  volume  = {795},

  pages   = {95},

  doi     = {10.1088/0004-637X/795/1/95},

  eprint  = {1401.2985}

}

@ARTICLE{Bovy2016,
       author = {{Bovy}, Jo and {Bahmanyar}, Anita and {Fritz}, Tobias K. and {Kallivayalil}, Nitya},
        title = "{The Shape of the Inner Milky Way Halo from Observations of the Pal 5 and GD--1 Stellar Streams}",
      journal = {\apj},
         year = 2016,
        month = dec,
       volume = {833},
       number = {1},
          eid = {31},
        pages = {31},
          doi = {10.3847/1538-4357/833/1/31},
archivePrefix = {arXiv},
       eprint = {1609.01298},
 primaryClass = {astro-ph.GA},
       adsurl = {https://ui.adsabs.harvard.edu/abs/2016ApJ...833...31B}
}

@ARTICLE{Nibauer2022,
       author = {{Nibauer}, Jacob and {Belokurov}, Vasily and {Cranmer}, Miles and {Goodman}, Jeremy and {Ho}, Shirley},
        title = "{Charting Galactic Accelerations with Stellar Streams and Machine Learning}",
      journal = {\apj},
         year = 2022,
        month = nov,
       volume = {940},
       number = {1},
          eid = {22},
        pages = {22},
          doi = {10.3847/1538-4357/ac93ee},
archivePrefix = {arXiv},
       eprint = {2205.11767},
 primaryClass = {astro-ph.GA},
       adsurl = {https://ui.adsabs.harvard.edu/abs/2022ApJ...940...22N}
}

@ARTICLE{NibauerBonaca2025,
       author = {{Nibauer}, Jacob and {Bonaca}, Ana},
        title = "{Galactic Accelerations from the GD-1 Stream Suggest a Tilted Dark Matter Halo}",
      journal = {\apjl},
         year = 2025,
        month = may,
       volume = {985},
       number = {1},
          eid = {L22},
        pages = {L22},
          doi = {10.3847/2041-8213/add0a9},
archivePrefix = {arXiv},
       eprint = {2504.07187},
 primaryClass = {astro-ph.GA},
       adsurl = {https://ui.adsabs.harvard.edu/abs/2025ApJ...985L..22N}
}

@ARTICLE{PriceWhelan2014,
       author = {{Price-Whelan}, Adrian M. and {Hogg}, David W. and {Johnston}, Kathryn V. and {Hendel}, David},
        title = "{Inferring the Gravitational Potential of the Milky Way with a Few Precisely Measured Stars}",
      journal = {\apj},
         year = 2014,
        month = oct,
       volume = {794},
       number = {1},
          eid = {4},
        pages = {4},
          doi = {10.1088/0004-637X/794/1/4},
archivePrefix = {arXiv},
       eprint = {1405.6721},
 primaryClass = {astro-ph.GA},
       adsurl = {https://ui.adsabs.harvard.edu/abs/2014ApJ...794....4P}
}

@ARTICLE{Reino2021,
       author = {{Reino}, Stella and {Rossi}, Elena M. and {Sanderson}, Robyn E. and {Sellentin}, Elena and {Helmi}, Amina and {Koppelman}, Helmer H. and {Sharma}, Sanjib},
        title = "{Galactic potential constraints from clustering in action space of combined stellar stream data}",
      journal = {\mnras},
         year = 2021,
        month = apr,
       volume = {502},
       number = {3},
        pages = {4170-4193},
          doi = {10.1093/mnras/stab304},
archivePrefix = {arXiv},
       eprint = {2007.00356},
 primaryClass = {astro-ph.GA},
       adsurl = {https://ui.adsabs.harvard.edu/abs/2021MNRAS.502.4170R}
}

@ARTICLE{LawMajewski2010,
       author = {{Law}, David R. and {Majewski}, Steven R.},
        title = "{The Sagittarius Dwarf Galaxy: A Model for Evolution in a Triaxial Milky Way Halo}",
      journal = {\apj},
         year = 2010,
        month = may,
       volume = {714},
       number = {1},
        pages = {229-254},
          doi = {10.1088/0004-637X/714/1/229},
archivePrefix = {arXiv},
       eprint = {1003.1132},
 primaryClass = {astro-ph.GA},
       adsurl = {https://ui.adsabs.harvard.edu/abs/2010ApJ...714..229L}
}

@ARTICLE{Vasiliev2021,
       author = {{Vasiliev}, Eugene and {Belokurov}, Vasily and {Erkal}, Denis},
        title = "{Tango for three: Sagittarius, LMC, and the Milky Way}",
      journal = {\mnras},
         year = 2021,
        month = feb,
       volume = {501},
       number = {2},
        pages = {2279-2304},
          doi = {10.1093/mnras/staa3673},
archivePrefix = {arXiv},
       eprint = {2009.10726},
 primaryClass = {astro-ph.GA},
       adsurl = {https://ui.adsabs.harvard.edu/abs/2021MNRAS.501.2279V}
}

@ARTICLE{BonacaHogg2018,
       author = {{Bonaca}, Ana and {Hogg}, David W.},
        title = "{The Information Content in Cold Stellar Streams}",
      journal = {\apj},
         year = 2018,
        month = nov,
       volume = {867},
       number = {2},
          eid = {101},
        pages = {101},
          doi = {10.3847/1538-4357/aae4da},
archivePrefix = {arXiv},
       eprint = {1804.06854},
 primaryClass = {astro-ph.GA},
       adsurl = {https://ui.adsabs.harvard.edu/abs/2018ApJ...867..101B}
}

@ARTICLE{Mateu2023,
       author = {{Mateu}, Cecilia},
        title = "{galstreams: A library of Milky Way stellar stream footprints and tracks}",
      journal = {\mnras},
         year = 2023,
        month = apr,
       volume = {520},
       number = {4},
        pages = {5225-5258},
          doi = {10.1093/mnras/stad321},
archivePrefix = {arXiv},
       eprint = {2204.10326},
 primaryClass = {astro-ph.GA},
       adsurl = {https://ui.adsabs.harvard.edu/abs/2023MNRAS.520.5225M}
}

@ARTICLE{Gibbons2014,
       author = {{Gibbons}, S.~L.~J. and {Belokurov}, V. and {Evans}, N.~W.},
        title = "{`Skinny Milky Way please', says Sagittarius}",
      journal = {\mnras},
         year = 2014,
        month = dec,
       volume = {445},
       number = {4},
        pages = {3788-3802},
          doi = {10.1093/mnras/stu1986},
archivePrefix = {arXiv},
       eprint = {1406.2243},
 primaryClass = {astro-ph.GA},
       adsurl = {https://ui.adsabs.harvard.edu/abs/2014MNRAS.445.3788G}
}

@ARTICLE{Erkal2019,
       author = {{Erkal}, D. and {Belokurov}, V. and {Laporte}, C.~F.~P. and {Koposov}, S.~E. and {Li}, T.~S. and {Grillmair}, C.~J. and {Kallivayalil}, N. and {Price-Whelan}, A.~M. and {Evans}, N.~W. and {Hawkins}, K. and {Hendel}, D. and {Mateu}, C. and {Navarro}, J.~F. and {del Pino}, A. and {Slater}, C.~T. and {Sohn}, S.~T. and {Orphan Aspen Treasury Collaboration}},
        title = "{The total mass of the Large Magellanic Cloud from its perturbation on the Orphan stream}",
      journal = {\mnras},
         year = 2019,
        month = aug,
       volume = {487},
       number = {2},
        pages = {2685-2700},
          doi = {10.1093/mnras/stz1371},
archivePrefix = {arXiv},
       eprint = {1812.08192},
 primaryClass = {astro-ph.GA},
       adsurl = {https://ui.adsabs.harvard.edu/abs/2019MNRAS.487.2685E}
}

@ARTICLE{Jing2002,
       author = {{Jing}, Y.~P. and {Suto}, Yasushi},
        title = "{Triaxial Modeling of Halo Density Profiles with High-Resolution N-Body Simulations}",
      journal = {\apj},
         year = 2002,
        month = aug,
       volume = {574},
       number = {2},
        pages = {538-553},
          doi = {10.1086/341065},
archivePrefix = {arXiv},
       eprint = {astro-ph/0202064},
 primaryClass = {astro-ph},
       adsurl = {https://ui.adsabs.harvard.edu/abs/2002ApJ...574..538J}
}

@ARTICLE{gebhard2024_atmosphere,
  title = {Flow matching for atmospheric retrieval of exoplanets: Where reliability meets adaptive noise levels},
  author = {Gebhard, Timothy D and Wildberger, Jonas and Dax, Maximilian and Kofler, Annalena and Angerhausen, Daniel and Quanz, Sascha P and Sch{\"o}lkopf, Bernhard},
  journal = {Astronomy \& Astrophysics},
  volume = {693},
  pages = {A42},
  year = {2025},
  publisher = {EDP Sciences},
}

@ARTICLE{Tollet2016,
       author = {{Tollet}, Edouard and {Macci{\`o}}, Andrea V. and {Dutton}, Aaron A. and {Stinson}, Greg S. and {Wang}, Liang and {Penzo}, Camilla and {Gutcke}, Thales A. and {Buck}, Tobias and {Kang}, Xi and {Brook}, Chris and {Di Cintio}, Arianna and {Keller}, Ben W. and {Wadsley}, James},
        title = "{NIHAO - IV: core creation and destruction in dark matter density profiles across cosmic time}",
      journal = {\mnras},
         year = 2016,
        month = mar,
       volume = {456},
       number = {4},
        pages = {3542-3552},
          doi = {10.1093/mnras/stv2856},
archivePrefix = {arXiv},
       eprint = {1507.03590},
 primaryClass = {astro-ph.GA},
       adsurl = {https://ui.adsabs.harvard.edu/abs/2016MNRAS.456.3542T}
}

@ARTICLE{McClure-Griffiths_2016,
doi = {10.3847/0004-637X/831/2/124},
url = {https://doi.org/10.3847/0004-637X/831/2/124},
year = {2016},
month = {nov},
publisher = {The American Astronomical Society},
volume = {831},
number = {2},
pages = {124},
author = {McClure-Griffiths, N. M. and Dickey, John M.},
title = {MILKY WAY KINEMATICS. II. A UNIFORM INNER GALAXY H i TERMINAL VELOCITY CURVE},
journal = {The Astrophysical Journal}
}

@ARTICLE{Kuijken1991,
       author = {{Kuijken}, Konrad and {Gilmore}, Gerard},
        title = "{The Galactic Disk Surface Mass Density and the Galactic Force K Z at Z = 1.1 Kiloparsecs}",
      journal = {\apjl},
         year = 1991,
        month = jan,
       volume = {367},
        pages = {L9},
          doi = {10.1086/185920},
       adsurl = {https://ui.adsabs.harvard.edu/abs/1991ApJ...367L...9K}
}

@ARTICLE{Dillamore2025,
       author = {{Dillamore}, Adam M. and {Sanders}, Jason L. and {Belokurov}, Vasily and {Zhang}, Hanyuan},
        title = "{Dynamical streams in the local stellar halo}",
      journal = {\mnras},
         year = 2025,
        month = jul,
       volume = {541},
       number = {1},
        pages = {214-233},
          doi = {10.1093/mnras/staf965},
archivePrefix = {arXiv},
       eprint = {2503.02926},
 primaryClass = {astro-ph.GA},
       adsurl = {https://ui.adsabs.harvard.edu/abs/2025MNRAS.541..214D}
}

@article{kuhmichel2026bayesflow,
  title={{BayesFlow} 2: Multi-backend amortized {B}ayesian inference in Python},
  author={Kühmichel, Lars and Huang, Jerry M and Pratz, Valentin and Arruda, Jonas and Olischläger, Hans and Habermann, Daniel and Kucharsky, Simon and Elsemüller, Lasse and Mishra, Aayush and Bracher, Niels and Jedhoff, Svenja and Schmitt, Marvin and Bürkner, Paul-Christian and Radev, Stefan T},
  journal={arXiv preprint arXiv:2602.07098},
  year={2026}
}

@inproceedings{akiba2019optuna,
  title={{O}ptuna: A Next-Generation Hyperparameter Optimization Framework},
  author={Akiba, Takuya and Sano, Shotaro and Yanase, Toshihiko and Ohta, Takeru and Koyama, Masanori},
  booktitle={The 25th ACM SIGKDD International Conference on Knowledge Discovery \& Data Mining},
  pages={2623--2631},
  year={2019}
}

@ARTICLE{adamw_2017,
       author = {{Loshchilov}, Ilya and {Hutter}, Frank},
        title = "{Decoupled Weight Decay Regularization}",
      journal = {arXiv e-prints},
         year = 2017,
        month = nov,
          eid = {arXiv:1711.05101},
        pages = {arXiv:1711.05101},
          doi = {10.48550/arXiv.1711.05101},
archivePrefix = {arXiv},
       eprint = {1711.05101},
 primaryClass = {cs.LG},
       adsurl = {https://ui.adsabs.harvard.edu/abs/2017arXiv171105101L}
}

@ARTICLE{Salimans_2022,
       author = {{Salimans}, Tim and {Ho}, Jonathan},
        title = "{Progressive Distillation for Fast Sampling of Diffusion Models}",
      journal = {arXiv e-prints},
         year = 2022,
        month = feb,
          eid = {arXiv:2202.00512},
        pages = {arXiv:2202.00512},
          doi = {10.48550/arXiv.2202.00512},
archivePrefix = {arXiv},
       eprint = {2202.00512},
 primaryClass = {cs.LG},
       adsurl = {https://ui.adsabs.harvard.edu/abs/2022arXiv220200512S}
}

@ARTICLE{Sante2026,
       author = {{Sante}, Andrea and {Font}, Andreea S. and {Kawata}, Daisuke and {Makinen}, T. Lucas and {Grand}, Robert J.~J.},
        title = "{A simulation-based inference of the Milky Way merger history}",
      journal = {\mnras},
         year = 2026,
        month = jun,
          doi = {10.1093/mnras/stag1162},
       adsurl = {https://ui.adsabs.harvard.edu/abs/2026MNRAS.tmp.1097S}
}

@ARTICLE{Ruzza2026,
       author = {{Ruzza}, Alessandro and {Lodato}, Giuseppe and {Rosotti}, Giovanni and {Armitage}, Philip and {Facchini}, Stefano and {Andrews}, Sean M. and {Bae}, Jaehan and {Barraza-Alfaro}, Marcelo and {Benisty}, Myriam and {Curone}, Pietro and {Fasano}, Daniele and {Hall}, Cassandra and {Hilder}, Thomas and {Izquierdo}, Andr{\'e}s F. and {Longarini}, Cristiano and {M{\'e}nard}, Fran{\c{c}}ois and {Pinte}, Christophe and {Stadler}, Jochen and {Teague}, Richard and {Terry}, Jason and {Wilner}, David J. and {Winter}, Andrew J. and {Yoshida}, Tomohiro C. and {Zawadzki}, Brianna},
        title = "{exoALMA. XXIII. Estimating Disk and Planet Properties from Dust Morphologies with DBNets 2.0}",
      journal = {\apjl},
         year = 2026,
        month = mar,
       volume = {1000},
       number = {1},
          eid = {L16},
        pages = {L16},
          doi = {10.3847/2041-8213/ae434c},
archivePrefix = {arXiv},
       eprint = {2603.13149},
 primaryClass = {astro-ph.EP},
       adsurl = {https://ui.adsabs.harvard.edu/abs/2026ApJ..1000L..16R}
}

@ARTICLE{CASBI_2024,
       author = {{Viterbo}, Giuseppe and {Buck}, Tobias},
        title = "{CASBI -- Chemical Abundance Simulation-Based Inference for Galactic Archeology}",
      journal = {arXiv e-prints},
         year = 2024,
        month = nov,
          eid = {arXiv:2411.17269},
        pages = {arXiv:2411.17269},
          doi = {10.48550/arXiv.2411.17269},
archivePrefix = {arXiv},
       eprint = {2411.17269},
 primaryClass = {astro-ph.GA},
       adsurl = {https://ui.adsabs.harvard.edu/abs/2024arXiv241117269V}
}

@ARTICLE{Lovell2025,
       author = {{Lovell}, Christopher C. and {Starkenburg}, Tjitske and {Ho}, Matthew and {Angl{\'e}s-Alc{\'a}zar}, Daniel and {Dav{\'e}}, Romeel and {Gabrielpillai}, Austen and {Iyer}, Kartheik G. and {Matthews}, Alice E. and {Roper}, William J. and {Somerville}, Rachel S. and {Sommovigo}, Laura and {Villaescusa-Navarro}, Francisco},
        title = "{Learning the Universe: cosmological and astrophysical parameter inference with galaxy luminosity functions and colours}",
      journal = {\mnras},
         year = 2025,
        month = dec,
       volume = {544},
       number = {4},
        pages = {3949-3979},
          doi = {10.1093/mnras/staf1888},
archivePrefix = {arXiv},
       eprint = {2411.13960},
 primaryClass = {astro-ph.GA},
       adsurl = {https://ui.adsabs.harvard.edu/abs/2025MNRAS.544.3949L}
}

\appendix
\section{Prior and posterior predictive checks}
\label{sec:appendix_predictive}

We show in this appendix, for each stream, the prior and posterior predictive checks in the full Gaia observational space, complementing the summary shown in Fig.~\ref{fig:ppc_combined}. For each stream we show the prior predictive check (top) and the corresponding posterior corner plot (bottom).

\begin{figure}[h]
    \centering
    \includegraphics[width=\linewidth, trim = 0 0 0 50, clip]{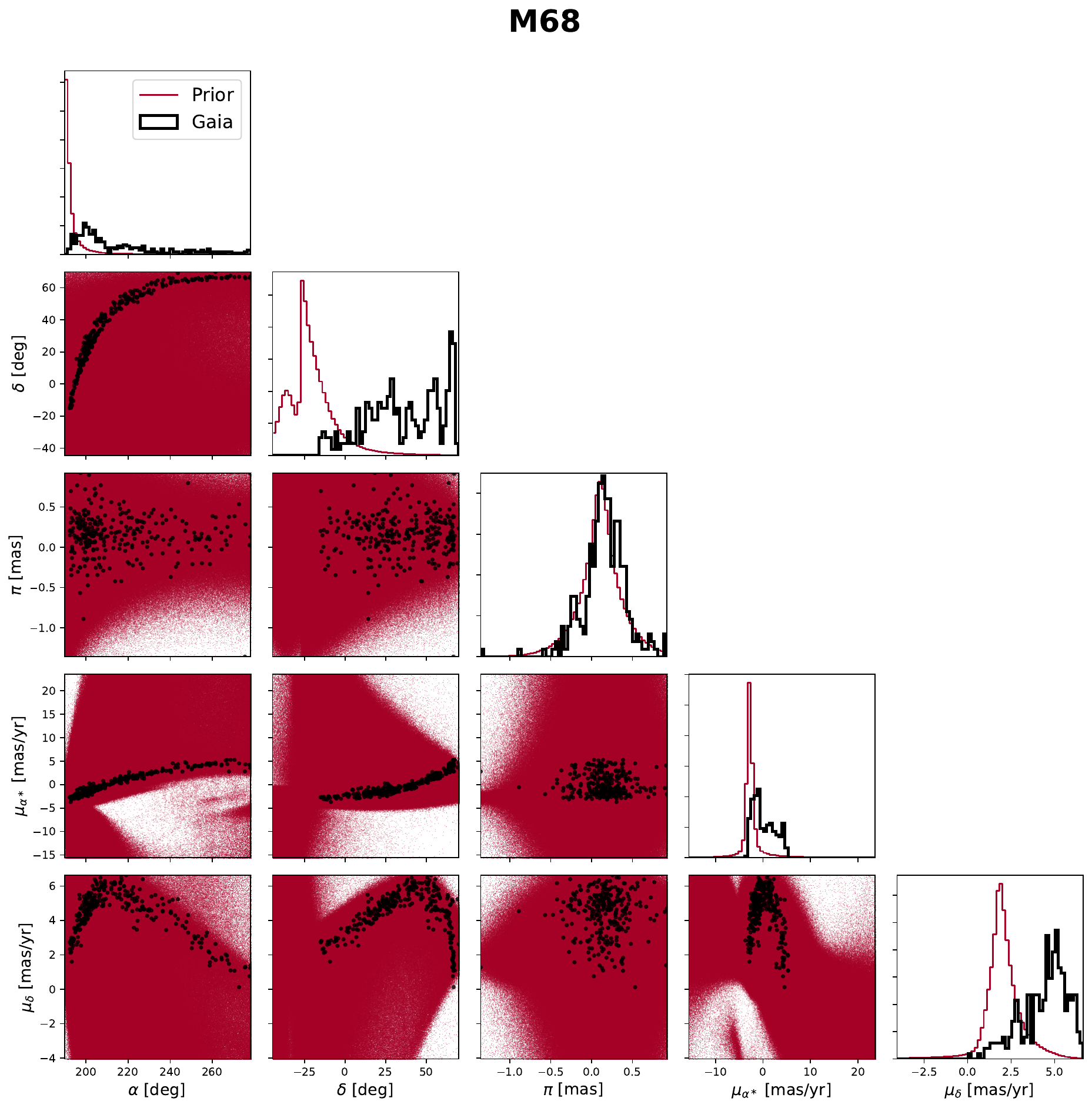}
    \\[1em]
    \includegraphics[width=\linewidth, trim = 0 0 0 50, clip]{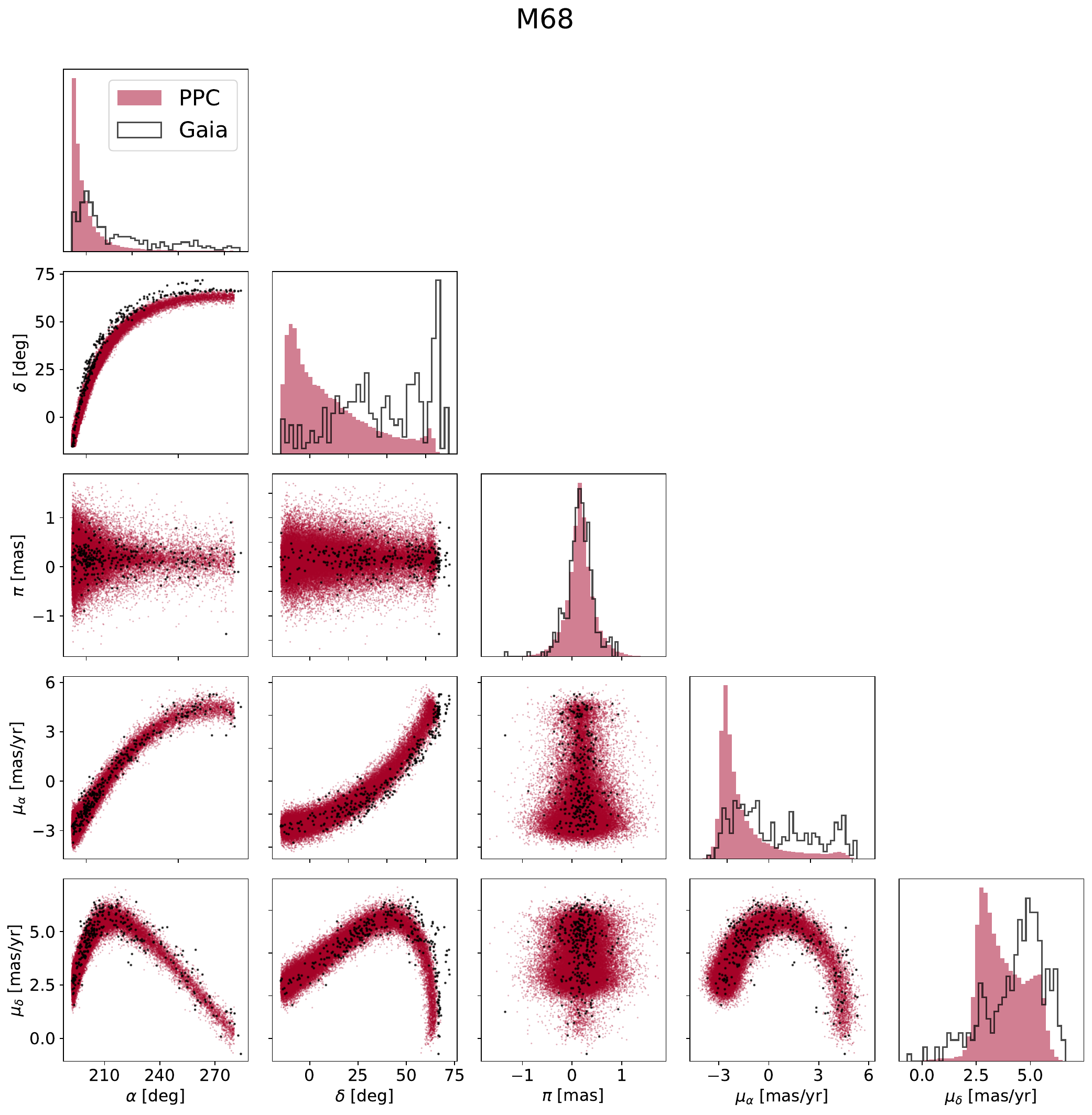}
    \caption{Prior (upper panel) and posterior (lower panel) predictive checks M~68}
    \label{fig:M68_PPC_vs_prior}
\end{figure}

\begin{figure}[h]
    \centering
    \includegraphics[width=\linewidth,  trim = 0 0 0 50, clip]{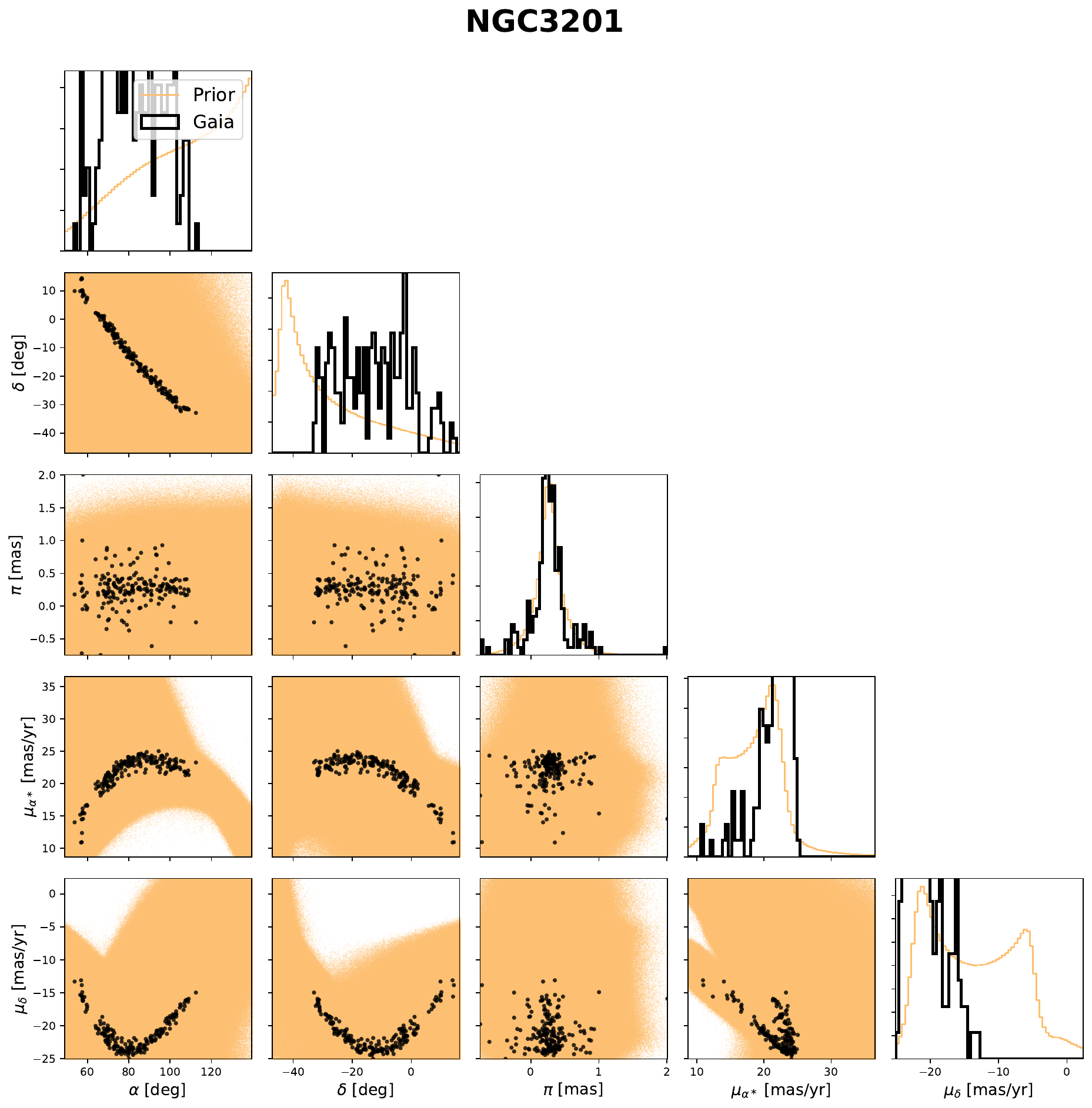}
    \\[1em]
    \includegraphics[width=\linewidth,  trim = 0 0 0 50, clip]{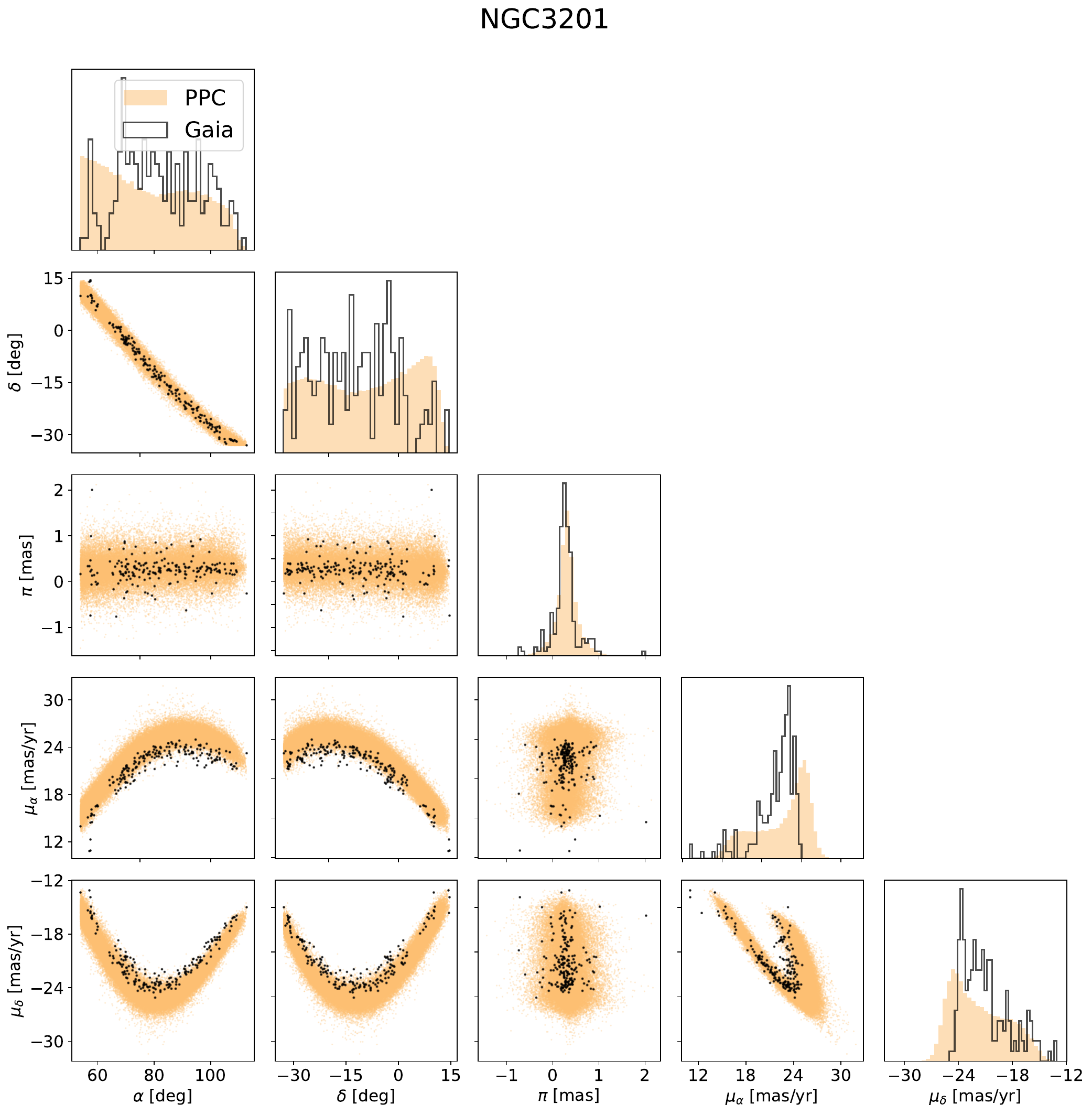}
    \caption{Prior (upper panel) and posterior (lower panel) predictive checks NGC~3201}
    \label{fig:NGC3201_PPC_vs_prior}
\end{figure}

\begin{figure}[h]
    \centering
    \includegraphics[width=\linewidth,  trim = 0 0 0 50, clip]{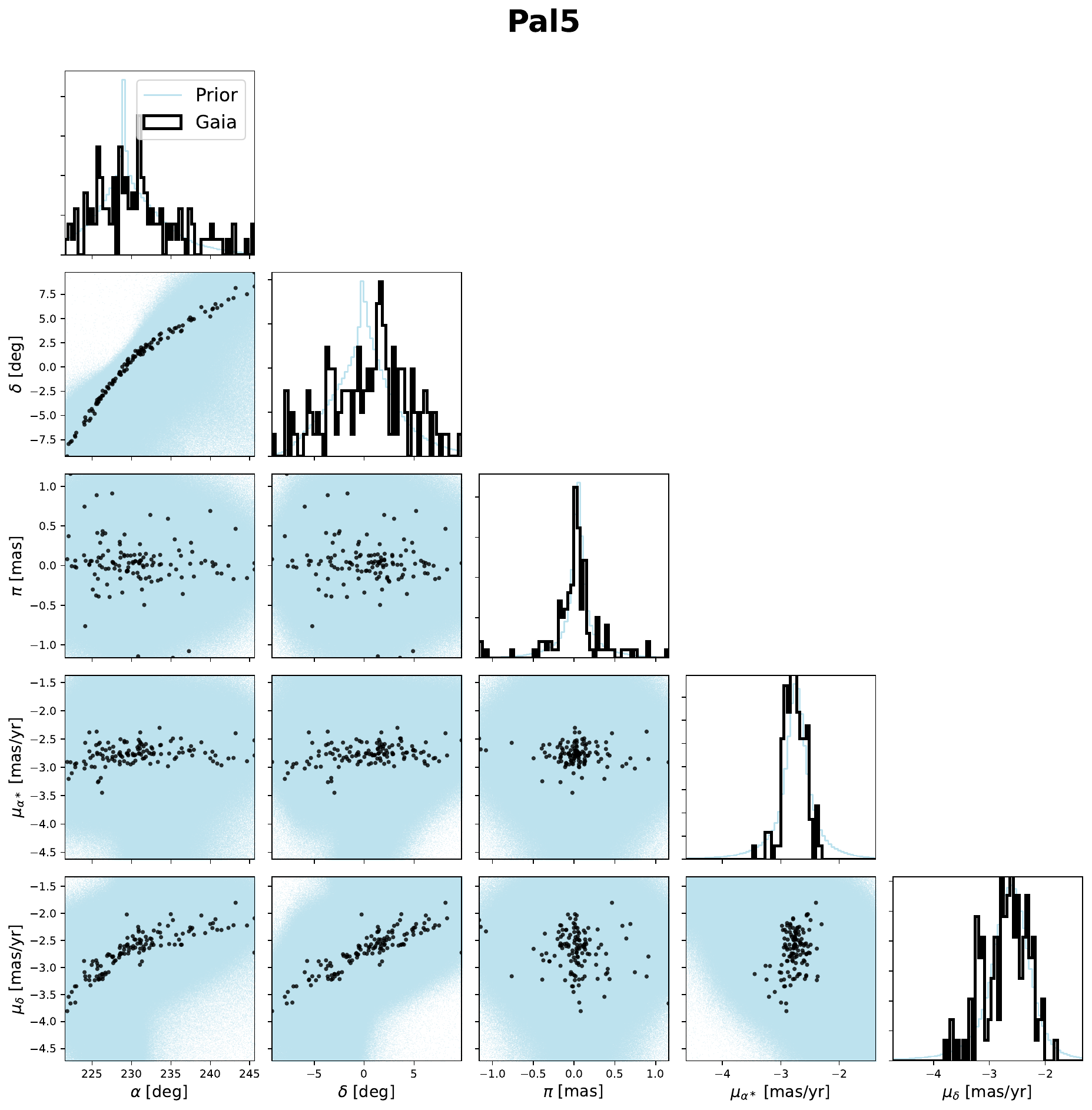}
    \\[1em]
    \includegraphics[width=\linewidth,  trim = 0 0 0 50, clip]{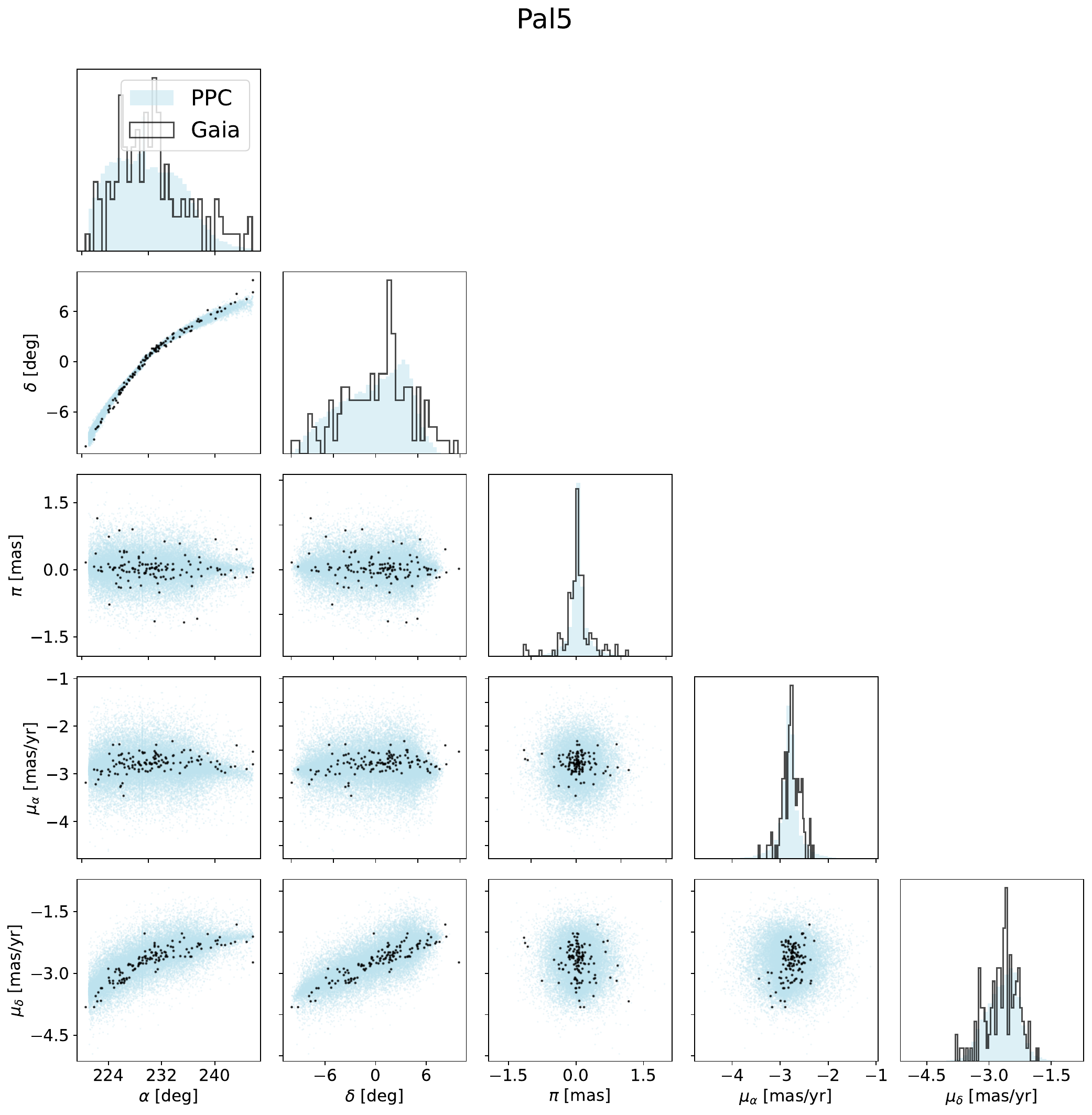}
    \caption{Prior (upper panel) and posterior (lower panel) predictive checks Pal~5}
    \label{fig:Pal5_PPC_vs_prior}
\end{figure}

\clearpage

\section{Local parameter calibration}

In this appendix we report the empirical coverage of the local per-stream parameters.

\begin{figure}[h]
    \centering
    \includegraphics[width=.8\linewidth]{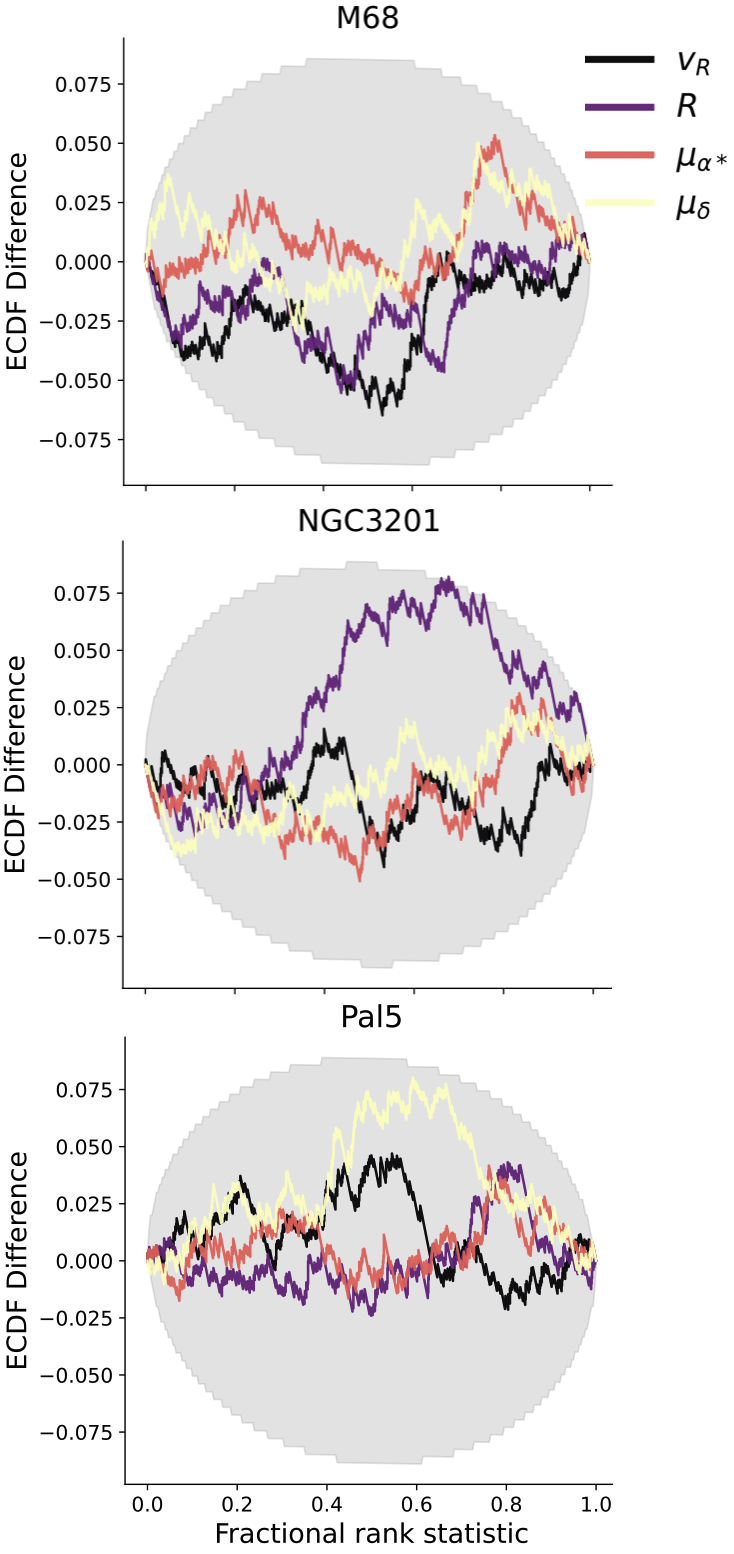}
    \caption{ECDFs of the local parameters
        $\boldsymbol{\theta}_j$, shown per stream as a deviation from
        uniformity. The shaded regions are confidence band.}
    \label{fig:local_PP_plot}
\end{figure}


\end{document}